\documentclass[twoside,twocolumn,11pt]{article}
\usepackage{extsizes}
\usepackage[super,sort&compress,comma]{natbib} 
\usepackage[version=3]{mhchem}
\usepackage[left=1.5cm, right=1.5cm, top=1.785cm, bottom=2.0cm]{geometry}
\usepackage{balance}
\usepackage{times,mathptmx}
\usepackage{sectsty}
\usepackage{graphicx} 

\usepackage{wrapfig}
\usepackage{bm} 
\usepackage{amsmath,amssymb}
\usepackage{subfig}

\usepackage{lastpage}
\usepackage[format=plain,justification=justified,singlelinecheck=false,font={stretch=1.125,small,sf},labelfont=bf,labelsep=space]{caption}
\usepackage{float}
\usepackage{fancyhdr}
\usepackage{fnpos}
\usepackage[english]{babel}
\usepackage{array}
\usepackage{droidsans}
\usepackage{charter}
\usepackage[T1]{fontenc}
\usepackage[usenames,dvipsnames]{xcolor}
\usepackage{setspace}
\usepackage[compact]{titlesec}
\usepackage{hyperref}

\usepackage{lipsum}
\usepackage{widetext}
\usepackage{comment}

\renewcommand{\v}{{\bm v}}

\newcommand{\av}[1]{\left\langle #1 \right\rangle}

\definecolor{cream}{RGB}{222,217,201}

\begin{document}

\pagestyle{fancy}
\thispagestyle{plain}
\fancypagestyle{plain}{
\renewcommand{\headrulewidth}{0pt}
}

\makeFNbottom
\makeatletter
\renewcommand\LARGE{\@setfontsize\LARGE{15pt}{17}}
\renewcommand\Large{\@setfontsize\Large{12pt}{14}}
\renewcommand\large{\@setfontsize\large{10pt}{12}}
\renewcommand\footnotesize{\@setfontsize\footnotesize{7pt}{10}}
\makeatother

\renewcommand{\thefootnote}{\fnsymbol{footnote}}
\renewcommand\footnoterule{\vspace*{1pt}%
\color{cream}\hrule width 3.5in height 0.4pt \color{black}\vspace*{5pt}} 
\setcounter{secnumdepth}{5}

\makeatletter 
\renewcommand\@biblabel[1]{#1}            
\renewcommand\@makefntext[1]%
{\noindent\makebox[0pt][r]{\@thefnmark\,}#1}
\makeatother 
\renewcommand{\figurename}{\small{Fig.}~}
\sectionfont{\sffamily\Large}
\subsectionfont{\normalsize}
\subsubsectionfont{\bf}
\setstretch{1.125} 
\setlength{\skip\footins}{0.8cm}
\setlength{\footnotesep}{0.25cm}
\setlength{\jot}{10pt}
\titlespacing*{\section}{0pt}{4pt}{4pt}
\titlespacing*{\subsection}{0pt}{15pt}{1pt}

\fancyfoot{}
\fancyfoot[LO,RE]{\vspace{-7.1pt}\includegraphics[height=9pt]{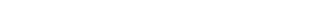}}
\fancyfoot[RO]{\footnotesize{\sffamily{1--\pageref{LastPage} ~\textbar  \hspace{2pt}\thepage}}}
\fancyfoot[LE]{\footnotesize{\sffamily{\thepage~\textbar\hspace{2pt} 1--\pageref{LastPage}}}}
\fancyhead{}
\renewcommand{\headrulewidth}{0pt} 
\renewcommand{\footrulewidth}{0pt}
\setlength{\arrayrulewidth}{1pt}
\setlength{\columnsep}{6.5mm}
\setlength\bibsep{1pt}

\makeatletter 
\newlength{\figrulesep} 
\setlength{\figrulesep}{0.5\textfloatsep} 

\newcommand{\topfigrule}{\vspace*{-1pt}%
\noindent{\color{cream}\rule[-\figrulesep]{\columnwidth}{1.5pt}} }

\newcommand{\botfigrule}{\vspace*{-2pt}%
\noindent{\color{cream}\rule[\figrulesep]{\columnwidth}{1.5pt}} }

\newcommand{\dblfigrule}{\vspace*{-1pt}%
\noindent{\color{cream}\rule[-\figrulesep]{\textwidth}{1.5pt}} }

\makeatother

\twocolumn[
\vspace{1em}
\sffamily
\begin{tabular}{m{0cm} p{17.5cm} }

 & \noindent\LARGE{\textbf{Quasi-two-dimensional dispersions of Brownian particles with competitive interactions: Dynamical clustering, non-Gaussianity and hydrodynamic correlations}} \\
\vspace{0.3cm} & \vspace{0.3cm} \\

& \noindent\large{Zihan Tan,$^{\ast}$\textit{$^{abc}$}, Vania Calandrini,\textit{$^{c}$} Jan K. G. Dhont\textit{$^{bd}$} and Gerhard N\"agele\textit{$^{bd}$}} \\
\vspace{0.3cm} & \vspace{0.3cm} \\
 & \noindent\normalsize{\textit{}
	
} \\

\end{tabular}


  ]

\renewcommand*\rmdefault{bch}\normalfont\upshape
\rmfamily
\section*{}
\vspace{-1cm}

\footnotetext{\textit{$^{a}$~Fachgruppe of Theoretische Physik, Institut für Physik und Astronomie,
		 Technische Universit\"at Berlin, Hardenbergstr$\beta$e 36, 10623 Berlin, Germany; E-mail: zihan.tan@tu-berlin.de}}
\footnotetext{\textit{$^{b}$~Biomacromolecular Systems and Processes, Institute of Biological Information Processing, Forschungszentrum J\"ulich, 52428 J\"ulich, Germany}}
\footnotetext{\textit{$^{c}$~Computational Biomedicine, Institute for Advanced Simulation, Forschungszentrum J\"ulich, 52428 J\"ulich, Germany}}
\footnotetext{\textit{$^{d}$~Department of Physics, Heinrich-Heine Universit\"at D\"usseldorf, D-40225 D\"usseldorf, Germany}}




\section*{Abstract}
\textit{\fontsize{9}{9}\selectfont We conduct a comprehensive dynamical analysis of quasi-two-dimensional (Q2D) dispersions of Brownian particles with competing short-range attractive (SA) and long-range repulsive (LR) interactions using Langevin dynamics (LD) and multiparticle collision dynamics (MPC). As the attractive interaction is strengthened, self-diffusion is significantly suppressed, and clustering gives rise to pronounced subdiffusive behavior. We find that cluster lifetimes are influenced more strongly by attraction strength than by particle concentration. Two dynamical criteria for the transition from non-clustered to clustered phases are identified in terms of the mean cluster lifetime and the relaxation time of local hexagonal order, respectively. Moreover, clustered Q2D-SALR systems exhibit pronounced non-Gaussian dynamics. In particular, the self-van Hove function in the equilibrium-cluster phase displays an approximately exponential form, consistent with an underlying diffusing-diffusivity mechanism. Importantly, MPC simulations reveal the critical role of hydrodynamic interactions (HIs) in collective dynamics. We observe that the anomalously enhanced large-scale collective diffusion characteristic of hydrodynamically interacting Q2D systems is qualitatively preserved in Q2D-SALR dispersions. However, this enhancement suppresses the intermediate-range-order peak in the hydrodynamic function compared to its three-dimensional counterpart. Furthermore, by analyzing the time-dependent evolution of hydrodynamic function and the sound mode in hydrodynamic correlations, we find that clustering in Q2D-SALR systems leads to an earlier onset of HIs than in Q2D hard-sphere reference systems, implying HIs become relevant already on inertial timescales.}
\section{Introduction}
\label{sec:intro}

Brownian particles interacting via competing short-range attraction (SA) and long-range repulsion (LR), commonly referred to as SALR systems, have attracted sustained attention over the past two decades~\cite{yliu:2019,ruiz:2021}. The competition between attraction and repulsion across separated length scales leads to the formation of finite-sized clusters and other microphase structures~\cite{strad:2004,tan04,Destainville2018,wasnik:2015,sieber:2007,gurry:2009}. These systems exhibit rich equilibrium and nonequilibrium phase behavior, with clustering emerging as a hallmark feature.

Structural transformations in SALR systems is often accompanied by nontrivial dynamics, including cluster formation, restructuring, and particle exchange. Alongside extensive investigations of static properties in both three-dimensional (3D) and (quasi-)two-dimensional (Q2D) systems~\cite{godfrin:2014,tan04, zhuang:2016,charbonneau:2016,Litniewski:2025,archer:2008pre, Imperio:2004,Imperio:2006}, considerable effort has been devoted to understanding their dynamical behavior~\cite{sciortino:2004,camp:2005prl,klix:2010, schw:2016}. Dynamic studies are essential not only to complement structural analyses, but also to elucidate how microphase separation evolves in time and how emerging microstructures influence transport properties, including short- and long-time diffusion and rheological response~\cite{barhoum:2010,yearley:2013,godfrin:2015,year:2014,godf:2016}.

Depending on the characteristic cluster lifetime $\tau_b$, clusters are typically classified as transient, dynamic, or permanent~\cite{porcar:2010,yliu:2011,godf:2016,nawrocki:2017,das2018clustering,godfrin:2018,perdomo:2022}. When $\tau_b \ll \tau_D$, where $\tau_D$ denotes the time for a particle to diffuse over its diameter, clusters are transient. For $\tau_b \gtrsim \tau_D$, dynamic clusters emerge with finite but persistent lifetimes, while permanent clusters survive over the experimental or numerical observation window. Time-scale dependent clustering has been reported for a variety of SALR protein systems, including Keggin-type heteropolyanions~\cite{bera:2016}, villin headpiece proteins~\cite{nawrocki:2017}, insulin~\cite{erlkamp:2014}, bovine serum albumin, and $\gamma$-globulin proteins~\cite{balbo:2013}.

Despite this progress, the role of hydrodynamic interactions (HIs) in SALR clustering has only been addressed in a limited number of three-dimensional (3D) studies~\cite{porcar:2010,yliu:2011,varga:2016,godf:2016,das2018clustering,godfrin:2018}. These works indicate that the coupling between electrostatic long-range repulsion and the intrinsically long-ranged HIs can significantly influence cluster stability and collective transport. However, a systematic understanding of this interplay remains incomplete. Beyond average transport coefficients, clustering in SALR systems is expected to induce pronounced dynamic heterogeneity~\cite{Kob:1999,Reichhardt:2003,SchnyderDullens:2017}. Particles residing within clusters experience restricted mobility, while particles in the dilute phase remain comparatively mobile~\cite{KaurDas:2003}. Such coexistence of distinct dynamical environments leads to deviations from Gaussian displacement statistics, even when the mean-squared displacement (MSD) appears diffusive. The non-Gaussian parameter therefore provides a sensitive probe of heterogeneous dynamics and intermittent particle trapping. In confined Q2D systems, where hydrodynamic interactions generate long-range correlated motion, the interplay between clustering and hydrodynamics might further modify these deviations from Gaussian behavior. Understanding how non-Gaussianity evolves across structural phases (states) thus offers complementary insight into the microscopic mechanisms underlying dynamic clustering.

Our recent work~\cite{tan04} examined the structural properties of quasi-two-dimensional (Q2D) SALR systems, where particles are confined to a monolayer. We identified four distinct equilibrium and nonequilibrium phases and proposed empirical structural criteria, such as the intermediate-range order (IRO) peak, to characterize the transition between dispersed and clustered states. Compared to their 3D counterparts, Q2D systems exhibit both qualitative similarities and notable differences arising from confinement. Some of these observations are further corroborated by a recent study of a colloidal monolayer with depletant-mediated attraction and dipolar repulsion~\cite{Yeh:2025}.

Hydrodynamics in Q2D geometries introduces additional complexity and is directly relevant to protein diffusion in cellular membranes. A prominent example is lateral diffusion during G-protein-coupled receptor (GPCR) signaling, where a $G_\alpha$ protein, released upon receptor activation, diffuses along the membrane while remaining membrane-associated and triggers downstream biochemical cascades~\cite{hilger2018}. Membrane hydrodynamics is commonly described by the Saffman–Delbr\"uck (S-D) model~\cite{Saffman1975,ramadurai2009}, which treats the membrane as a thin viscous 2D incompressible fluid layer embedded in a bulk solvent. However, recent large-scale simulations indicate that lipid composition can shift diffusion behavior from S-D-like to effectively Stokesian~\cite{hu:2025pnas,Vattulainen:2017}. Even in generic Q2D colloidal dispersions without competing interactions, the long-wavelength collective diffusion is anomalously enhanced due to hydrodynamic coupling mediated by the surrounding three-dimensional fluid~\cite{naegele2000wall,naegele2001jcp,naegele2002,bleibel2017onset,pelaez2018hydrodynamic,Panzuela18,Alice:2024,alice:2025,Chamorro-Burgos:2026}. This enhancement manifests as a divergence of the wavenumber-dependent collective diffusion coefficient $D(q)$ and the hydrodynamic function $H(q)$ as $q \to 0$. The physical origin lies in the coupling between out-of-plane fluid momentum and in-plane density fluctuations, leading to an effective osmotic compressibility despite the incompressibility of the bulk solvent. This anomalous behavior occurs for both rigid and soft confinement~\cite{panzuela2017collective} and contrasts with classical membrane models such as Saffman's theory~\cite{Saffman1975}, which assume strictly in-plane incompressibility and therefore predict no such divergence.

Most theoretical descriptions of Q2D collective diffusion assume instantaneous HIs. At very short times, however, fluid inertia and memory effects become relevant, even in the single-particle case~\cite{franosch2011resonances,winkler2011backtracking}. The time-dependent hydrodynamic correlations can be split into a transverse momentum-diffusion (shear) mode and a longitudinal density-coupled (sound) mode~\cite{hansen:2013,ladd1995pre,ladd1995prl,bakker2002role}. The former is associated with vorticity diffusion and viscous momentum transport, whereas the latter corresponds to propagating pressure fluctuations~\cite{hansen:2013}. Domínguez and co-workers identified two characteristic length scales governing this evolution: a hydrodynamic scale below which HIs are negligible and a crossover scale beyond which inertial retardation dominates~\cite{dominguez2014signature}. How such time-dependent hydrodynamics interacts with SALR-induced microstructures remains largely unexplored. In Q2D-SALR systems, the interplay between competing interactions, confinement-induced anomalous collective diffusion, and hydrodynamic retardation raises several open questions. How does clustering modify the divergence of $H(q)$ at small $q$? Does the intermediate-range order peak observed in 3D systems leave a fingerprint in the Q2D hydrodynamic function? How do short-time inertial effects influence the buildup of collective correlations in clustered phases?

To address these questions, we extend our previous structural study of Q2D-SALR systems to their dynamical behavior. We employ Langevin dynamics (LD), where HIs are absent, and multiparticle collision dynamics (MPC), which resolves fluctuating hydrodynamics and captures both inertial and overdamped regimes. In particular, we analyze self-diffusion, non-Gaussianity, cluster lifetimes, hexagonal order relaxation, and the spatio-temporal development of the shear (transversal model) and sound (longitudinal mode) hydrodynamic correlations~\cite{hansen:2013}. Through this combined structural and dynamical perspective, we aim to clarify how competing interactions and hydrodynamic coupling jointly shape clustering and collective transport in confined SALR dispersions.

This paper is structured as follows. Section~\ref{sec:methods} describes the numerical methods, namely LD, MPC, and how they are implemented for the Q2D-SALR system.  Subsequently, section~\ref{sec:results} presents our main findings. We begin by revisiting the discussion of the generalized phase behavior and thermalization. The dynamical properties are then analyzed through the short- and long-time self-diffusion coefficients. Then, we classify the cluster dynamics by analyzing the cluster lifetime function, the time-dependent hexagonal order parameter, and the non-Gaussianness. Next, we investigate the hydrodynamically induced anomalous collective diffusion at short times and the resulting divergence of hydrodynamic function $H(q)$. To this end, we examine the spatio-temporal evolution of HIs via a time-dependent generalization of the hydrodynamic function $H(q,t)$, where $H(q) = H(q,\infty)$. Particular emphasis is placed on the distinct component, $H_d(q,t)$, which captures spatio-temporal cross-correlations between SALR particles. In addition, we elucidate the role of sound wave propagation in concentrated Q2D-SALR systems. Finally, section~\ref{sec:sum+conc} summarizes and concludes our findings.
\section{Theoretical background and simulation methods}
\label{sec:methods}
We hereby outline the employed Q2D-SALR model system of Brownian spheres and the analysis tools used to characterize the simulation data. We skip the Langevin dynamics (LD) scheme as it is detailed in Ref.~\cite{tan04}.

\subsection{Q2D-SALR Brownian spheres}
\label{subsec:SALR-Potential}
We consider $N_p$ identical spherical particles of mass $M$ and diameter $\sigma$, suspended in a three-dimensional Newtonian solvent. The solvent remains fully 3D, while particle motions are confined to be planar~\cite{tan04}. The considered SALR pair potential $u(r)$ combines a steep generalized Lennard–Jones attraction with a screened electrostatic repulsion~\cite{char:2007,mani:2014,das2018clustering}:
\begin{equation}
	u(r)=
	4\epsilon \left[ \left( \frac{\sigma}{r} \right)^{100}
	-\left( \frac{\sigma}{r} \right)^{50} \right]
	+ k_B T \ell_B Z_{\text{eff}}^2 \frac{e^{-r/\lambda_D}}{r}.
	\label{eq:ljy}
\end{equation}
Here, the center-to-center distance $r=r_{ij}=|{\bm R_i-\bm R_j}|$ between particle $i$ and $j$ is restricted to in-plane, with $\bm R_i$ denoting the position vector of particle $i$. The first term provides a hard-sphere-like core and a narrow attractive well of depth $\epsilon$, while the second term represents a Yukawa-type screened Coulomb repulsion, characterized by the Bjerrum length $\ell_B$, effective valency $Z_{\text{eff}}$, and Debye screening length $\lambda_D$. The potential is truncated and shifted at $5\sigma$.

In this work, the attraction strength is varied in the range $\epsilon^\ast=\epsilon/k_BT \in [2,20]$, corresponding to a dimensionless temperature
\begin{equation}
	T^\ast = \frac{k_BT}{\epsilon},
\end{equation}
between $0.05$ and $0.5$. The reduced screening length and electrostatic strength are fixed to $\lambda_D/\sigma = 1.794$ and $\ell_B Z_{\text{eff}}^2/\sigma = 3.588$, respectively. For colloidal particles of diameter $\sigma \approx 100\,\mathrm{nm}$, the chosen $\lambda_D/\sigma$ corresponds approximately to a dilute aqueous 1:1 electrolyte of concentration $\sim 3\,\mu\mathrm{M}$~\cite{mani:2014,das2018clustering}.
\subsection{Time evolution of bond correlation and local hexagonal order}
The important quantity for characterizing the bond and cluster dynamics is the bond correlation function
\begin{align}
	B(t)=\frac{\left<\sum_{i<j}b_{ij}(t)b_{ij}(0)\right>}{N_c(0)}.
\end{align}
Here $b_{ij}(t)$ is $1$ if two particles remain bonded, i.e. $r_{ij}<x_0$ with $x_0$ the local maximum of the potential in Eq.~\eqref{eq:ljy}, and 0 otherwise, and $N_c(0)=\left<\sum_{i<j}b_{ij}(0)\right>$ is the number of bonds at $t=0$ such that $B(0)=1$. Therefore, the correlation function $B(t)$ gives the fraction of particle-pair bonds present at $t = 0$ that remain intact after the lag time $t$. This correlation function is typically used to characterize the attraction driven dynamic arrest in colloidal systems~\cite{zacca:2005}. 

The temporal persistence of local hexagonal ordering is quantified, e.g., by the normalized time correlation function, $g_6(t)$~\cite{Zahn:2000,Haghgooie:2005,Lin:2006,Kelleher:2017,alice:2018}, of the local hexagonal order parameter $q_6(t)$ defined by 
\begin{align}
	g_6(t)= \text{Re}\Big< \frac{q_6(t)\cdot q_6^\ast(0)}{|q_6(0)|^2} \Big>
	\label{eq:cq6q6}
\end{align}
where
\begin{align}
	q_6(t)= \frac{1}{6} \sum_j \exp\{i 6\alpha_j(t)\}
\end{align}
is the complex-valued amplitude of the local hexagonal ordering parameter, $|q_6|^2$, of a considered (central) particle, with the sum extending over its six nearest neighbors. Moreover, $\alpha_j$ is the angle between the vector pointing from the central particle to a neighboring particle $j$, and an arbitrary axis stretching out from the center of the neighboring particle~\cite{halpe:1978,Nelson1979,bialk:2012,zottl2014,theers2018clustering,Pasupalak:2020}. The bracket indicates an average over all particles in the simulation box and over the time origin.
In a crystal, $g_6(t)$ decays to a non-zero constant, whereas in a two-dimensional hexatic or isotropic liquid phase, respectively, an algebraic or exponential decay is observed. 
\subsection{Non-Gaussian parameters}
At colloidal (Brownian) short and long-times, respectively, the fluid-state MSD is linear in time which is referred to as Fickean diffusion behavior. As discussed by Granick et al. in \cite{Granick:2012}, a Brownian system can reveal Fickean diffusion behavior (i.e., a linear time dependence of MSD) yet showing non-Gaussian behavior. Deviations from the Gaussian form of the \textit{self-van Hove function} $G(r,t)$ ($r=|{\bm R}|$, cf. Appendix) or the in $q^2$ exponential form of its Fourier transform signal, e.g., correlated inter-particle motions with location of particles on many different length scales or the presence of multiple relaxation times. Non-Gaussian deviations are indicative, e.g., of dynamic heterogeneity, i.e. of a heterogeneous distribution of regions with particularly mobile (relative to a Gaussian distribution) and relatively immobile particles that pertain to have different relaxation times \cite{KaurDas:2003,Kob:1999,Reichhardt:2003,SchnyderDullens:2017}. The mobile particles are not randomly distributed but form spatially correlated groups (clusters) which move typically along stringlike paths. Dynamic heterogeneity is observed both in three-dimensional supercooled fluids \cite{Kob:1999} and two-dimensional systems near to an order-disorder transition \cite{Reichhardt:2003,SchnyderDullens:2017}.   

To quantify dynamic heterogeneity arising from clustering in Q2D-SALR systems, we analyze the respective non-Gaussianity. The so-called standard time-dependent non-Gaussian factors (see Appendix for details), 
\begin{align}
	\alpha_n^{(d)}(t) = \frac{\langle r^{2n}(t)\rangle_d}{c_n^{(d)}\;\!\left(\langle r^{2}(t)\rangle_d\right)^n}-1\,,\;\; n=2,3,...
\end{align}
which measure successive deviations of the self part of the van Hove function $G(r,t)$ from Gaussian behavior. For a purely Gaussian process, the even moments obey $\langle r^{2n}(t)\rangle_d = c_n^{(d)} \left(\langle r^{2}(t)\rangle_d\right)^n$, such that all $\alpha_n^{(d)}(t)=0$. For isotropic systems, all odd vectorial displacement moments vanish, irrespective of Gaussianity.

The values for the coefficients $c_n^{(d)}$ depend on the spatial dimension and read for $d=2$,
\begin{align}
	c_n^{(d=2)}& = \frac{(2n)!!}{2^n}=n!
\end{align}
so that $c_2^{(d=2)}=2$. Thus the leading (2nd order) standard non-Gaussian factor is given in two dimensions by (with $\langle ...\rangle_2=\langle ...\rangle$ for short)
\begin{align}
	\alpha_2^{(d=2)}(t) = \frac{\langle r^{4}(t)\rangle}{2\;\!\left(\langle r^{2}(t)\rangle\right)^2}-1 \geq -1/2\,,
	\label{eq:alpa_2}
\end{align}
which relates to the ratio of the mean-quartic displacement and mean-squared displacement (MSD). According to Fuchs et al. in \cite{Fuchs:1998}, $\alpha_2(t)<0$ means that the probability for the tagged particle to move very far apart during time $t$ is suppressed relative to the one for a Gaussian process.
\subsection{Short-time collective diffusion and hydrodynamic function}
The short-time ($\tau_h \ll t \ll \tau_D$) collective dynamics can be probed experimentally via the dynamic structure factor $S(q,t)$ using dynamic light scattering (DLS) or neutron spin-echo (NSE) spectroscopy. Here, $\tau_D = R^2/d_0$ is the single-particle diffusion time ($d_0$: Stokes–Einstein diffusion coefficient. $R$: particle radius.), and $\tau_h$ denotes the time beyond which hydrodynamic interactions (HIs) can be treated as instantaneous. For monodispersed isotropic systems, the short-time decay of $S(q,t)$ is exponential~\cite{hansen:2013},
\begin{equation}
	\begin{aligned}
		S(q,t\ll\tau_D) &=\frac{1}{N_p}\left< \sum_{i=1}^{N_p} \sum_{j=1}^{N_p} \exp \left[i\bm q \cdot (\bm R_i(t)-\bm R_j(0))\right] \right>\\
		&=S(q)\exp\left(-q^2D(q)t\right),
		\label{sqt}
	\end{aligned}
\end{equation}
with collective diffusion coefficient
\begin{equation}
	D(q)=\frac{d_0H(q)}{S(q)}.
	\label{Dq}
\end{equation}
Here, $S(q)$ denotes the static structure factor and $H(q)$ the hydrodynamic function characterizing the influence of HIs, and $\bm R_i(t)$ the position of the $i^\mathrm{th}$ Brownian particle at time $t$. In the case that HIs are excluded, $H(q)=1$. Eqs.~\eqref{sqt} and~\eqref{Dq} allow us to directly evaluate $S(q)$, $D(q)$, and $H(q)$ from the simulations.

Formally, $H(q)$ is expressed as the hydrodynamic mobility tensor $D_{ij}$ in its reciprocal space,
\begin{equation}
	\begin{aligned}
		H(q)=\frac{k_BT}{N_pd_0q^2}\left<\sum_{i,j=1}^{N_p}\bm{q}D^{-1}_{ij}\bm{q}\exp\left\{i\bm{q}\cdot(\bm{R}_i-\bm{R}_j)\right\}\right>.
	\end{aligned}
	\label{Hq}
\end{equation}
In particular for quasi-two-dimensional (Q2D) colloidal systems, N\"agele~\cite{naegele2002} analyzed the role of direct interaction at low area fractions and analytically derived a striking feature, namely the divergence of the collective diffusion coefficient $D(q)$ and the hydrodynamic function $H(q)$ as $q \to 0$. A point force approximation, i.e. the Oseen tensor, was used for computing the hydrodynamic tensor $\bm{D}$. The corresponding $H(q)$ in Q2D system is therefore given by~\cite{naegele2002}
\begin{equation}
	\begin{aligned}
		H(q)=&1+\frac{3}{2}\frac{\phi_{2D}}{qR}+\frac{3}{2}\frac{\phi_{2D}}{R}\,\\
		&\int^{\infty}_0\text{d}r r\left[g(r)-1\right]\left[2J_0(qr)-\frac{J_1(qr)}{qr}\right],
	\end{aligned}
	\label{Q2DHq}
\end{equation}
where $J_n$ denotes the Bessel function of the first kind and $g(r)$ the radial distribution function. The term $\sim 1/(qR)$ produces a first-order pole at $q\to 0$, implying a divergence of $H(q)$ and thus of $D(q)$. For $q\sigma<1$, Eq.~\eqref{Q2DHq} reduces approximately to~\cite{banchio1999diffusion}
\begin{equation}
	\begin{aligned}
		H(q)\approx&1+\frac{3}{2}\frac{\phi_{2D}}{qR}+\frac{9}{4}\frac{\phi_{2D}}{R}\int^{\infty}_0\text{d}r \left[g(r)-1\right].
	\end{aligned}
	\label{Q2DHqS}
\end{equation}
Eq.~\eqref{Q2DHqS} postulates that at large (length) scales, where direct SALR forces become subdominant, density fluctuations homogenize much faster than normal single-particle diffusion (cf. Eq.~\eqref{Dq}). The resulting long-range hydrodynamic correlations generate the $q^{-1}$ divergence of $H(q)$ and consequently of $D(q)$.
\subsection{Time-dependent hydrodynamic function $H(q,t)$ and longitudinal current-current correlation function $J_d(q,t)$}
To gain more insights on how the HIs build up at short times, we calculate time-dependent hydrodynamic function $H(q,t)$. The collective dynamics in the regime $t<\tau_h$ under Q2D confinement, where inertia is still relevant, has only recently received attention~\cite{dominguez2014signature,panzuela2017collective}. Following Ladd and co-workers~\cite{ladd1995pre,ladd1995prl}, $H(q,t)$ is defined as
\begin{equation}
	\begin{aligned}
		H(q,t)=&\frac{1}{2N_pD_0q^2t}\left< \sum_{i=1}^{N_p}\sum_{j=1}^{N_p}\bm{q}\cdot\left[\bm{R}_i(t)-\bm{R}_i(0)\right] \right.\\
		&\left. \vphantom{\sum_{j=1}^{N_p}}\times \left[\bm{R}_j(t)-\bm{R}_j(0)\right]\cdot\bm{q}\exp\left[ i\bm{q}\cdot\left(\bm{R}_i-\bm{R}_j\right)\right]\right>.
	\end{aligned}
	\label{eq:Hqt}
\end{equation}

Statistically, it is the particle displacement correlations in Fourier space, normalized by an isolated Brownian particle diffusion constant. 
Furthermore, the hydrodynamic function can be split into parts containing self-correlations and cross-correlations, i.e.,
\begin{equation}
	\begin{aligned}
		H(q,t)=&H_s(q,t)+H_d(q,t).
	\end{aligned}
	\label{eq:Hsdqt}
\end{equation}
$H_s(q,t)$ is directly associated with the MSD rescaled by the self-diffusion coefficient $D_0$. $H_d(q,t)$, whose explicit expression is Eq.~\eqref{eq:Hqt} when $i\neq j$, characterizes the HIs from the other particles. Note that when $t\to\infty$, $H(q,t)$ is reduced to $H(q)$. 

Note that the distinct hydrodynamic function $H_d(q,t)$ is related to a time-integrated velocity cross-correlation and therefore primarily probes the buildup of transverse momentum correlations (shear modes)~\cite{hansen:2013}.

To take a closer look at the sound wave propagation in concentrated suspensions, we probe the longitudinal collective mode, the longitudinal current-current correlation function $J(q,t)$, which is defined as~\cite{ladd1995pre,ladd1995prl,bakker2002role}
\begin{equation}
	\begin{aligned}
		J(q,t)=&\frac{1}{N_pq^2}\left< \sum_{i=1}^{N_p}\sum_{j=1}^{N_p}\bm{q}\cdot\bm{V}_i(t)\bm{V}_j(0)\right.\\
		&\left. \vphantom{\sum_{j=1}^{N_p}}\cdot\bm{q}\exp\left[ i\bm{q}\cdot\left(\bm{R}_i-\bm{R}_j\right)\right]\right>.
	\end{aligned}
	\label{eq:Jqt}
\end{equation}
Akin to $H(q,t)$, $J(q,t)$ can further be split into self and distinct parts,
\begin{equation}
	J(q,t)=C_{vv}+J_d(q,t).
	\label{eq:JJiqt}
\end{equation}
The self part is the velocity auto-correlation function (VCF) $C_{vv}$ and the distinct part $J_d(q,t)$ accounts for the collective cross-correlations for which $i\neq j$. 
\subsection{Multiparticle collision dynamics for SALR systems}
\subsubsection{Solvent}To account for hydrodynamic interactions between SALR particles in a viscous solvent, we employ multiparticle collision dynamics (MPC)~\cite{kap99,kap00,kapral_review,MPCD,zoettl:2018,tan05}, which explicitly models the solvent by point particles of mass $m$. The algorithm consists of alternating streaming and collision steps. During streaming, solvent particles move ballistically for a time interval $h$ within a cubic simulation box of length $L$ under periodic boundary conditions. In the collision step, we use the stochastic rotation dynamics (SRD) variant~\cite{huang2015,tan01,tan02,tan03}. Solvent particles are sorted into cubic cells of size $a$, where their velocities relative to the cell center-of-mass velocity are rotated by a fixed angle $\alpha$~\cite{MPCD,kapral_review,ihl01}. Additionally, a cell-level Maxwell–Boltzmann scaling of the relative velocities $\tilde{\v}_i$ is applied to ensure proper thermalization~\cite{huang2015}. This procedure conserves mass, momentum, angular momentum, and energy within each collision cell. To enhance computational efficiency, the MPC algorithm is parallelized and implemented on a graphics processing unit (GPU)~\cite{Westphal:2014}. Further implementation details can be found in Ref.~\cite{Qi:2020}.
\subsubsection{SALR particles}
The $N_p$ interacting SALR particles are coupled to the MPC solvent as neutrally buoyant, no-slip spheres, following Ref.~\cite{theers16friction,tan03}. During the streaming step, each colloid advances with center-of-mass velocity $\bm V_p(t)$ for time span $h$. Solvent particles that penetrate a colloid are backtracked to the collision time $t+h_i$ determined by $|\bm r_i(t) - \bm R_p(t) + h_i (\v_i - \bm V_p)|^2 = \sigma^2/4$. They then elastically collide with a virtual colloid at $\bm R_p(t) + h_i \bm V_p(t)$, transferring momentum $\bm p_i$ and updating their velocities to $\bm v_i' = \bm v_i(t + h_i)$. The updated velocities are~\cite{tan03}:
\begin{align}
	\v_i' &= \v_i(t) - \bm p_i/m , \\
	\bm V_p(t+h) &= \bm V_p(t) + \sum_i \bm p_i / M , \\
	\bm \Omega_p(t+h) &= \bm \Omega_p(t) + 0.5\sigma \sum_i (\bm n_i \times \bm p_i) / I ,
\end{align}
where $m$ is the mass of the solvent particles colliding with the colloid, $\bm n_i = (\bm r_i - \bm R_p)/|\bm r_i - \bm R_p|$ and $I =\chi M \sigma^2$ with $\chi = 1/10$.

No-slip boundary conditions are imposed using a bounce-back rule~\cite{padding2005stick,zottl2014,theers2016modeling,tan03}. To maintain thermal equilibrium and further enforce the hydrodynamic boundary condition, phantom particles are introduced inside each colloid~\cite{Lamura01}. 

In practice, we approximate the collision time by $h_i = h/2$ for all solvent–colloid collisions to improve computational efficiency. This simplification has been shown to be accurate for sufficiently small time steps~\cite{hecht2005,padding2005stick,theers2016modeling}.
\subsection{Simulation parameters} \label{sec:sim_para}
\paragraph*{LD simulations}As stated in Ref.~\cite{tan04}, the single-particle friction coefficient is selected as $\gamma\approx3\pi\eta\sigma=80.1 \sqrt{Mk_BT}/\sigma$ with associated single-particle momentum relaxation time $\tau_M=M/\gamma=0.0125\sigma\sqrt{M/k_BT}$. both numerical methods can be meaningfully compared. The LD simulations integrate the Langevin equation using a small time step of approximately $\Delta t\approx0.000078\sigma \sqrt{M/k_BT}$. The simulations are performed with fixed number of particles $N_p=1024$ within cubic simulation box sizes with various linear dimension to adjust desired area fraction. All the LD simulations run for at least $10^8$ time steps. This is at least above $6\times10^3$ times the particle momentum relaxation time. We have checked systems with $N_p=4096$ and observe no qualitative changes of the measured properties.

\paragraph*{MPC simulations}The dynamics of our Q2D-SALR model with hydrodynamic interaction effects are simulated by MPC and viewed on different time and length scales. Based on the MPC units for length of $a$, mass of $m$, and energy of the thermal energy $k_{B}T$, we therefore use the units
\begin{align}
	t_0=\sqrt{ma^2/(k_{B}T)}\approx0.0078\sigma \sqrt{M/k_BT} , \nonumber \\
	v_0 = v_{th}= \sqrt{k_BT/m}\approx25.58\sqrt{k_BT/M} , \\
	\hspace{-5mm}
	\eta_0=\sqrt{m k_BT}/a^2 \approx0.977\sqrt{M k_BT}/\sigma^2\nonumber
\end{align}
for time, $t_0$, velocity, $v_0$, and viscosity, $\eta_0$, respectively. Note that $t_0$ is equal to the ratio of cell size $a$ and thermal velocity $v_{th}=\sqrt{k_BT/m}$. In these units, the speed of sound of MPC fluid is equal to one. The rotation angle for SRD is set to $\alpha=130^\circ$. To keep the same solvent viscosity parameter $\eta$ and mass density of the SALR sphere in both LD and MPC simulations, the average number of particles per collision cell is selected as $\av{N_c}=10$, and the collision time steps are taken as $h=0.05\times t_0$. With the mass density $\rho= \langle N_c\rangle\;\!m/a^3$, the corresponding kinematic viscosity is $\nu/\nu_0=0.81$, where $\nu_0 = \eta_0a^3/m$. The properties of the embedded SALR spheres, with a diameter of $\sigma=5a$, are considered neutral buoyant with mass $M=654.5m$. The related dimensionless Schmidt number is $Sc =\nu/d_0=17$, expressing that the viscous diffusion of (transversal) momentum in the fluid is distinctly faster than diffusive mass transport, with the latter characterized by the mass diffusion coefficient of the MPC particles. MPC Simulations are performed using periodic boundary conditions to a cubic simulation box of length $L=200a$ or specified otherwise. 

These parameter settings ensure that the SALR particle properties and solvent viscosity are identical in the LD and MPC simulations. The LD simulations are nevertheless typically run for longer than the MPC simulations, though both runs are sufficiently long for the results presented here.
\begin{figure}[!h]
	\centering
	\subfloat{		
		\begin{picture}(100,180)
			\put(-80,-8){\includegraphics[height=0.365\textwidth]{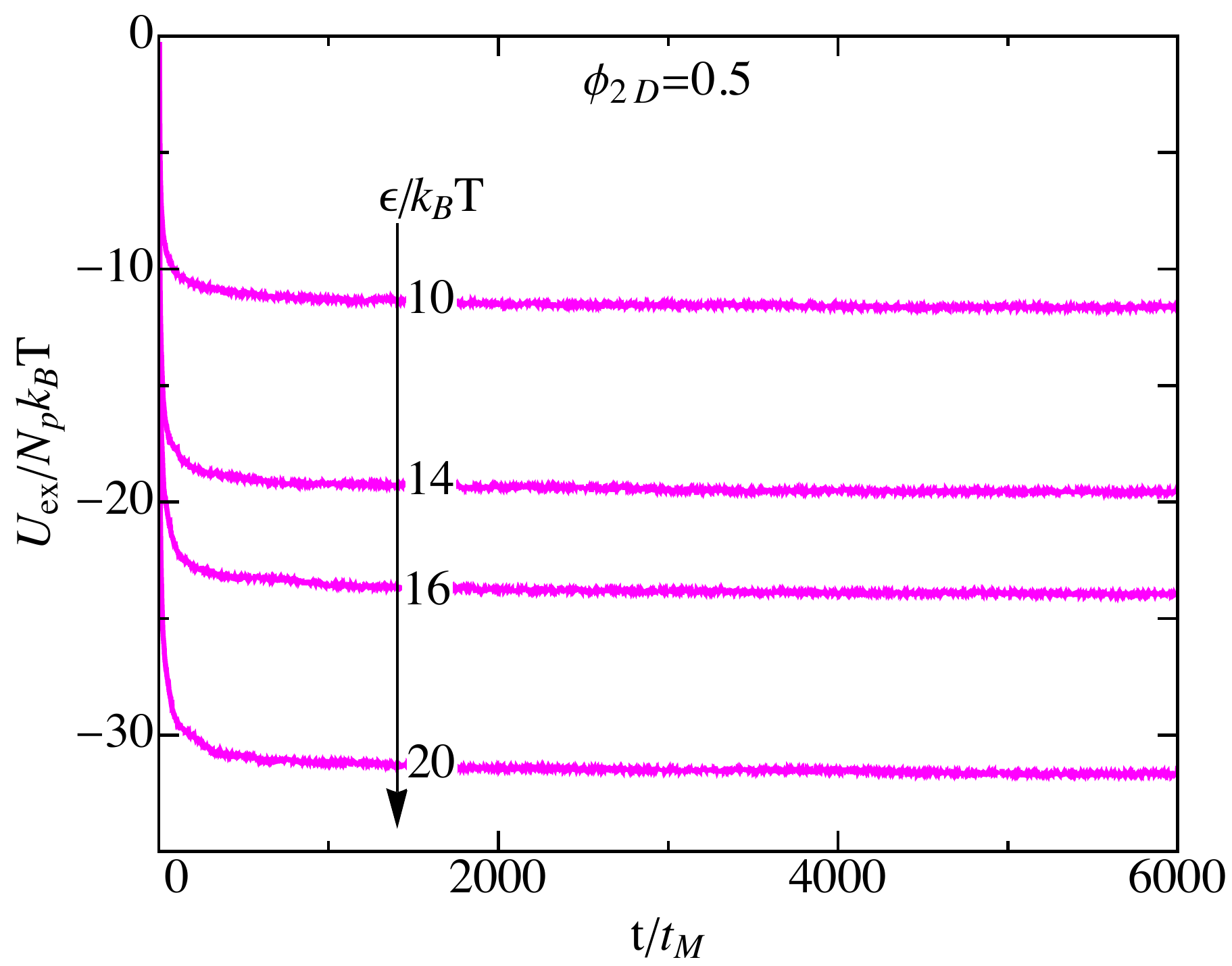}}
			\put(-33,156){(a)}
		\end{picture}
		\label{fig:Uex_b}
	}\\
	\vspace{0mm}
	\subfloat{		
		\begin{picture}(100,180)
			\put(-77,-8){\includegraphics[width=0.44\textwidth,trim={0 0 0 0},clip]{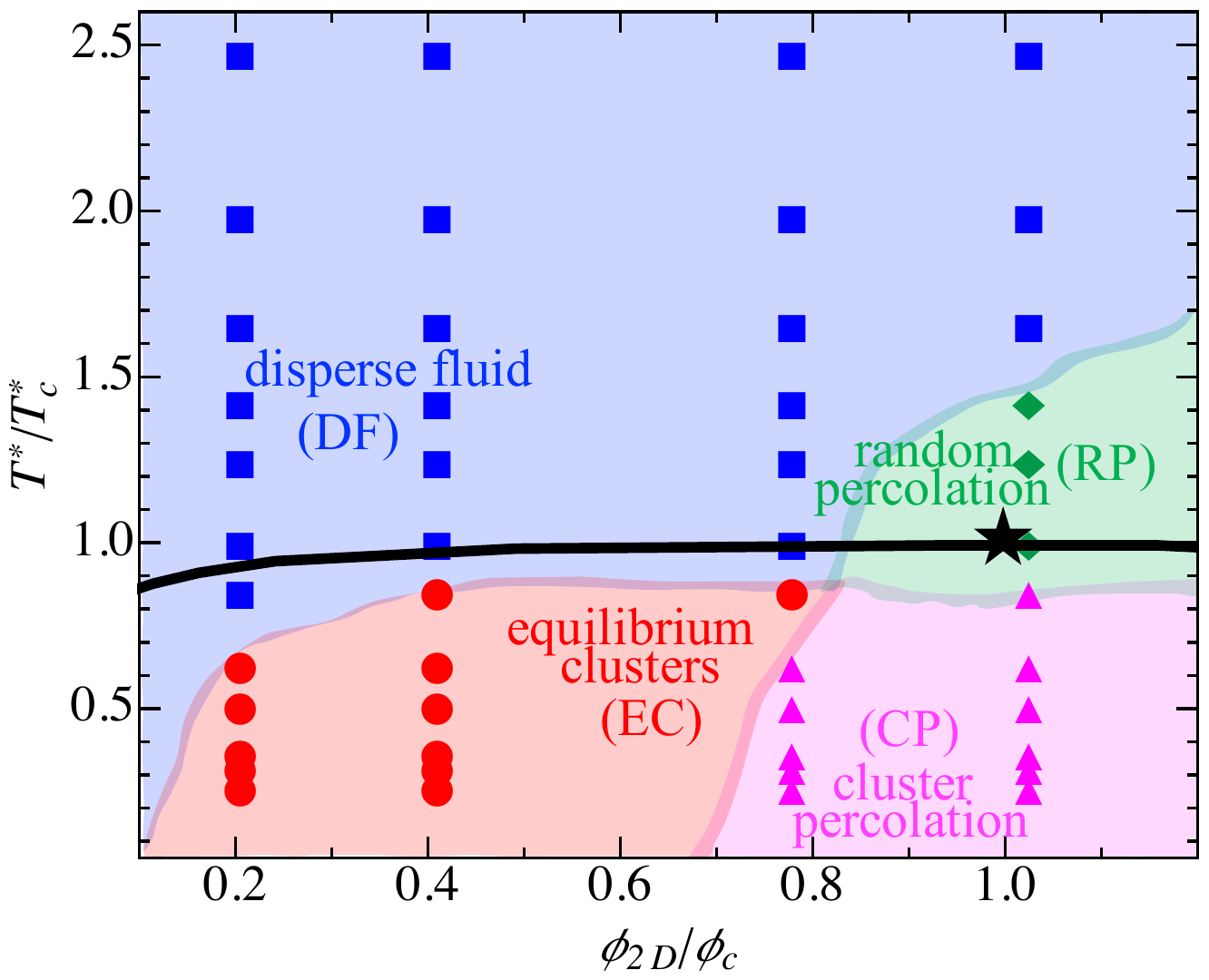}}
			\put(-33,156){(b)}
		\end{picture}
		\label{fig:phases}
	}\\
	\vspace{2mm}
	\caption{\label{fig:statics}(a)~Excess internal energy, $U_\text{ex}$ per particle and $k_BT$ for Q2D-SALR cluster percolation phase at area fraction $\phi_\text{2D}=0.5$ for various attraction strength values $\epsilon$, as function of LD simulation time $t$, expressed in units of $t_M\approx0.0125\sigma \sqrt{M/k_BT}$. The vertical arrow points towards increasing values of $\epsilon$. (b)~Generalized Q2D-SALR phase diagram determined by structural analysis for forming clusters, namely by the shape of the cluster size distribution function~\cite{godfrin:2014}. All $48$ state points are classified in four generalized phases: disperse fluid (blue squares), equilibrium clusters (red disks), random percolation (green diamonds), and cluster percolation (magenta triangles). The solid line denotes the attractive binodal based on the first-order perturbation calculation of square-well potential with same attraction strength.}
\end{figure}
\section{Results}
\label{sec:results}
\subsection{Thermalization and generalized phase diagram}
The first dynamical process considered here is the thermalization towards equilibrium. In our simulations, systems are prepared at different area fractions with initial particle positions assigned according to a uniform distribution, and velocities initialized according to a Maxwellian distribution at an initial temperature one order of magnitude larger than the target value. The system is quenched subsequently, and the equilibration process is monitored by calculating  the average potential energy, $U_\text{ex}$, per particle and $k_BT$, as a function of the LD simulation time $t$. Fig.~\ref{fig:statics}\subref{fig:Uex_b} shows representative LD simulation results for $U_\text{ex}$ at area fraction $\phi_\text{2D}=0.5$, and for values of the reduced attraction strength $\epsilon/(k_BT)$  (i.e., inverse reduced effective temperature) in the range from $10$ up to $20$. As noticed from the figure, a stationary state is reached overall within the simulation time window. We have ascertained that for fixed system parameters, the same stationary state is reached, independent of the initial conditions. For very low effective temperatures (strong short-range attraction) and small area fractions, after a first rapid decrease of the average potential energy towards an apparent equilibrium state, the according system exhibits a slow energy drift. A similar energy drift has been reported in Refs.~\cite{Imperio:2004,toledano:2009,Brito:2020}.

As reported in Ref.~\cite{tan04}, four categories of (generalized) phases are identified by analyzing the cluster distribution function, characterizing the fraction of particles forming clusters of size $s$. Fig.~\ref{fig:statics}\subref{fig:phases} displays the computed $\phi_{2D}-T^*$ phase diagram normalized by the critical point values, with critical temperature $T_c^*=0.203$ and critical area fraction $\phi_c=0.488$ (cf. asterisks), of the square-well attractive reference system with attraction range parameter $\lambda=1.06$~\cite{tan04}. These four phases are: dispersed fluid (DF), equilibrium clusters (EC), random percolation (RP), and cluster percolation (CP), with respectively distinct colors (Fig.~\ref{fig:statics}\subref{fig:phases}). The thick black line represents the binodal curve of the corresponding attractive system, obtained from second-order perturbation theory calculations. See Reference~\cite{tan04} for technical details.

Note that the displayed quantities in the following figures and plots use the same (or similar) color shown in Fig.~\ref{fig:statics}\subref{fig:phases}. In other words, the blue curves (points) denote the system in the DF phase, the red ones correspond to the EC phase, the green ones to the RP phase, and the magenta ones to the CP phase.

\subsection{Self diffusion}
\begin{figure}[!h]
	\subfloat{		
		\begin{picture}(100,180)
			\put(10,0){\includegraphics[width=0.44\textwidth,trim={0 0 0 0},clip]{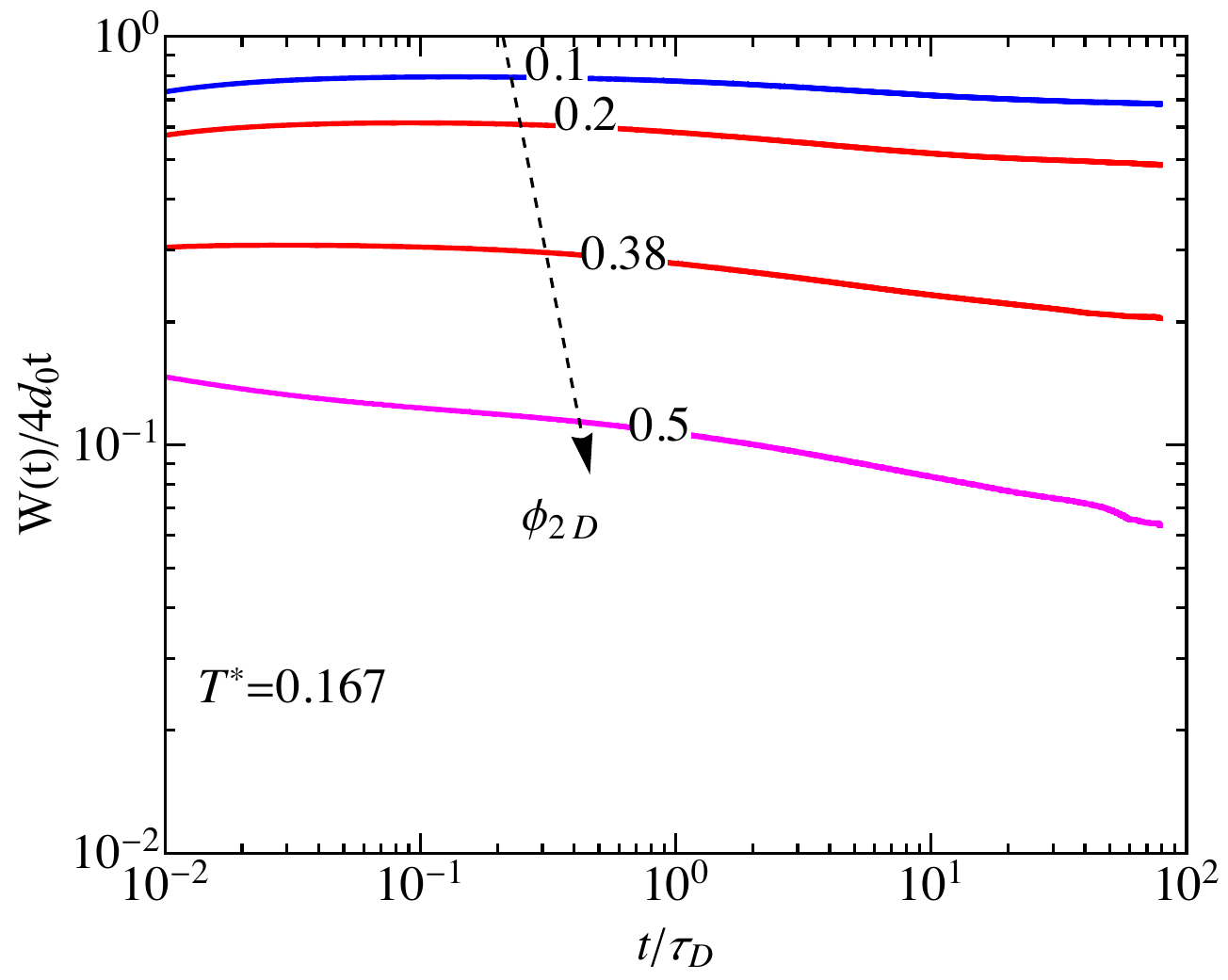}}
			\put(215,28){(a)}
		\end{picture}
		\label{fig:smax}
	}\\
	\subfloat{		
		\begin{picture}(100,180)
			\put(10,0){\includegraphics[width=0.44\textwidth,trim={0 0 0 0},clip]{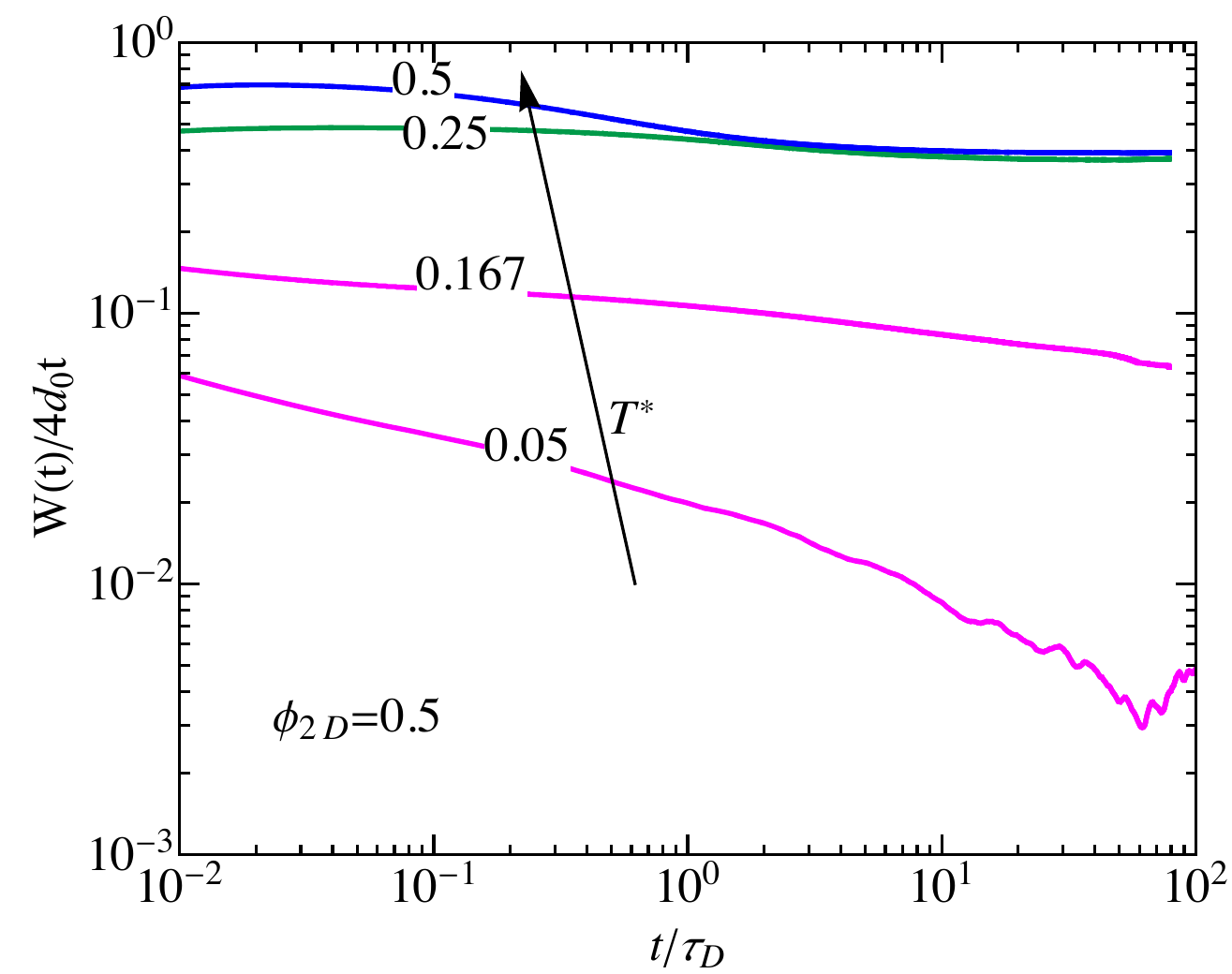}}
			\put(215,28){(b)}
		\end{picture}
		\label{fig:zb_T}
	}
	\caption{\label{fig:msd} Self-diffusive properties of Q2D-SALR particles for different temperature and area fractions. Shown are rescaled mean-squared displacement, $\langle|\bm R(t)-\bm R(0)|^2\rangle/4td_0$, of a Q2D-SALR system for different $(\phi_{2D}, T^*)$ values. (a)~$T^*=0.167$ and (b)~$\phi_{2D}=0.5$ for varying $T^*=1/\epsilon$. The color of the curves encodes their respective generalized phases that are used in Fig.~\ref{fig:statics}\protect\subref{fig:phases}.}
\end{figure}
We study self-diffusion by analyzing the normalized MSD of SALR particles, expressed as a dimensionless, time-dependent diffusion coefficient, $W(t)=\langle|\bm R(t)-\bm R(0)|^2\rangle/(4td_0)$, as a function of dimensionless time (Fig.~\ref{fig:msd}). The colors of the curves correspond to the different phases shown in Fig.~\ref{fig:statics}\subref{fig:phases}. Here, $d_0$ is the diffusion constant of a single Brownian particle, and $t$ is the correlation time (normalized by the single-particle diffusion time $\tau_D$). Only the time region beyond full momentum relaxation is shown, i.e., $t \gg M/\gamma = \tau_M$, the non-inertial timescale. Intuitively, clustering, which becomes more pronounced with increasing $\epsilon$ and $\phi_{2D}$, tends to suppress particle self-diffusion. Systems in different phases exhibit distinct diffusive behaviors.

Figure~\ref{fig:msd}(a) shows results at $T^*=0.167$. For the DF phase (blue curve), short- and long-time plateaus are observed, corresponding to the short- and long-time self-diffusion coefficients, $d_s$ and $d_l$, where $d_s > d_l$. Although the system is classified as a ``dispersed fluid", transient clusters formed by short-range attractions still slow down diffusion. As the area fraction increases, in the equilibrium clusters (EC) phase, dynamics are further slowed, with both $d_s$ and $d_l$ reduced. At $\phi_{2D}=0.38$, the long-time diffusion decreases to about $20\%$. In the CP phase, a two-step relaxation emerges, corresponding to two subdiffusive regimes.

Figure~\ref{fig:msd}(b) presents the diffusion dynamics at $\phi_{2D}=0.5$. Interestingly, the RP phase exhibits short- and long-time Brownian diffusion similar to the DF phase, although slightly slower at long times. Since the EC phase is absent at this higher area fraction, diffusion slows drastically in the CP phase. Moreover, a monotonically decreasing $W(\tau)$ is observed when clusters percolate (green curve in Fig.~\ref{fig:msd}(b)), revealing a two-step relaxation associated with two distinct subdiffusive regimes~\cite{berg:2019}.

Hence, the Q2D-SALR particles at fluid phases (DF, EC and RP phases) show distinctive short- and long-time Brownian diffusive behavior due to clustering rather than to the caging. The two-sub-diffusive feature coincides with the gel formation when the system is percolated. 
\subsection{Bond correlation function and cluster lifetime $\tau_b$}
\begin{figure}[!h]
	\subfloat{		
		\begin{picture}(100,180)
			\put(-3,-8){\includegraphics[width=0.44\textwidth,trim={0 0 0 0},clip]{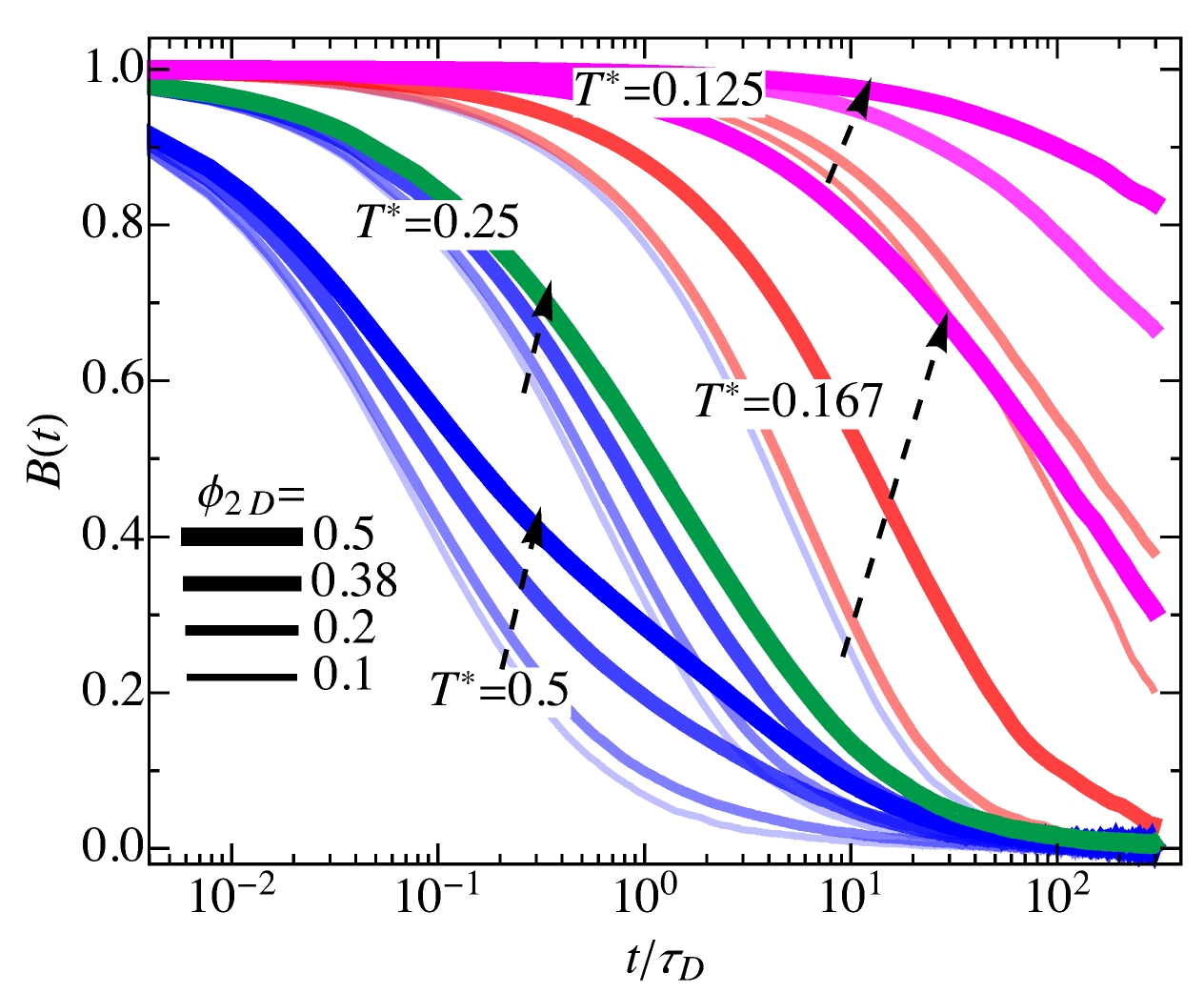}}
			\put(210,165){(a)}
		\end{picture}
		\label{fig:bcf}
	}\\
	\vspace{3mm}
	\subfloat{		
		\begin{picture}(100,180)
			\put(-5,-10){\includegraphics[width=0.44\textwidth,trim={0 0 0 0},clip]{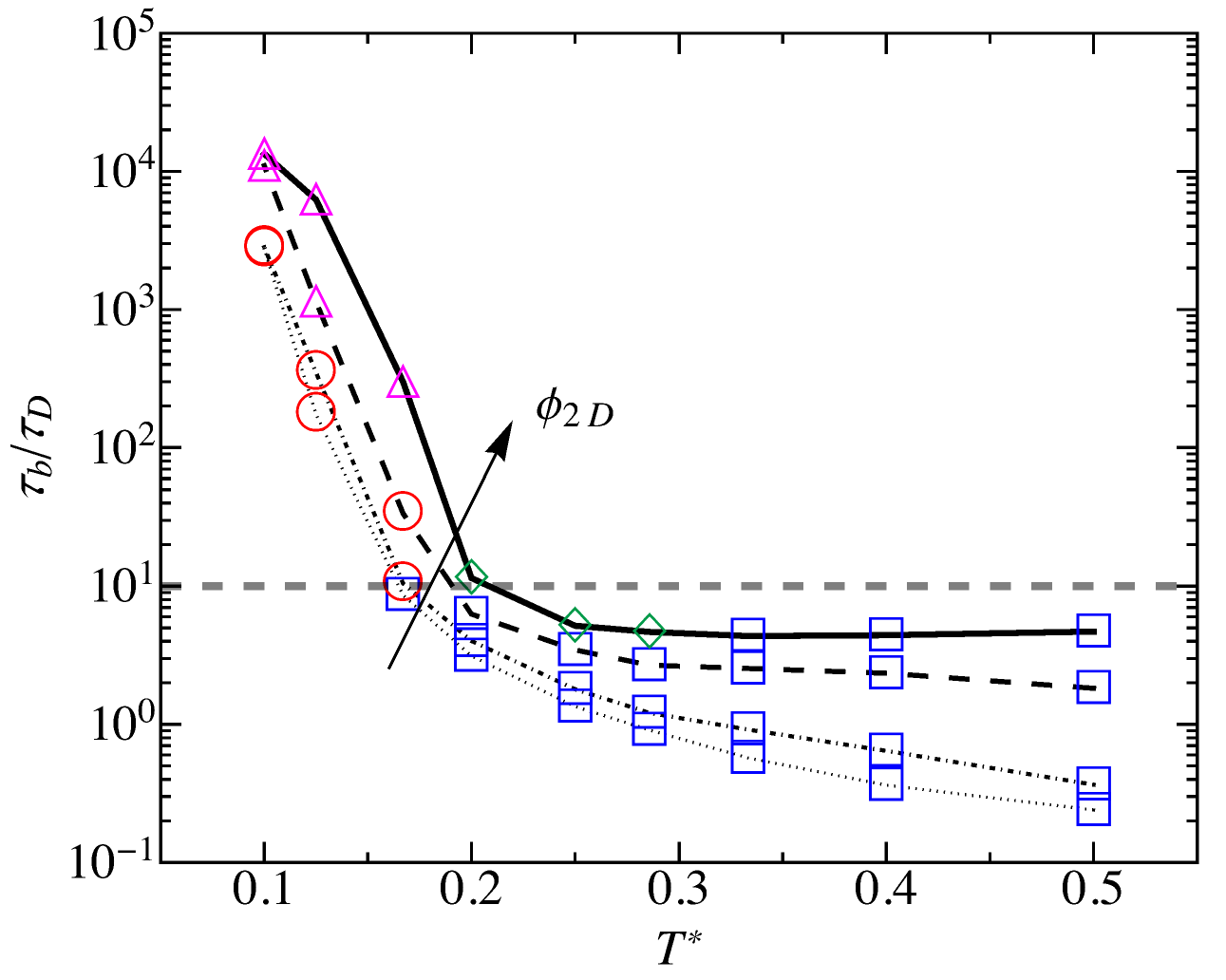}}
			\put(208,162){(b)}
		\end{picture}
		\label{fig:blt}
	}
	\caption{\label{fig:bcflt}(a)~Time-evolution of  bond correlation function $B(t)$ represented as lines for various temperatures and densities. The line thickness implies its $\phi_{2D}$ values (marked by the dashed arrows in ascending order) as displayed in the legend. Each dashed arrow on top of the curves groups the system belonging to the same $T^*$ as indicated. The color coding is as in Fig.~\ref{fig:statics}\protect\subref{fig:phases} (b)~Cluster lifetime $\tau_b$ normalized by diffusion timescale $\tau_D=R^2/d_0$ as a function of effective temperature. The color and the shape of the open symbols encode their respective generalized phases that used in Fig.~\ref{fig:statics}\protect\subref{fig:phases} with $\beta_b\in(0.3,0.75)$. The horizontal transparent dash line indicates the critical $\tau_b$ value separating the clustered (EC and CP) phases from DF and RP phases. Note that for better illustration, the state points for $T^*<0.1$ corresponding to $\tau_b$ values exceeding $10^5\tau_D$ are not presented.}
\end{figure}
We now analyze the relaxation of forming bonds during clustering. As a representative example, Fig.~\ref{fig:bcflt}\subref{fig:bcf} shows $B(t)$ for four area fractions and four attraction strengths. The colors encode the static phases (cf. Fig.~\ref{fig:statics}\subref{fig:phases}), while the line thickness indicates the corresponding values of $\phi_{2D}$. Overall, $B(t)$ in the dispersed fluid (DF) and random percolation (RP) phases decays much faster than in the clustered phases (EC and CP). This is expected, since particles remain relatively mobile in the DF and RP phases and particles bond with neighbors for only a brief time, resulting in short-lived clusters. In contrast, in the EC phase and partially in the CP phase, $B(t)$ does not decay to zero within the simulated time window, indicating long-lived or more persistent clusters. Interestingly, curves at the same temperature are grouped together and preserve similar shapes. In comparison, the influence of density is less pronounced, albeit higher $\phi_{2D}$ generally leads to longer-lasting correlations. This suggests that temperature $T^*$ (or equivalently the inverse attraction strength $1/\epsilon$) plays a more dominant role than particle density (which forms clusters via packing effects) in controlling dynamic clustering in SALR particles. 

To estimate the cluster lifetime $\tau_b$, the bond correlation function $B(t)$ is modeled with high accuracy as a stretched exponential function $A\exp(-t/\tau^*_b)^{\beta_b}$. The obtained stretching exponent $\beta_b$ depends on the attraction strength $\epsilon$ and $T^*$. Thus, the characteristic cluster lifetime $\tau_b$, essentially the bond correlation decay time, is estimated by $\tau_b=\tau^*_b/\beta_b\Gamma(1/\beta_b)$ with $\Gamma(x)$ the Gamma function. 

Fig.~\ref{fig:bcflt}\subref{fig:blt} shows the characteristic cluster lifetime $\tau_b$ as a function of effective temperature $T^*$ for various area fractions, with the fitting parameter $\beta_b$ ranging from $0.3$ to $0.75$. First, $\tau_b$ in the fluid phases (DF and EC) is significantly shorter than in the clustered phases. The variation of $\tau_b$ induced by $\phi_{2D}$ spans less than two orders of magnitude. In contrast, the attraction strength has a more dominant effect, more prominently reflected by the exponential-like increase of $\tau_b$ with decreasing attraction $\epsilon$ for $T^* < 0.2$. At very low temperatures ($T^* < 0.07$), the dynamics becomes extremely slow, and bonds remain essentially unbroken within the simulation time window. As suggested in $3D$~\cite{porcar:2010,yliu:2011,godf:2016,nawrocki:2017,das2018clustering,godfrin:2018,perdomo:2022}, the dynamic clusters can be classified by $\tau_b$ as transient ($\tau_b<\tau_D$), dynamic ($\tau_b>\tau_D$ with finite value) and permanent ($\tau_b\gg\tau_D$). Applying this 3D dynamic criterion on our Q2D-SALR system, most of the state points (blue squares) in the DF phase actually fall into the dynamic cluster regime, only for a few low-density, high-temperature cases ($\phi_{2D}=0.1$ and $0.2$ at $T^*\approx0.3\-- 0.5$), transient clusters are identified. Hence, the structure characterization of the clusters must be compensated by analyzing the lifetime of the clusters.

Intriguingly, we observed that a threshold value of $\tau_b/\tau_D\approx 10$ separates the DF and EC phases, suggesting a dynamic criterion for the DF--EC transition. The state points presented in Fig.~\ref{fig:bcflt}\subref{fig:blt} for both clustered EC and CP phases are characterized as dynamic clusters. More interestingly, as temperature $T^*$ increases, the cluster lifetime drops faster at equilibrium cluster regime, and then undergoes an apparent slowing decay of $\tau_b$ values. This two-stage change of $\tau_b$ in DF and EC phases is similar to what was discussed in Ref.~\cite{das2018clustering}. The cluster lifetime at DF regime is consistent with two-particle Smoluchowski theory calculated dissociation time (or binary escape time) $\tau^D_c$~\cite{Chan:1984}; whilst in the EC phase, it increases more strongly with decreasing temperature, aligning better with the so-called first passage time $\tau^K_c$ according to Kramers' theory for barrier crossing~\cite{Kramers:1940}.

\subsection{Dynamics of local hexagonal order}
\begin{figure}[!h]
	\vspace{5mm}
	\subfloat{		
		\begin{picture}(100,180)
			\put(0,5){\includegraphics[width=0.44\textwidth,trim={0 0 0 0},clip]{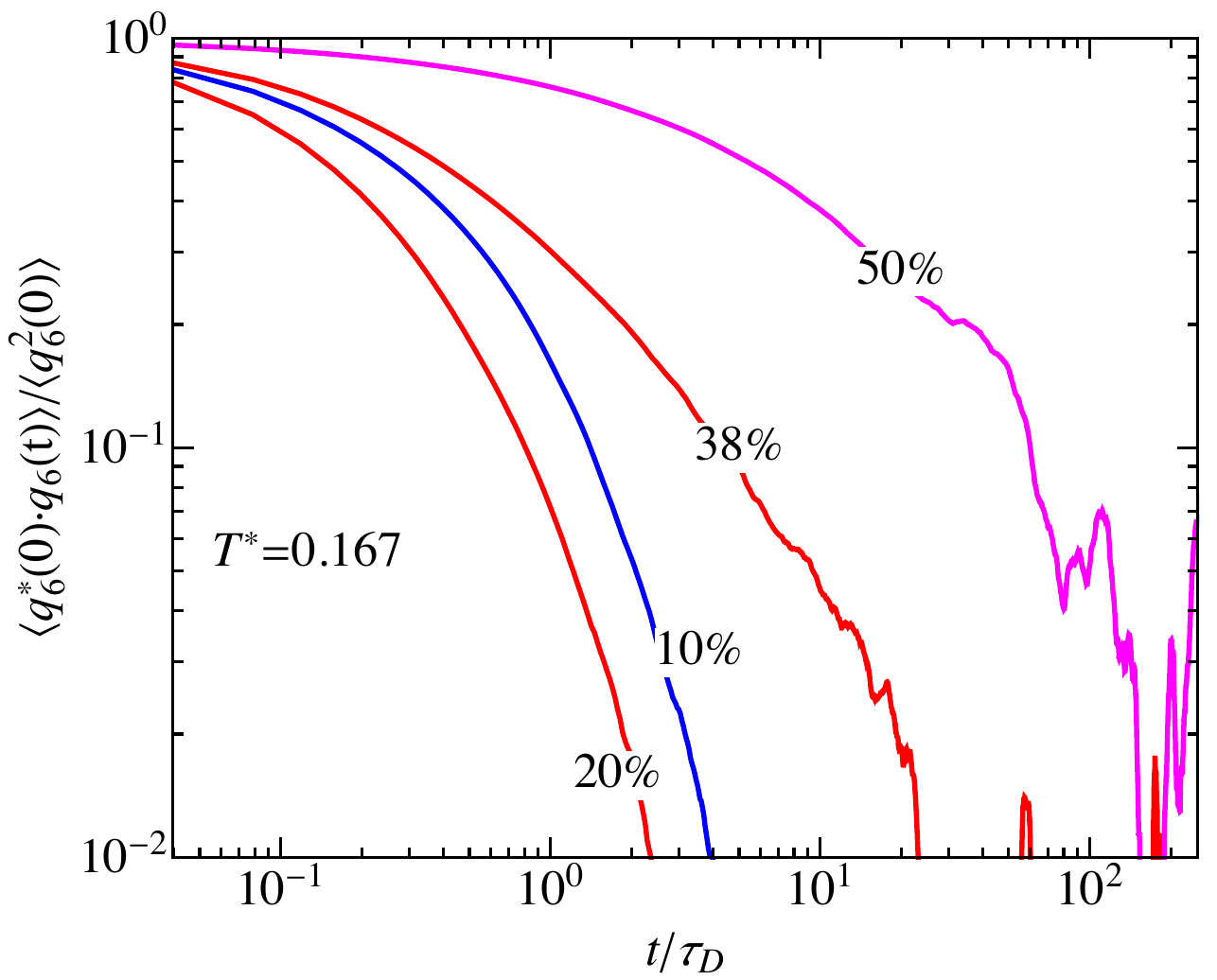}}
			\put(35,40){(a)}
		\end{picture}
		\label{fig:hex1}
	}\\
	\vspace{-1mm}
	\subfloat{		
		\begin{picture}(100,180)
			\put(0,5){\includegraphics[width=0.44\textwidth,trim={0 0 0 0},clip]{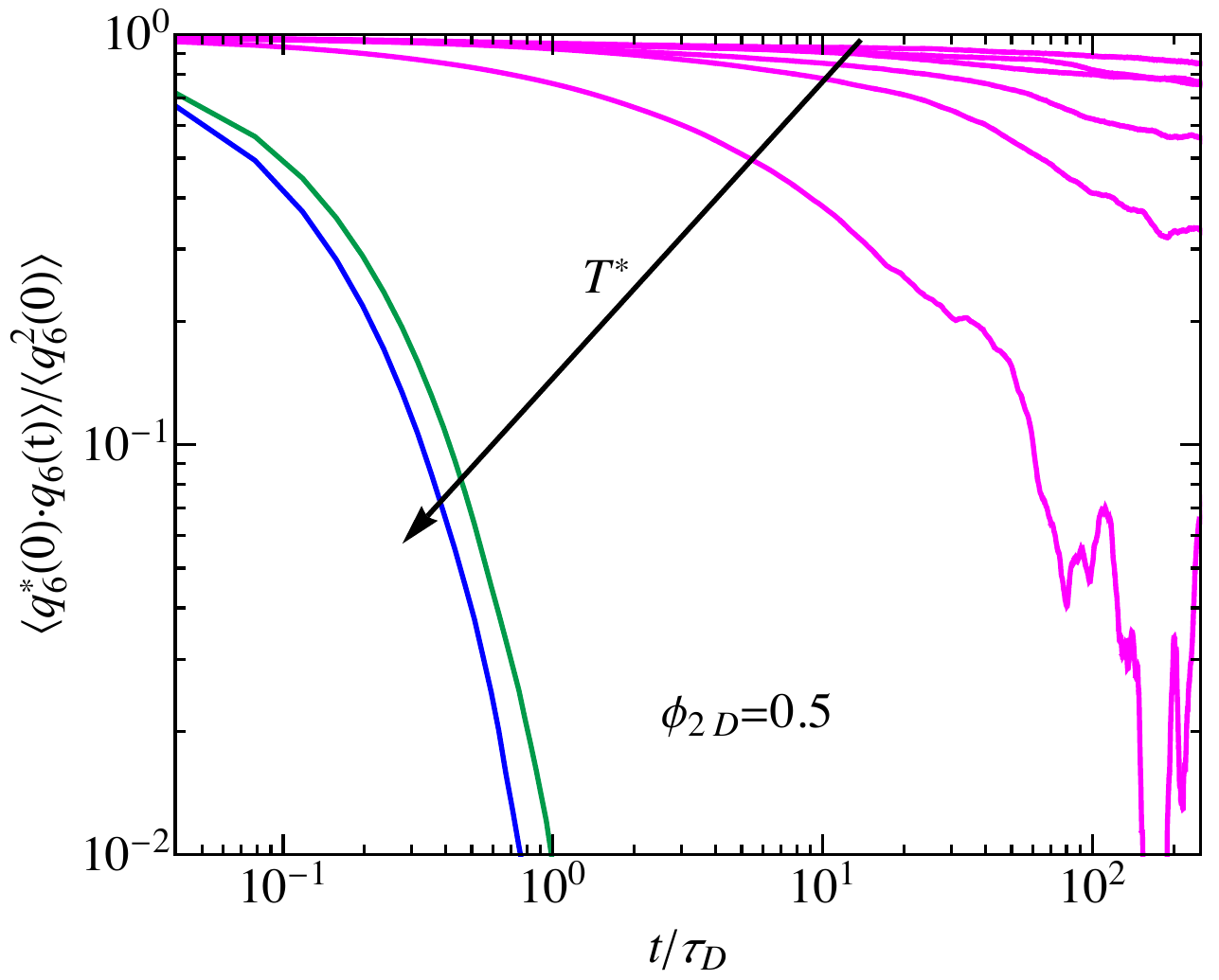}}
			\put(35,40){(b)}
		\end{picture}
		\label{fig:hex2}
	}
	\caption{\label{fig:hexx}Langevin dynamics simulation results for the normalized hexagonal order parameter correlation function (a) at fixed reduced temperature $T^*=0.167$ and area concentrations $\phi_{2D}$ as displayed, and (b) for fixed area fraction $\phi_{2D}=0.5$ but varying $T^\ast$, respectively. Color coding of considered phases as in Fig.~\ref{fig:statics}\protect\subref{fig:phases}.}%
\end{figure}
\begin{figure}[!h]
	\begin{picture}(100,180)
		\put(0,5){\includegraphics[width=0.44\textwidth,trim={0 0 0 0},clip]{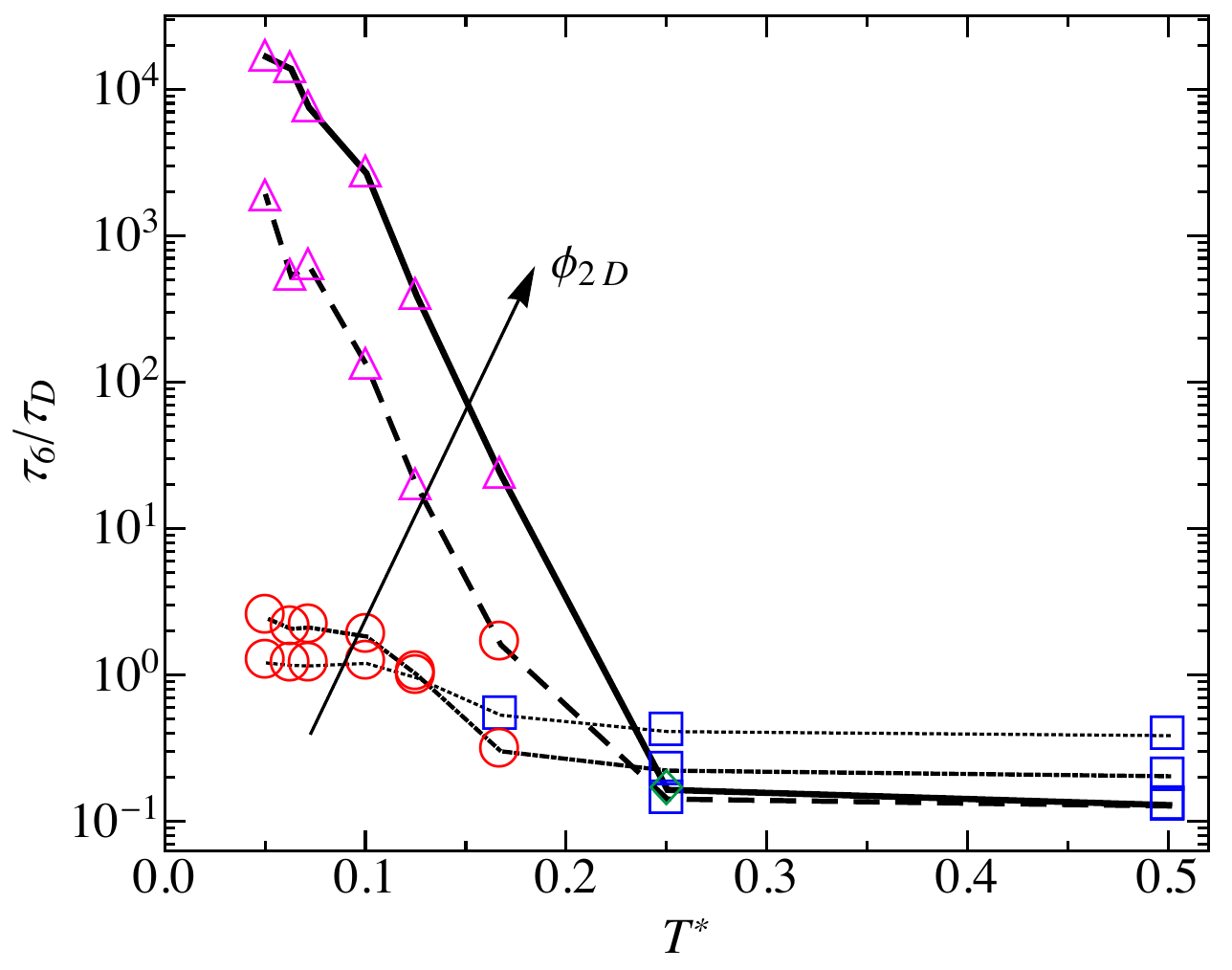}}
	\end{picture}
	\caption{\label{fig:tq6}~Hexagonal order relaxation time, $\tau_6$, obtained from a stretched exponential fit of Eq.~\eqref{eq:cq6q6} with exponent $\beta_6\approx0.4\sim0.8$, normalized by diffusion time scale $\tau_D$ as function of $T^*$. The arrows indicate $\phi_{2D}$ and $T^*$ in ascending order. Color coding of considered phases as in Fig.~\ref{fig:statics}\protect\subref{fig:phases}.}%
\end{figure}
The competing attractive and repulsive interactions in Q2D-SALR systems give rise to the formation of clusters exhibiting local hexagonal ordering~\cite{tan04}.

Figs.~\ref{fig:hexx}\subref{fig:hex1} and \ref{fig:hexx}\subref{fig:hex2} display LD simulation results for $g_6(t)$ (colored solid lines) at fixed temperature $T^*=0.167$ for varying $\phi_{2D}$, and at fixed $\phi_{2D}=0.5$ for varying $T^\ast$, respectively. The figures show that $g_6(t)$ in the unclustered DF and random percolation (RP) phases generally decays significantly faster than in the EC and cluster percolation (CP) phases. However, this trend is not universal, as illustrated in Fig.~\ref{fig:hexx}\subref{fig:hex1}: the $g_6(t)$ curve of the DF system at $\phi_{2D}=0.1$ decays more slowly than that of the EC system at $\phi_{2D}=0.2$. The latter state point lies close to the DF–EC phase boundary. Compared to $\phi_{2D}=0.1$, particles at $\phi_{2D}=0.2$ form small clusters with shorter interparticle distances, allowing for increased rotational motion around cluster centers, which accelerates decorrelation.

Interestingly, consistent with our static analysis in Ref.~\cite{tan04}, the persistence of hexagonal ordering in the RP phase (Fig.~\ref{fig:hexx}\subref{fig:hex2}) is similarly as weak as in the DF phase. This can be attributed to the absence of stable cluster formation and, consequently, of pronounced bond-orientational ordering in both phases.

The $g_6(t)$ curves for all four phases can be well described by a stretched exponential function,
\begin{align} g_6(t)=\exp{(-t/\tau_6)^{\beta^*_6}}\,. \label{eq:cq6q6_fit} \end{align}
Here, the characteristic relaxation time of local hexagonal order is given by $\tau_6=\tau_6^*/\beta_6^*\,\Gamma(1/\beta_6^*)$, with $\beta_6^*$ the stretching exponent. The dependence of $\tau_6$ on $\phi_{2D}$ and $T^*$ is shown in Fig.~\ref{fig:tq6}, using the same color coding as in Fig.~\ref{fig:statics}\subref{fig:phases}.

First, the relaxation times $\tau_6$ in the unclustered DF and RP phases are clearly smaller than the single-particle diffusion time $\tau_D$, with DF systems at higher concentrations exhibiting slightly shorter relaxation times. In contrast, $\tau_6$ in the EC phase can exceed $\tau_D$, but remains of the same order of magnitude. For CP systems, the hexagonal time correlations are strongly sensitive to both attraction strength $\epsilon$ (i.e., temperature $T^*$) and concentration. As shown in Fig.~\ref{fig:tq6}, $\tau_6$ in the CP phase is typically orders of magnitude larger and decreases approximately exponentially with increasing effective temperature. Interestingly, one can notice that the $\tau_6 \approx \tau_D$ marks the DF and EC phase crossover in the $T^*-\tau_6$ plot.
\subsection{Non-Gaussian dynamics}
\begin{figure}[!h]
	\subfloat{		
		\begin{picture}(100,200)
			\put(10,0){\includegraphics[width=0.44\textwidth,trim={0 0 0 0},clip]{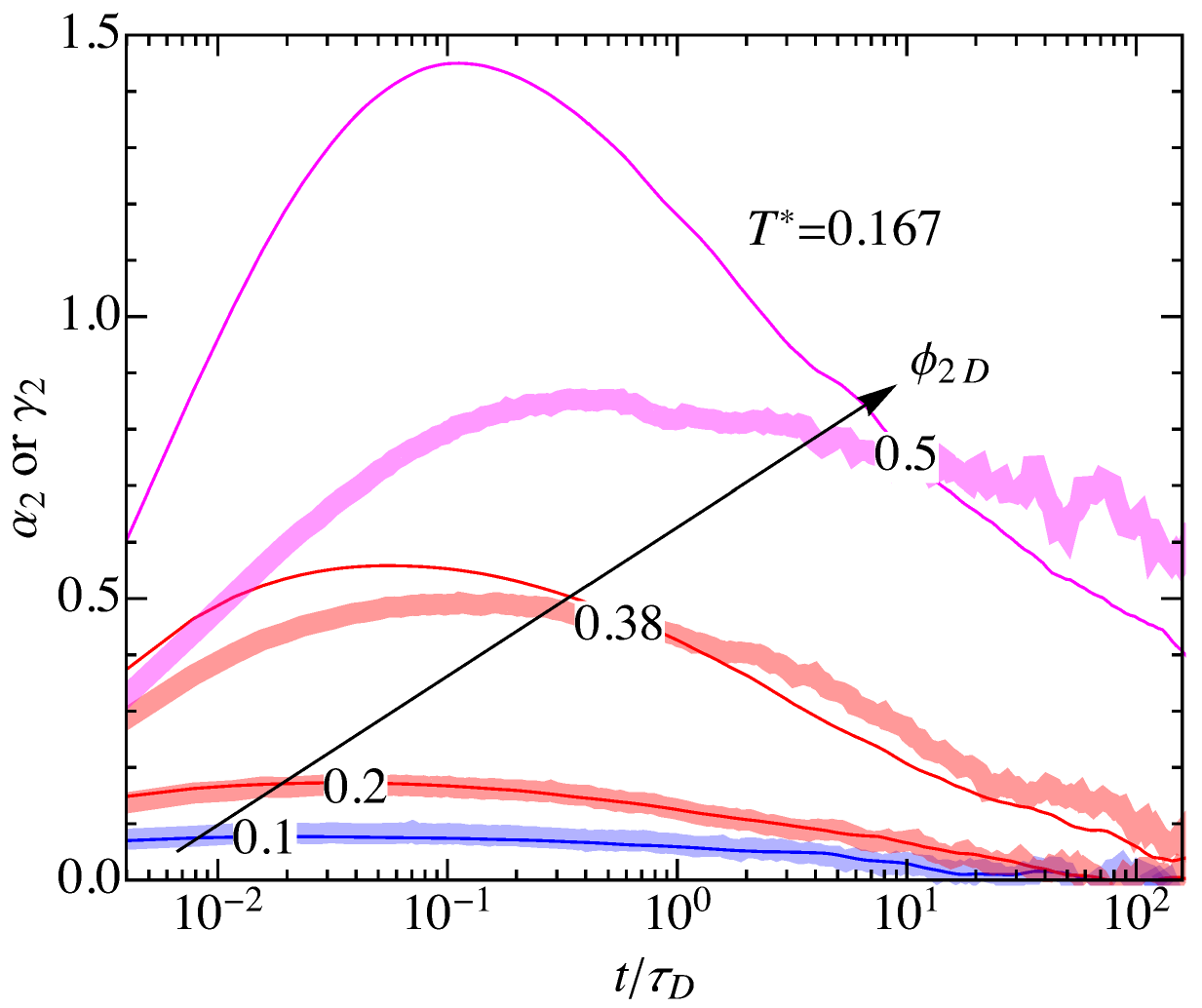}}
			\put(38,176){(a)}
		\end{picture}
		\label{fig:nonGauss1}
	}\\
	\subfloat{		
		\begin{picture}(100,200)
			\put(10,0){\includegraphics[width=0.44\textwidth,trim={0 0 0 0},clip]{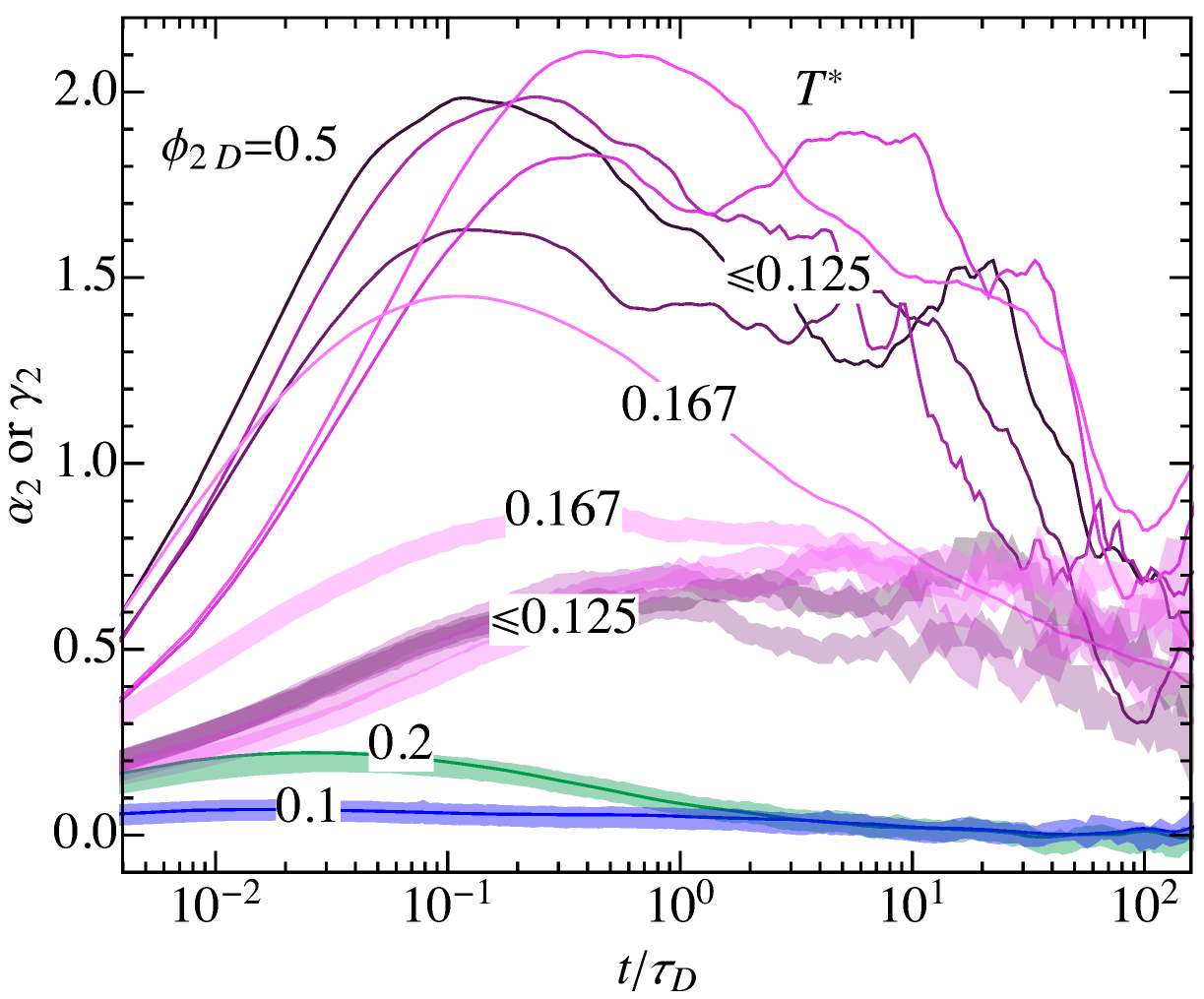}}
			\put(38,176){(b)}
		\end{picture}
		\label{fig:nonGauss2}
	}
	\caption{\label{fig:nonGauss} Non-Gaussian parameters for different $(\phi_{2D}, T^*)$ values. (a)~$T^*=0.167$ and (b)~$\phi_{2D}=0.5$ for varying densities and $T^*=1/\epsilon$, respectively. The thin curves correspond to the non-Gaussian parameter $\alpha_2$ defined in Eq.~\eqref{eq:alpa_2} and the bold transparent ones are $\gamma_2$ from Eq.~\eqref{eq:gamma_2}. Different values of $\phi_{2D}$ and $T^*$ are as indicated.  Note that the non-Gaussian effects become insensitive to temperature when $T^*\leq0.125$. The color of the curves encodes their respective generalized phases that used in Fig.~\ref{fig:statics}\protect\subref{fig:phases}.}
\end{figure}
\paragraph *{Non-Gaussian parameters} 
To quantify deviations from Gaussian diffusion, we evaluate the conventional non-Gaussian parameter $\alpha_2$ (cf. Eq.~\eqref{eq:alpa_2} and Appendix). 

In parallel, to render a more comprehensive understanding of non-Gaussianity and dynamic heterogeneity in our Q2D-SALR systems, we employ the alternative definition proposed by Flenner and Szamel~\cite{FlennerSzamel:2005,Salzmann:2006}, given in two dimensions by
\begin{align}
	\gamma_2^\text{{(d=2)}}(t)= \frac{1}{\pi}\;\!\big< r^2(t)\big>_2\;\!\left(\Big<\frac{1}{r(t)}\Big>_2\right)^2-1\,,
	\label{eq:gamma_2}
\end{align} 
which combines contributions from both the MSD and mean inverse displacement square $\left(\Big<\frac{1}{r(t)}\Big>_2\right)^2$. Note that in comparison with $\alpha_2$, according to Flenner and Szamel, $\gamma_2^{(d=3)}(t)$ in three dimensions weights strongly particles which have moved less than expected for the Gaussian distribution through its $1/r^2$ moment, and particles moving farther than Gaussian ones through the MSD. Salzman and Schweizer claim that different from $\alpha_2^{(d=3)}(t)$ this alternative non-Gaussian factor allows for detecting non-Fickean diffusion (i.e. a MSD non-linear in time) at relatively long times and hence long distances. In the same vein, we expect this argument to also be valid in 2D, as $\gamma_2^{(d=2)}(t)$ is only differed by weighting small distance via $\left(\Big<\frac{1}{r(t)}\Big>_2\right)^2$.

Fig.~\ref{fig:nonGauss}\subref{fig:nonGauss1} showcases the LD results of non-Gaussian parameters $\alpha_2$ (depicted as thin lines) at time $t\gg \tau_M$ and $T^*=0.167$ for different area fractions, with the color coding as in Fig.~\ref{fig:statics}\subref{fig:phases}. Clearly, $\alpha_2\approx0$ for $\phi_{2D}=0.1$ (blue curve, DF phase), showing negligible non-Gaussian effects at DF phase as expected. As area fraction grows, non-Gaussian peak arises with increasing amplitude (nearly $1.5$ for $\phi_{2D}=0.5$, CP phase), whilst the peak position seems to be less affected by the area fraction, which stagnates at intermediate time scale $t\approx 0.1\tau_D$. In parallel, the alternative non-Gaussian parameter $\gamma_2$ closely follows $\alpha_2$ in the DF and weak EC regimes, shown as the bold transparent blue ($\phi_{2D}=0.1$) and red ($\phi_{2D}=0.2$) curves, but significant differences emerge in strongly clustered (EC and CP) states, where the $\gamma_2$ (thick transparent) curves are more flattened with smaller amplitudes. In particular at CP cases (bold magenta transparent line), $\gamma_2$ exhibits reduced peak amplitudes (approximately half of $\alpha_2$), broader profiles, and a slight shift of the peak toward longer times ($t \sim 0.5\tau_D$). Moreover, $\gamma_2$ displays a long-time tail persisting up to $t \sim 100\tau_D$, indicating sustained dynamic heterogeneity. This behavior suggests that particles remain partially confined within transient cages and do not reach fully Gaussian or Fickian diffusion even at long times.

Figure~\ref{fig:nonGauss}\subref{fig:nonGauss2} presents the temperature dependence at fixed $\phi_{2D}=0.5$. Similarly, both $\alpha_2$ and $\gamma_2$ remain close to zero at high temperature ($T^*=0.25$, DF phase). Entering the RP regime ($T^*=0.5$), a weak non-Gaussian peak emerges at short times ($t \approx 0.03\tau_D$), indicating the onset of heterogeneity due to percolation. With increasing attraction strength (decreasing $T^*$), both parameters grow, but $\alpha_2$ consistently exhibits larger and earlier peaks than $\gamma_2$.

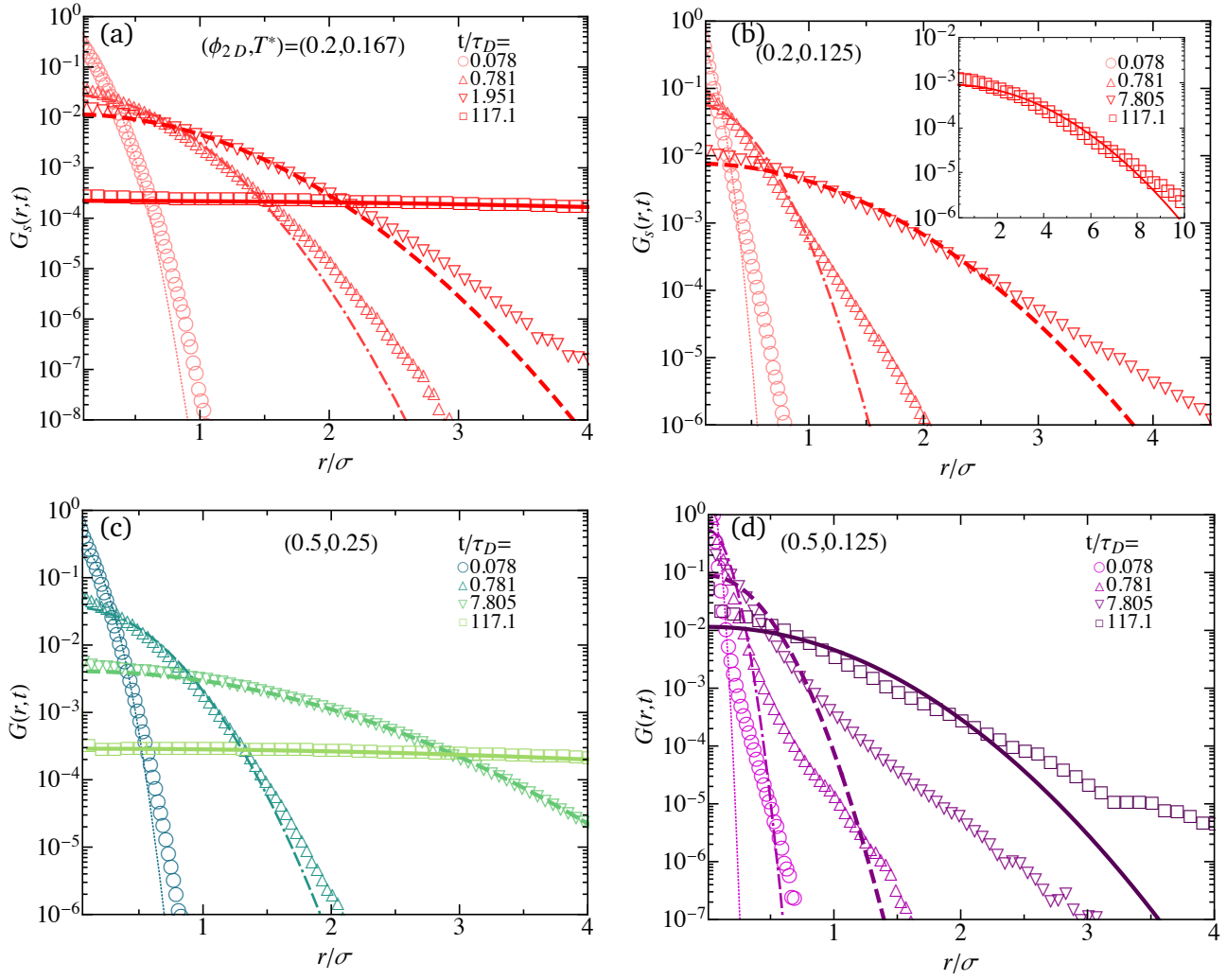
\begin{figure*}[!h]
	\centering
	\subfloat{		
		\begin{picture}(100,180)
			\put(-148,2){\includegraphics[width=0.45\textwidth,trim={0 0 0 0},clip]{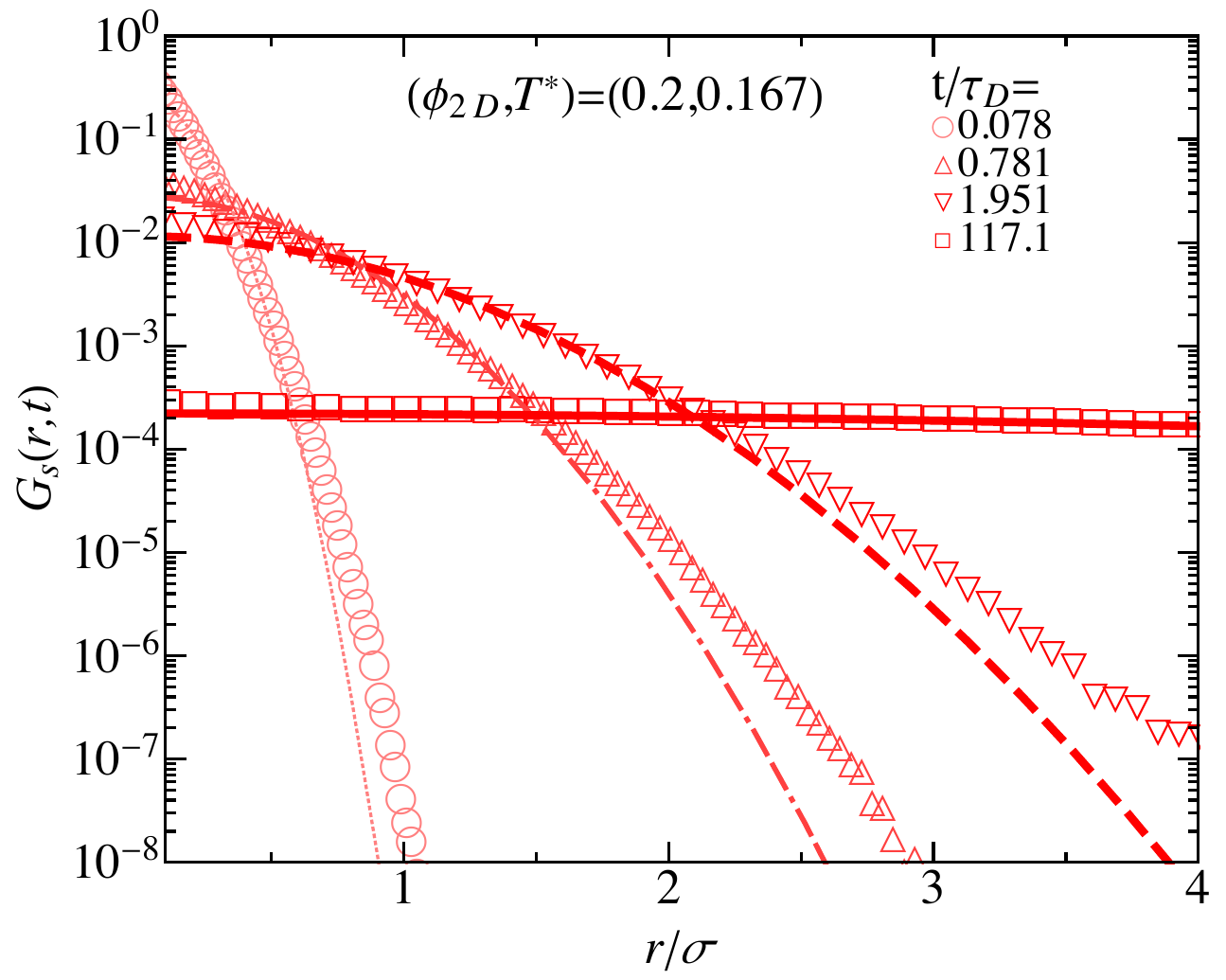}}
			\put(-110,175){(a)}
		\end{picture}
		\label{fig:vanHove1}
	}
	\subfloat{		
		\begin{picture}(100,180)
			\put(-5,0){\includegraphics[width=0.445\textwidth,trim={0 0 0 0},clip]{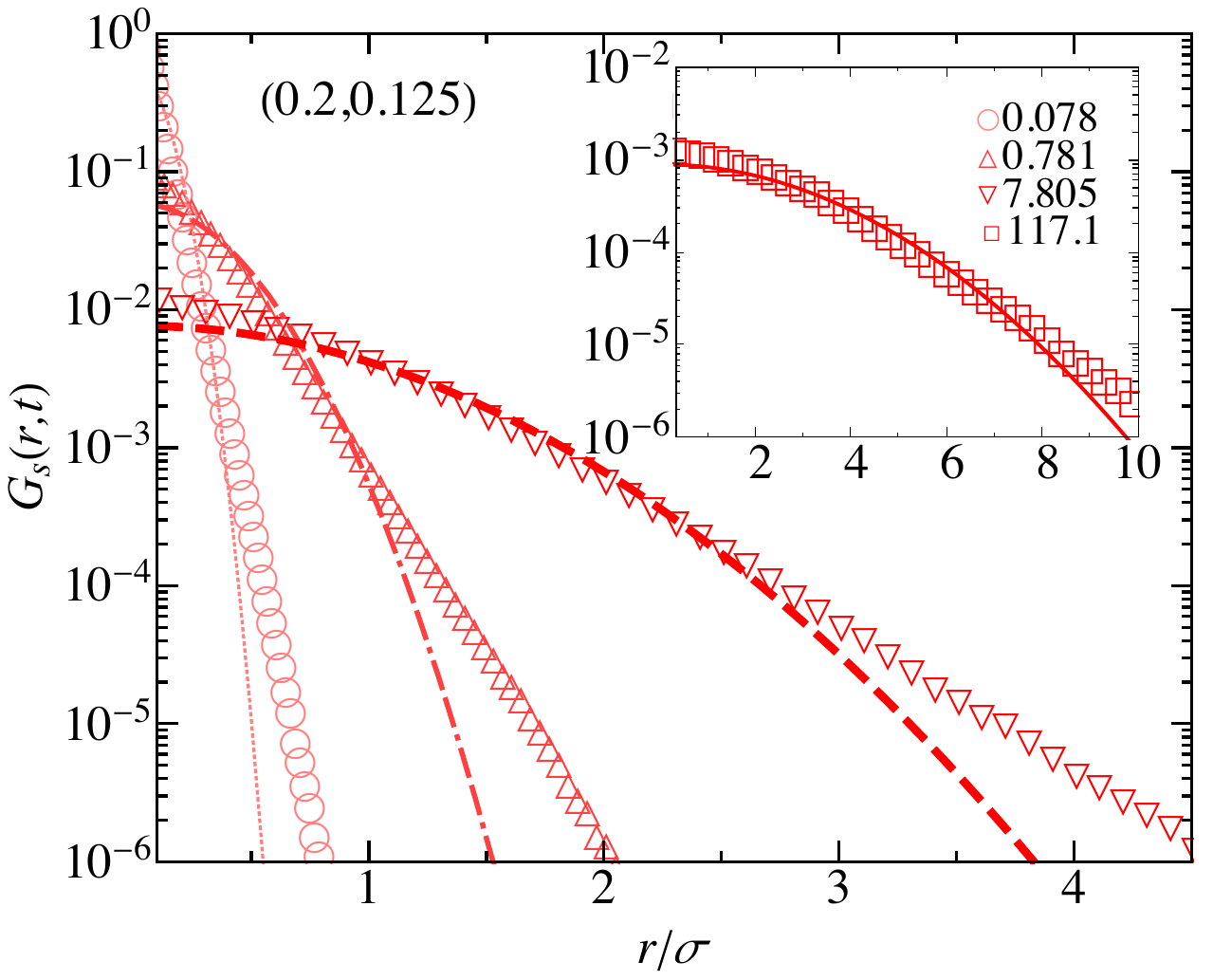}}
			\put(35,175){(b)}
		\end{picture}
		\label{fig:vanHove2}
	}\\
	\vspace{2mm}
	\subfloat{		
		\begin{picture}(100,180)
			\put(-148,2){\includegraphics[width=0.45\textwidth,trim={0 0 0 0},clip]{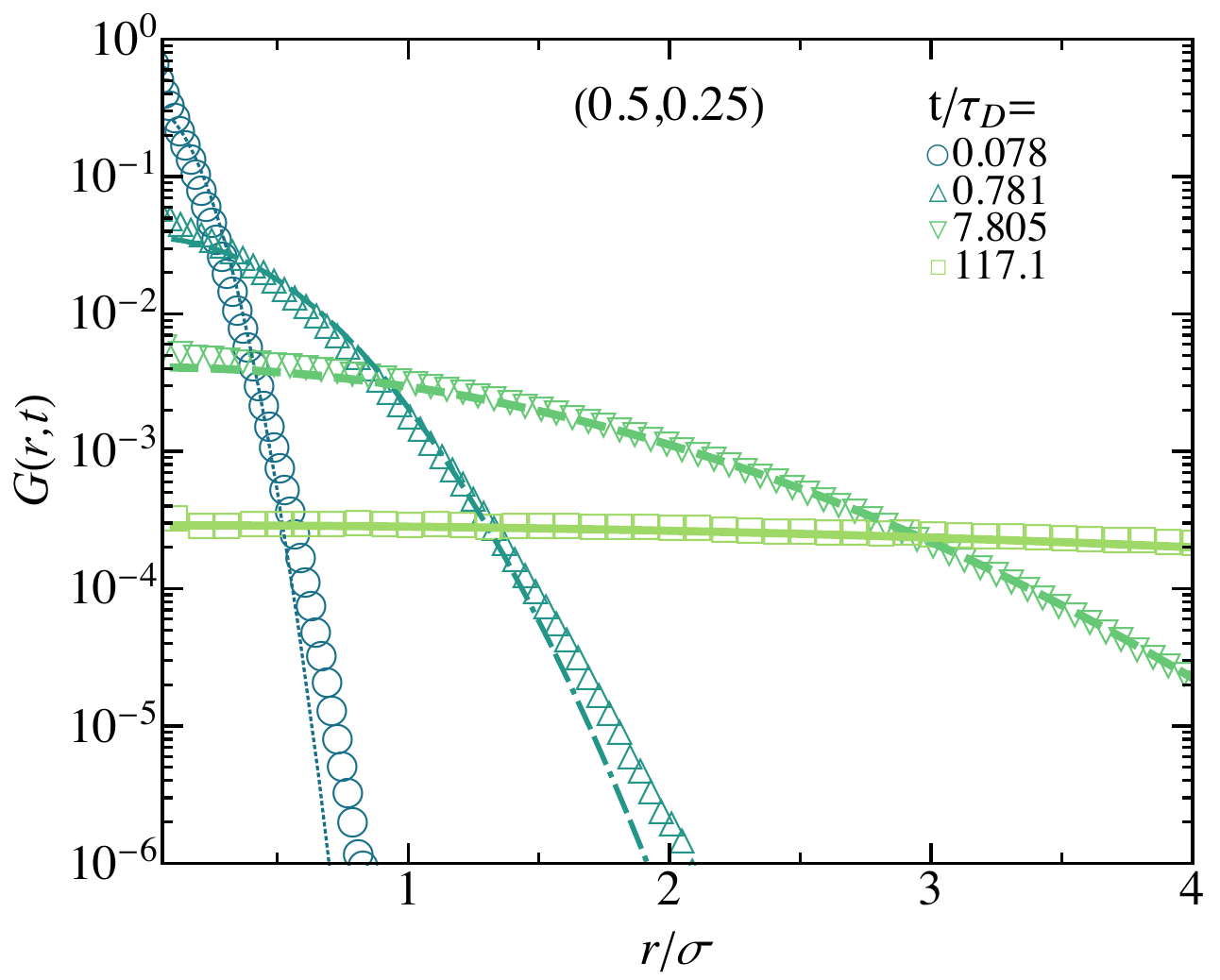}}
			\put(-110,175){(c)}
		\end{picture}
		\label{fig:vanHove3}
	}
	\subfloat{		
		\begin{picture}(100,180)
			\put(-5,0){\includegraphics[width=0.45\textwidth,trim={0 0 0 0},clip]{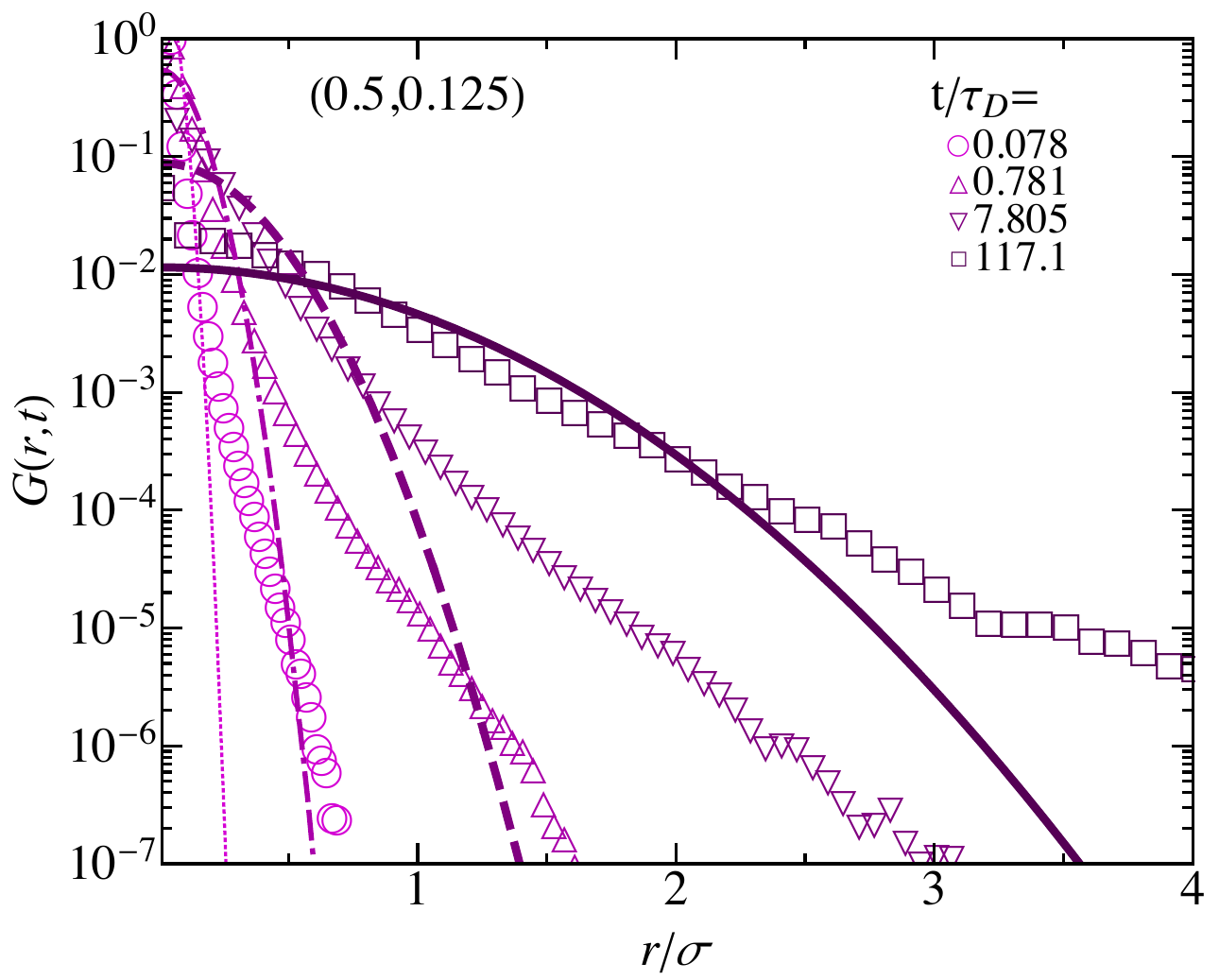}}
			\put(35,175){(d)}
		\end{picture}
		\label{fig:vanHove4}
	}
	\vspace{1mm}
	\caption{ Self-van Hove functions $G(r,t)$ for four representative state points at selected short and long times (in units of the diffusion time $\tau_D$), as indicated. The corresponding area fraction\--temperature pair values $(\phi_{2D},T^*)$ are given in brackets. Symbols denote LD simulation results obtained from Eq.~\eqref{eq:vanhof}, while lines represent the Gaussian approximation from Eq.~\eqref{eq:vanhof0}; different line and symbol styles correspond to different times $t$. Colors follow the phase classification used in Fig.~\ref{fig:statics}\protect\subref{fig:phases}, namely (a) EC phase near and (b) EC phase away from the DF-EC transition boundary, (c) RP and (d) CP phases, with varying intensities indicating different times.}
	\label{fig:vanhoves}
\end{figure*}
Intriguingly, at low temperatures $T^* \leq 0.125$, the non-Gaussian factor curves collapse for both $\alpha_2$ and $\gamma_2$, indicating a reduced sensitivity to further changes in attraction strength, consistent with the saturation behavior observed in structural quantities such as the mean cluster size and coordination number~\cite{tan04}. Again, the $\alpha_2$ measurement for the same temperature shows evidently higher values than $\gamma_2$ measurement. Surprisingly, the alternative Gaussian parameter $\gamma_2$ for $T^*\leq0.125$ at early and intermediate times ($t\leq$10$\tau_D$), within the measured time range, is even smaller than that for $T^*=0.167$. This can be attributed to the strong increase of  $\left(\Big<\frac{1}{r(t)}\Big>_2\right)^2$ due to enhanced localization, becoming significantly larger than the MSD and the mean-quartic displacement for computing $\alpha_2$ and $\gamma_2$, when temperature $T^*$ lowered from $0.167$ to $0.125$.

Interestingly, note that negative non-Gaussian parameters are not observed in our simulations, suggesting that the dynamic cage boundary around a particle is more fuzzy and the probability for moving far is enhanced.

\paragraph *{Self-van Hove function analysis} To further elucidate the origin of the observed non-Gaussian effects, we analyze the corresponding self-van Hove function, $G_s(r,t)$ (see Appendix for details) which characterizes the single-particle probability distribution of displacement over time. In contrast to the global measures provided by $\alpha_2$ and $\gamma_2$, $G_s(r,t)$ allows for a direct real-space identification of heterogeneous populations across the different Q2D-SALR phases. 

Fig.~\ref{fig:vanhoves} displays LD results (open symbols) for the self-van Hove function $G(r,t)$ in semi-logarithmic scale for representative area fraction–temperature state points $(\phi_{2D},T^*)$, at four selected times: $t/\tau_D = 0.078$ (circles), $0.781$ (upward triangles), $1.951$ (downward triangles), and $117.1$ (squares). These times correspond to regimes below, around, and far beyond the single-particle diffusion time $\tau_D$. The lines are Gaussian approximations obtained by inserting the corresponding MSD $W(t)$ into Eq.~\eqref{eq:vanhof0}, allowing a direct assessment of deviations from Gaussian diffusion. For the state point $(\phi_{2D},T^*)=(0.2,0.167)$, located close to the DF--EC transition boundary (cf. Fig.~\ref{fig:statics}\subref{fig:phases}), weak non-Gaussian behavior is already visible at short times, as shown in Fig.~\ref{fig:vanhoves}\subref{fig:vanHove1}. At long times ($t/\tau_D=117.1$), however, the Gaussian distribution is essentially recovered. This behavior is consistent with the corresponding non-Gaussian parameters shown in Fig.~\ref{fig:nonGauss}\subref{fig:nonGauss1}, where small but finite deviations occur at short times and vanish for $t\gg\tau_D$.

We next consider EC state points further away from the DF--EC boundary. One representative example is shown in Fig.~\ref{fig:vanhoves}\subref{fig:vanHove2} for $(\phi_{2D},T^*)=(0.2,0.125)$. At short and intermediate times, the LD results exhibit clear deviations from Gaussian behavior. Remarkably, $G(r,t)$ displays an approximately exponential decay over an extended range of displacements. Similar behavior is observed for all clustered state points sufficiently far from the phase boundary. Such exponential displacement distributions have been widely reported in Fickian yet non-Gaussian soft-matter and biological systems~\cite{wang:2009,Granick:2012,Barkai:2020,Pastore:2022}, and can be interpreted within the framework of \textit{superstatistics}~\cite{Metzler:2020} or related \textit{diffusing diffusivity}~\cite{Chubynsky:2014,Chechkin:2017} models. In this picture, particles diffuse in local environments (“patches”) characterized by different diffusivities $D$, distributed according to a probability distribution $p_D(D)$. In Q2D-SALR systems, different clusters effectively provide distinct local diffusivities. Averaging over these heterogeneous environments can give rise to a short-time displacement distribution with an approximately exponential (cusp-like) shape.

For the RP state point $(\phi_{2D},T^*)=(0.5,0.25)$, shown in Fig.~\ref{fig:vanhoves}\subref{fig:vanHove3}, $G(r,t)$ remains close to Gaussian even at short times. This indicates that, although the system is geometrically percolated, clustering-induced dynamic heterogeneity is still relatively weak. 

In contrast,  $G(r,t)$ in the CP phase, as shown in Fig.~\ref{fig:vanhoves}\subref{fig:vanHove4}, exhibits pronounced and more complex non-Gaussian behavior even for $t\gg\tau_D$. Here, $G(r,t)$ develops a stretched-exponential-like decay, reflecting strong dynamic heterogeneity, crowding, and persistent localization effects in the percolated clustered network.
\subsection{The role of short-time hydrodynamic correlations}
As discussed earlier, the collective dynamics of Q2D dispersions are governed by a strong interplay between structural correlations and hydrodynamic interactions (HIs), which become particularly pronounced at short times and large length scales~\cite{naegele2000wall,naegele2001jcp,naegele2002,Panzuela18,Alice:2024}. While the static structure factor $S(q)$ of SALR particles reflects the intermediate-range order (IRO) arising from competing interactions, the corresponding collective dynamics are strongly modified by Q2D hydrodynamic coupling. To this end, we present multiparticle collision dynamics (MPC) results for Q2D-SALR dispersions and discuss the associated steady-state and time-dependent hydrodynamic correlations in the following. We first examine the steady-state hydrodynamic function $H(q)$ and its dependence on area fraction and temperature. To elucidate the temporal development of hydrodynamic correlations, we then analyze the time-dependent hydrodynamic function $H(q,t)$ and the longitudinal current-current correlation function $J_d(q,t)$.
\subsubsection{Enhancement of large-scale hydrodynamic correlations}
\begin{figure}[!tp]
	\subfloat{		
		\begin{picture}(100,180)
			\put(-2,0){\includegraphics[width=0.47\textwidth,trim={0 0 0 0},clip]{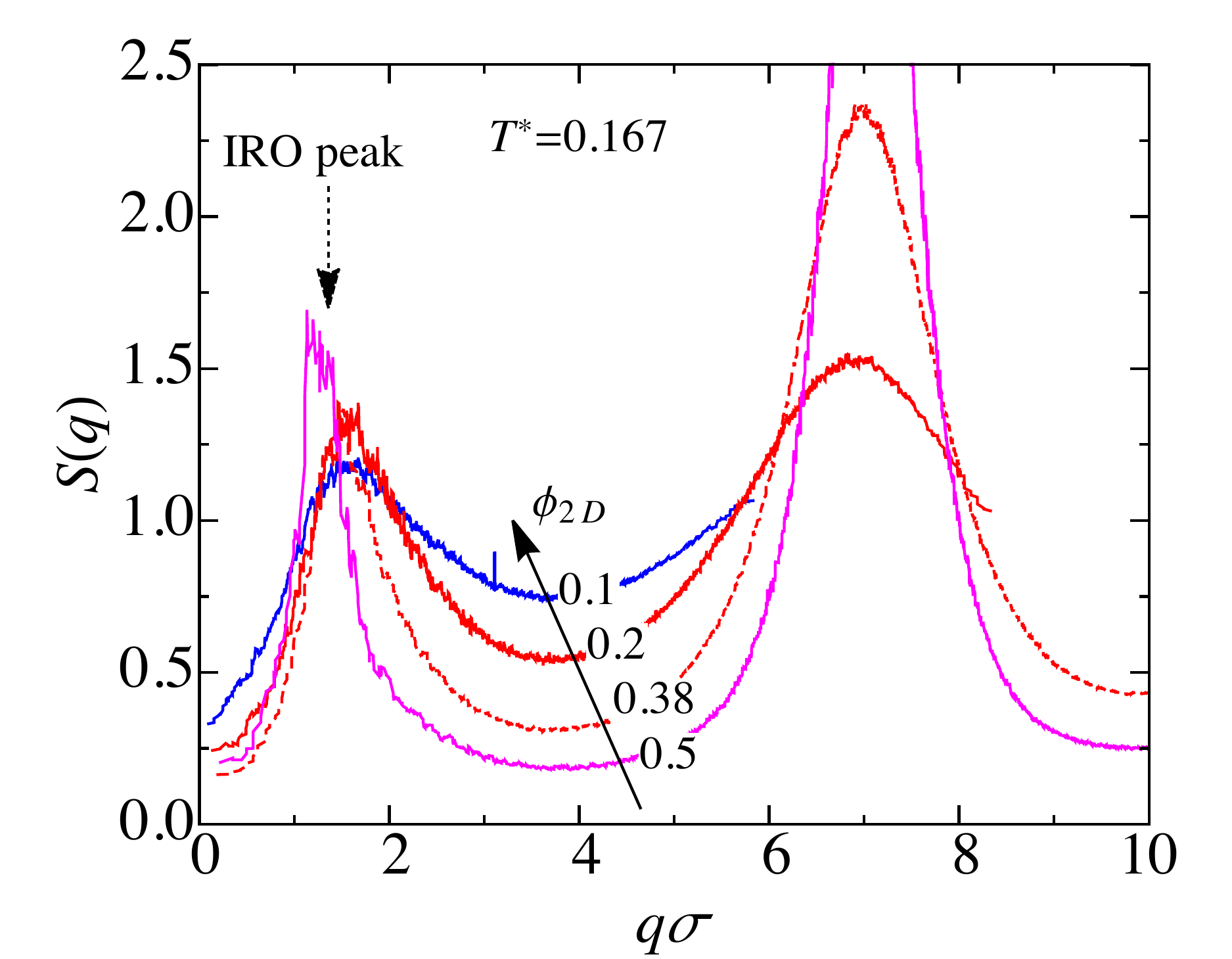}}
			\put(212,168){(a)}
		\end{picture}
		\label{fig:sq}
	}
	\\
	\subfloat{		
		\begin{picture}(100,180)
			\put(-2,0){\includegraphics[width=0.45\textwidth,trim={0 0 0 0},clip]{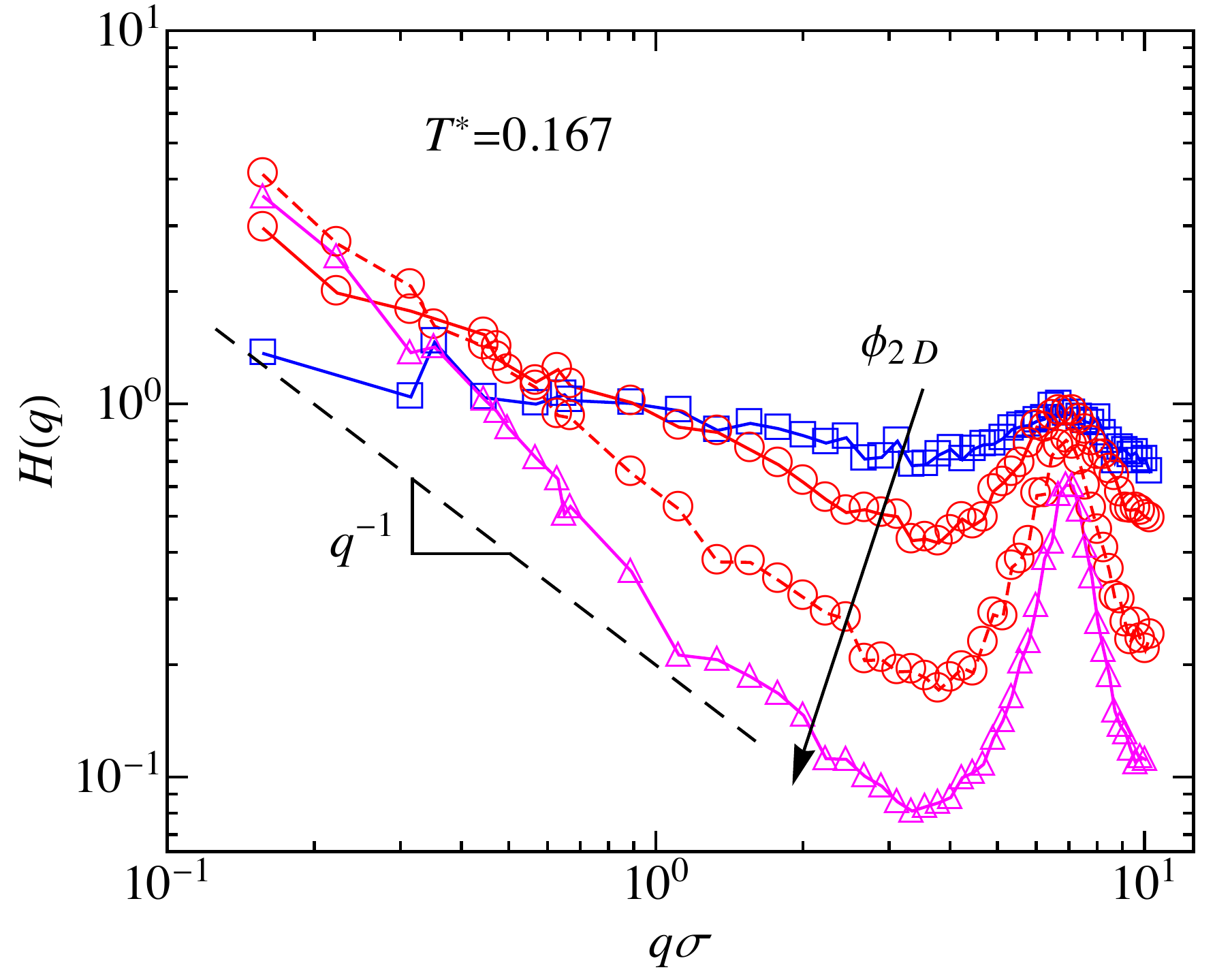}}
			\put(212,172){(b)}
		\end{picture}
		\label{fig:hq1}
	}\\
	\vspace{0mm}
	\subfloat{		
		\begin{picture}(100,180)
			\put(-2,0){\includegraphics[width=0.45\textwidth,trim={0 0 0 0},clip]{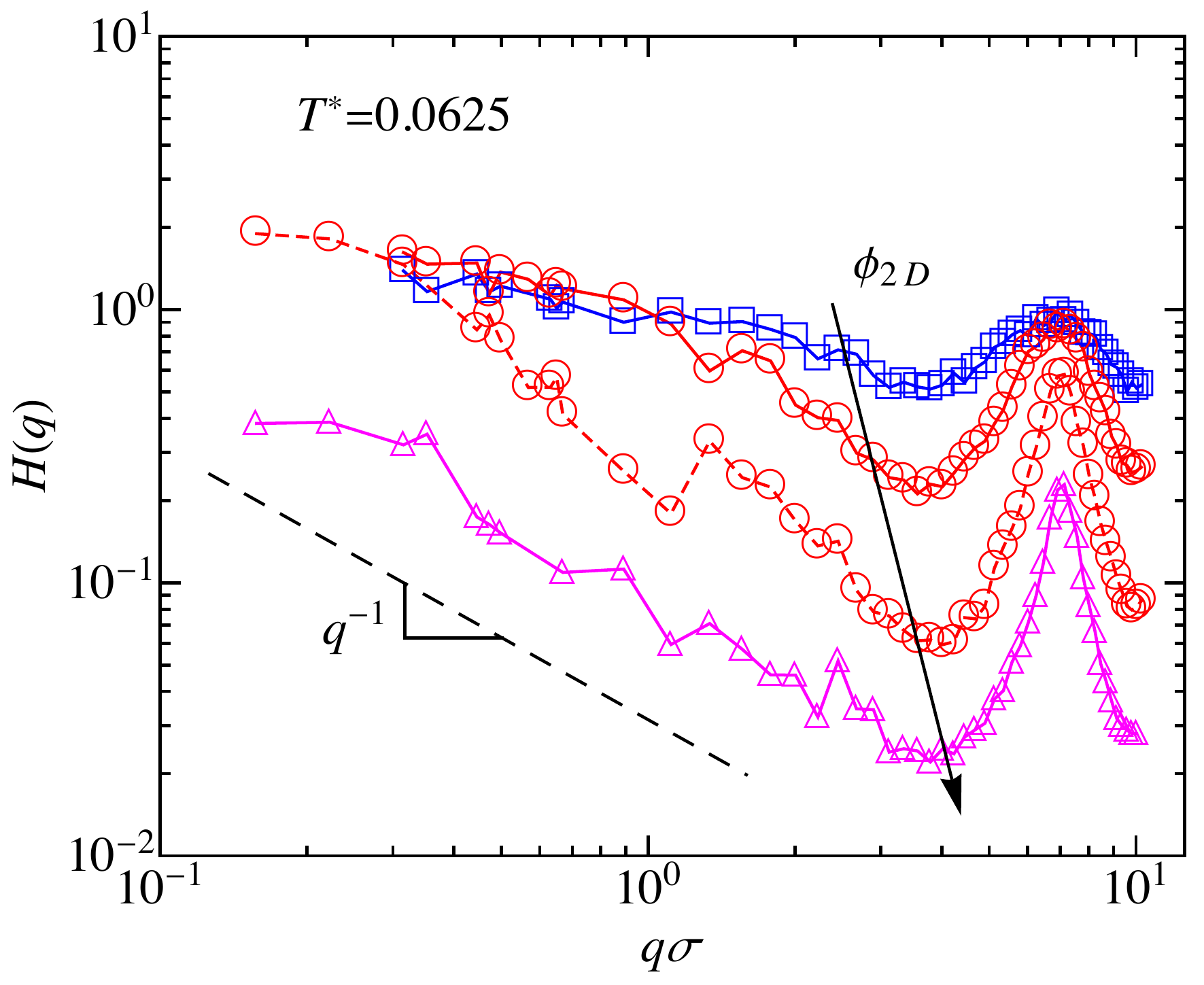}}
			\put(212,172){(c)}
		\end{picture}
		\label{fig:hq2}
	}
	\caption{\label{fig:hqs}(a)~Static structure factor of Q2D-SALR for different fraction $\phi_{2D}$ with pronounced IRO pre-peak.(b)~Hydrodynamic function $H(q)$ for different $\phi_{2D}$ values ($10\%$ with up-pointing triangles, $20\%$ down-pointing triangles, $38\%$ diamonds, and $50\%$ circles, respectively) at temperature $T^*=0.167$ and (c)~$0.0625$. The dashed line indicates the $q^{-1}$ divergence at small wave numbers based on Oseen (point particle) approximation in Eq.~\eqref{Q2DHqS}. The color of the curves encodes their respective generalized phases that used in Fig.~\ref{fig:statics}\protect\subref{fig:phases}.}
\end{figure}
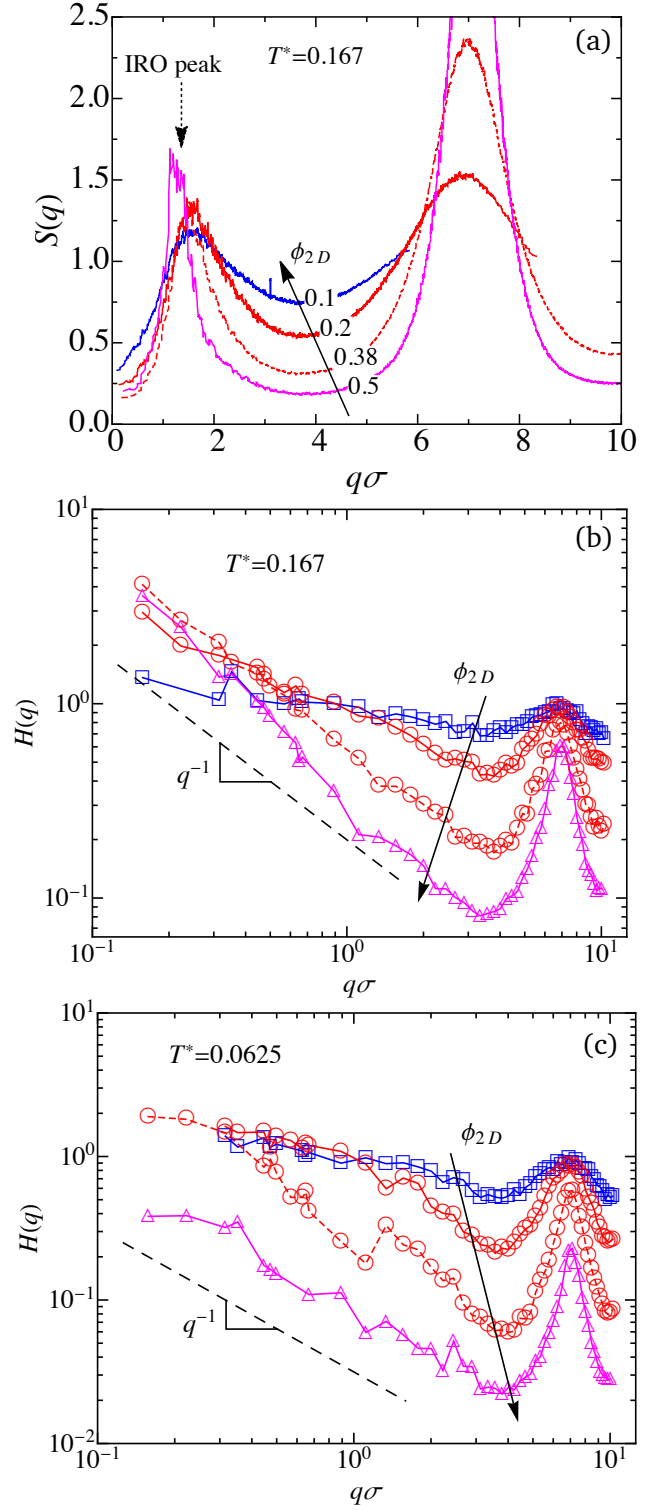
In the static structure factor $S(q)$ of Q2D-SALR particles, shown for comparison in Fig.~\ref{fig:hqs}\subref{fig:sq} (with the same color coding of phases as in the previous figures), the clustering-induced intermediate-range order (IRO) peak is closely associated with the thermal correlation length and the DF--EC transition, as discussed extensively in Ref.~\cite{tan04}. The IRO peak of $S(q)$ sharpens with increasing area fraction $\phi_{2D}$ and decreasing temperature, while its height increases accordingly. Akin to $S(q)$, an IRO peak in the hydrodynamic function $H(q)$ at approximately the same wavenumber is predicted theoretically for 3D-SALR systems~\cite{riest2015salr} and later confirmed experimentally~\cite{ries:2018}. This feature originates from the finite value of the collective diffusion coefficient $D(q)$ in three dimensions. Since $H(q)=D(q)S(q)/d_0$ (cf. Eq.~\eqref{Dq}), $H(q)$ closely follows the structure encoded in $S(q)$ and therefore exhibits a corresponding IRO peak.

MPC simulation results for the hydrodynamic function $H(q)$ of Q2D-SALR dispersions are shown in Fig.~\ref{fig:hqs} for several area fractions $\phi_{2D}$ at reduced temperatures $T^*=0.167$ (Fig.~\ref{fig:hqs}\subref{fig:hq1}) and $T^\ast=0.0625$ (Fig.~\ref{fig:hqs}\subref{fig:hq2}), respectively. The first observation is that all curves exhibit a nearest-neighbor peak at $q_m\approx2\pi/\sigma$. However, increasing area fraction and attraction strength suppress the magnitude of this peak, reflecting the suppression of HIs due to growing direct interactions.

Quite interestingly, the hydrodynamic function $H(q)$, and consequently the collective diffusion coefficient $D(q)$, is strongly enhanced at small wavenumbers. As indicated by the dashed lines in Fig.~\ref{fig:hqs}\subref{fig:hq1} and \subref{fig:hq2}, $H(q)$ grows approximately as $1/(q\sigma)$ upon decreasing $q$, typically for $q\sigma\lesssim1$. At higher area fractions and lower temperatures, this scaling extends up to $q\sigma\approx4$. As discussed earlier for generic Q2D dispersions, this $1/q$ enhancement originates from an out-of-plane backflow pattern that renders the in-plane hydrodynamics effectively quasi-compressible. Such a backflow mechanism, and the resulting small-$q$ divergence of $H(q)$ characteristic of Q2D systems~\cite{naegele2000wall,naegele2001jcp,naegele2002}, masks the IRO peak that would otherwise be observed in $H(q)$. Furthermore, an overall reduction in the magnitude of $H(q)$ is observed with increasing area fraction or attraction strength.
\begin{figure}[!h]
	\vspace{5mm}
	\subfloat{		
		\begin{picture}(100,180)
			\put(-2,0){\includegraphics[width=0.45\textwidth,trim={0 0 0 0},clip]{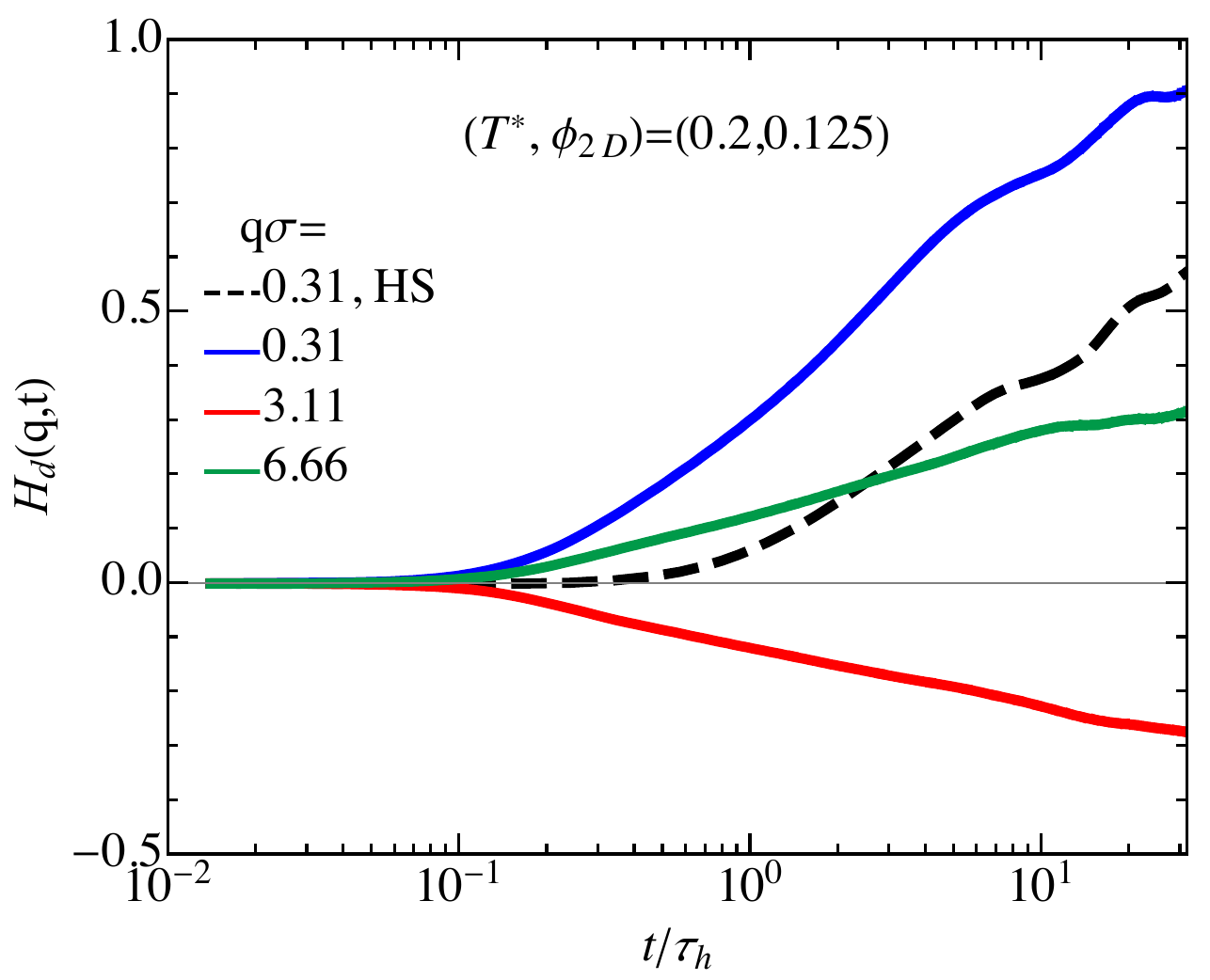}}
			\put(38,30){(a)}
		\end{picture}
		\label{fig:hdt}
	}\\
	\vspace{0mm}
	\subfloat{		
		\begin{picture}(100,180)
			\put(0,0){\includegraphics[width=0.45\textwidth,trim={0 0 0 0},clip]{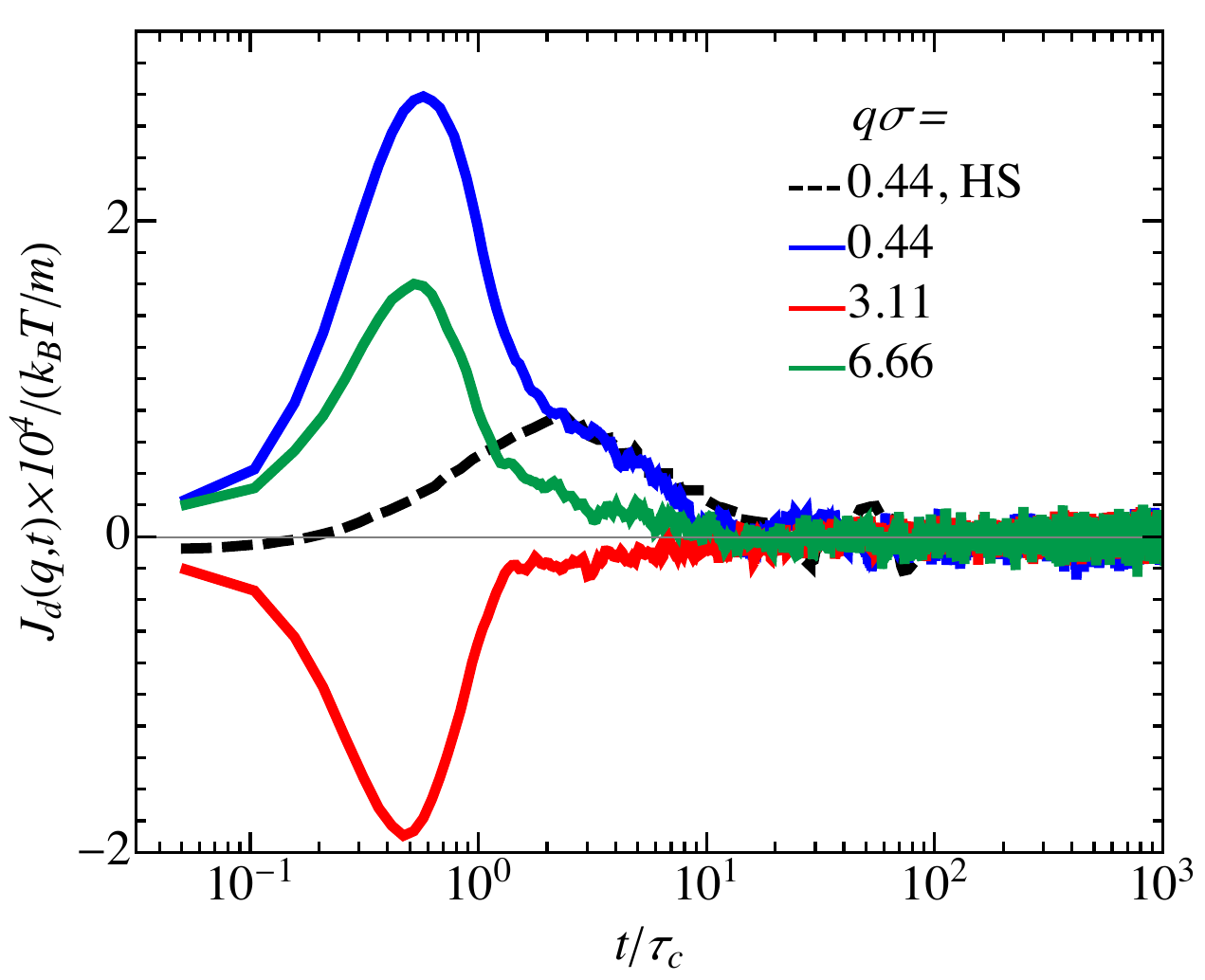}}
			\put(35,30){(b)}
		\end{picture}
		\label{fig:jdt}
	}
	\caption{\label{fig:hjdt}(a)~Time-evolution of the distinct hydrodynamic function $H_d(q,t)$ and (b)~corresponding distinct current-current correlation function $J_d(q,t)$ for a representative equilibrium cluster phase ($(\phi_{2D},T^*)=(0.2,0.125)$), at three distinct wave numbers (length scales) as displayed. Here, the MPC simulations are conducted in a cubic PBC box with linear size $100a$. The black dashed lines denote the hard-sphere (HS) reference system. For better illustration, only the small wave-number curves for HS system are shown. Note that the dimensionless time in $H_d(q,t)$ curves is $t$ normalized by the one particle vorticity relaxation time scale $t_h=R^2/\eta$, while in $J_d(q,t)$ curves is normalized by the sound wave relaxation time $t_c=R/v_\text{th}$.}
\end{figure}
\subsubsection{Time-dependent hydrodynamic correlation}
We now investigate how hydrodynamic interactions develop in time at different length scales. To quantify this, we first compute the time-dependent hydrodynamic function, $H(q,t)$, according to Eq.~\eqref{eq:Hqt}. Since the self part is essentially the normalized MSD, we focus here on the distinct contribution, $H_d(q,t)$, which measures the displacement cross-correlations between different particles (Eq.~\eqref{eq:Hsdqt}). The time is normalized by $\tau_h$, and the results are displayed in Fig.~\ref{fig:hjdt}\subref{fig:hdt}. We present $H_d(q,t)$ for the representative equilibrium-cluster state point $(\phi_{2D},T^*)=(0.2,0.125)$ at three wave numbers, $q\sigma=0.31$ (blue), $3.11$ (red), and $6.66$ (green). For comparison, the corresponding Q2D hard-sphere result at $q\sigma=0.31$ is shown by the dashed black curve.

For the Q2D hard-sphere system, the onset of hydrodynamic correlations occurs at $t\approx\tau_h$, where $H_d(q,t)$ becomes apparently non-zero, as expected. In contrast, for the Q2D-SALR system, this onset is shifted to significantly earlier times (ca. $0.2\tau_h$). We postulate this acceleration to the direct SALR interactions, which facilitate transient aggregation and clustering. The resulting clusters behave as effective hydrodynamic objects with an enlarged hydrodynamic radius, rendering hydrodynamic correlations relevant on shorter timescales.

More intriguingly, three characteristic hydrodynamic length scales can be identified from Fig.~\ref{fig:hjdt}\subref{fig:hdt}. First, at $q\sigma=0.31<1$, corresponding to the large-scale enhanced-hydrodynamic regime, $H_d(q,t)$ is positive, reflecting collective co-motion of particles. Second, at $q\sigma\approx2\pi$, corresponding approximately to the nearest-neighbor cage diameter $\sigma$, a moving SALR particle tends to drag its neighboring particles along the same direction, resulting again in positive correlations. Third, at the intermediate length scale $q\sigma=6.66$, corresponding roughly to a distance of order $2\sigma$, the motion of a particle induces backward flow that sweeps surrounding particles in the opposite direction, leading to negative correlations.

In a similar vein, to investigate sound-wave propagation in the same Q2D-SALR system, we calculate the distinct longitudinal current-current correlation function $J_d(q,t)$ (cf. Eqs.~\eqref{eq:Jqt} and \eqref{eq:JJiqt}). The results are shown in Fig.~\ref{fig:hjdt}\subref{fig:jdt}, where time is normalized by the sonic timescale $\tau_c$. Again, the cross-correlations emerge substantially earlier than that in the hard-sphere reference system, and persist for times considerably longer than $\tau_c$ before becoming overdamped, which coincide the observations in 3D colloidal suspensions~\cite{bakker2002role}. Similar to $H_d(q,t)$, the signs of the correlation peaks reveal the same three characteristic length scales. This suggests that, on sonic timescales, sound-wave propagation remains coherent with the underdamped hydrodynamic flow generated by the Brownian particles.

Given that $H_d(q,t)$ integrates the velocity fluctuations over time and directly isolates and highlights the shear mode (vorticity diffusion), our $H_d(q,t)$ and $J_d(q,t)$ results essentially demonstrate that both shear-wave-mediated and sound-wave-mediated spatio-temporal cross-correlations are enhanced and become observable at earlier times in clustered Q2D-SALR systems than in the corresponding hard-sphere reference system~\cite{dominguez2014signature,panzuela2017collective,bakker2002role}.

\section{Summary and Conclusions}
\label{sec:sum+conc}
In this work, we have comprehensively investigated the dynamic clustering properties, non-Gaussianness, self- and collective diffusion, and short-time collective hydrodynamic correlations of Q2D-SALR dispersions of Brownian particles. The analysis is carried out using Langevin dynamics (LD), which excludes hydrodynamic interactions (HIs), and multiparticle collision dynamics (MPC), which explicitly resolves fluctuating hydrodynamics and momentum transport over broad spatio-temporal scales.

The considered Q2D-SALR system exhibits a rich interplay between direct interactions and Q2D confinement. Our LD results demonstrate that clustering substantially modifies both short- and long-time self-diffusion across all four phases: dispersed fluid (DF), equilibrium clusters (EC), random percolation (RP), and cluster percolation (CP). In the DF, EC and RP phases, particles exhibit short-time diffusion followed by slower long-time Brownian diffusion, as they can escape from transient clusters or remain mobile in the percolated network. Diffusion in the EC phase is distinctly slower than in the DF and RP phases due to stable cluster formation. In the CP phase, slow dynamics and subdiffusion emerge, with particles remaining trapped in the percolated cluster network throughout the measured time window.

Analysis of the bond-correlation function reveals that the cluster lifetime $\tau_b$, well described by a stretched exponential, is primarily controlled by attraction strength rather than concentration. A threshold $\tau_b/\tau_D\approx10$ empirically separates the unclustered (DF, RP) from the clustered (EC, CP) phases, providing a dynamic criterion complementary to the static criteria of Ref.\cite{tan04}. The two-stage decay of $\tau_b$ with increasing effective temperature can be rationalized by two-particle Smoluchowski theory and Kramers' theory\cite{das2018clustering}.

The relaxation of local hexagonal order follows a similar phase-dependent trend. Systems in the DF and RP phases decay rapidly, consistent with the weak packing-induced ordering in the RP phase~\cite{tan04}. In the EC phase, hexagonal correlations persist beyond $\tau_D$ but are relatively insensitive to further increases in attraction strength, likely due to rotational motion of finite-sized clusters. In the CP phase, however, the relaxation time grows approximately exponentially with attraction strength, reaching up to $10^4\tau_D$.

The associated non-Gaussian dynamics are reflected in both standard and non-standard non-Gaussian parameters, $\alpha_2$ and $\gamma_2$, respectively. Attraction induced clustering leads to strong non-Gaussianity in the system at time $t\lesssim \tau_D$. For the EC phase, the nearly Fickian diffusion is recovered if correlation time exceeds around $100\tau_D$. In contrast, the CP phase exhibits even stronger non-Gaussian parameter values and persists beyond $100\tau_D$. The long-time (and therefore long distance) non-Gaussian behavior can be better elucidated by $\gamma_2$, which gives greater weight to particles with larger displacements at longer times. Accordingly, clustering induced dynamic heterogeneity leads to strong deviations of the self-van Hove function from Gaussian statistics. Most importantly, at EC states, the self-van Hove function exhibits approximately exponential spatial tails over intermediate times. Such behavior is consistent with a superstatistical picture where particles experience a broad spectrum of local diffusivities generated by transient and heterogeneous clusters induced by the competing SALR interactions.

More importantly, we investigated the effect of HIs through MPC simulations, which reveal the abnormal collective dynamics of Q2D systems. The hydrodynamic function $H(q)$ exhibits a strong long-wavelength enhancement scaling approximately as $H(q)\sim q^{-1}$ for $q\rightarrow0$, leading to anomalously enhanced collective diffusion. In contrast to three-dimensional SALR systems, where the IRO peak remains clearly visible in $H(q)$~\cite{riest2015salr,ries:2018}, the strong Q2D hydrodynamic enhancement masks the corresponding IRO structural feature. 

To elucidate the temporal development of hydrodynamic correlations, we analyzed the time-dependent hydrodynamic function, especially its distinct component $H_d(q,t)$. The results reveal that hydrodynamic correlations develop gradually from initially ballistic dynamics toward the fully established overdamped Q2D regime. Depending on wave number (corresponding to different length scales), the distinct component exhibits positive or negative values corresponding respectively to cooperative hydrodynamic drag and anticorrelated backflow motion. Furthermore, the longitudinal current-current correlation functions $J_d(q,t)$ demonstrate that sound-mediated momentum transport remains coherent over time scales substantially exceeding the sonic time $\tau_c$, indicating a coupling between propagating compressional modes and collective hydrodynamic flow. Interestingly, SALR-interaction-induced clustering leads to an earlier onset of collective cross-correlations in both the hydrodynamic shear mode and the sound mode.

Overall, the present results demonstrate that Q2D confinement fundamentally alters the collective dynamics of SALR dispersions through the coupling of competing interactions with hydrodynamic correlations. These findings contribute to the understanding of confined colloidal and protein-like systems where structural self-organization and hydrodynamic coupling coexist over multiple spatio-temporal scales. Our dynamic analysis suggests that the structural identification of clustered phases should be complemented by dynamic criteria based on cluster persistence and orientational relaxation, and corresponding hydrodynamic correlations. A natural extension of the current work is to gain dynamic insight into the clustering properties of more complex Q2D systems such as particles confined to a fluid–fluid interface with viscosity contrast~\cite{tan03}, which are biologically relevant to membrane protein aggregation effects at the intracellular level. A further extension of the present work concerns the hydrodynamic correlations of the employed Q2D-SALR model at extensive time scales, focusing on the EC and CP states. There, questions related to the Fickian versus non-Fickian and Gaussian versus non-Gaussian behaviors of Q2D-SALR particles can be addressed, in conjunction with the related theoretical concepts of superstatistics~\cite{Chechkin:2017,Metzler:2020,Granick:2012} and diffusing diffusivity~\cite{Chubynsky:2014}. 

\section*{Author contributions}
GN conceived the research project and derived the alternative non-Gaussian parameter in 2D. ZT developed the LD simulation code and partially the MPC code, conducted the simulation, and analyzed the obtained data. All authors together discussed the results and writing the manuscript.

\section*{Acknowledgements}
ZT and GN thank Roland G. Winkler (FZ J\"ulich) for sharing the MPC-GPU codes, Shibananda Das (Indian Institute of Science) for the help on cluster analysis. ZT thanks Adolfo J. Banchio (UN de Cordoba) for the discussion on the time-correlation of hexactic parameter function. The authors gratefully acknowledge computing time granted through JARA-HPC on the supercomputer JURECA at Forschungszentrum J\"ulich~\cite{jureca}.

\section*{Appendix: Self van Hove function, standard and non-standard non-Gaussian parameters}
\label{sec:apdx}
The self van Hove function $G(r,t)$ for an isotropic dispersion of $N$ equal spherical particles in $d$ dimensions is defined as~\cite{hansen:2013}
\begin{align}
	G(r,t) = \Big< \frac{1}{N}\sum_{i=1}^N \delta^{(d)}\left({\bm R}-\Delta{\bm r}_i(t)\right)\Big>_d, \quad G(r,0) = \delta^{(d)}({\bm R}),
	\label{eq:vanhof}
\end{align}
with normalization $\int d^dr\;G(r,t) = 1$. Here, $ \Delta{\bm R}_i(t) = {\bm R}_i(t) - {\bm R}_i(0)$, $i=1,2,...,N$, and $r=|{\bm R}|$. It is a measure of how the spatial correlations of a tagged particle evolve in time. For a semi-dilute colloidal system in the fluid regime away from glass/gel transitions, $G(r,t)$ is well approximated by the isotropic Gaussian form
\begin{align}
	G(r,t) \approx G^{(g)}(r,t) = \frac{1}{(4\pi W(t))^{d/2}} \exp\left(-\frac{r^2}{4W(t)}\right),
	\label{eq:vanhof0}
\end{align}
where $W(t) = \langle r^2(t) \rangle_d / (2d)$ is the mean-squared displacement (MSD). In the overdamped Brownian regime, $W(t)$ is monotonic and typically linear in time (Fickian), though non-Gaussian behavior may still occur \cite{Granick:2012}. 

The Gaussian self van Hove function $G^{(g)}(r,t)$ satisfies the diffusion-type equation:
\begin{align}
	\partial_t G^{(g)}(r,t) = \dot{W}(t) \Delta_{\bm R}^{(d)} G^{(g)}(r,t),
\end{align}
with initial condition $G^{(g)}(r,0) = \delta^{(d)}({\bm R})$.

The Fourier transform of $G(r,t)$, termed as \textit{self-intermediate scattering function} is
\begin{align}
	G(q,t) = \left< \frac{1}{N}\sum_{i=1}^N e^{i{\bm q}\cdot \Delta{\bm R}_i(t)} \right>_d, \quad G(q,0) = 1,
\end{align}
with wavevector ${\bm q}\in \mathbb{R}^D$ is dependent on $q=|{\bm q}|$ only for an isotropic system. For an ergodic system (no glass or gel) it is $G(q,t\to\infty)=0$ since a particle looses then its memory about its earlier positions during a (sufficiently) long time span $t$. For an isotropic Gaussian system, it is 
\begin{align}
	G^{(g)}(q,t) = \exp\left(-q^2 W(t)\right).
\end{align}
Expanding $G(q,t)$ around small $q$,
\begin{align}
	G(q,t) = G^{(g)}(q,t) \left[1 + \frac{1}{2}\alpha_2^{(d)}(t) (q^2 W(t))^2 + \mathcal{O}(q^6) \right].
\end{align}

Deviations from Gaussianity (e.g. multi-length scale relaxation, dynamic heterogeneity \cite{KaurDas:2003,SchnyderDullens:2017}) are captured by non-Gaussian parameters. The standard $2n$-th order parameter is
\begin{align}
	\alpha_n^{(d)}(t) = \frac{\langle r^{2n}(t) \rangle_d}{c_n^{(d)} \langle r^2(t) \rangle_d^n} - 1, \quad \alpha_n^{(d)}(t) = 0 \text{ for Gaussian } G(r,t),
\end{align}
with $c_n^{(d)}$ the Gaussian moment ratio coefficients. For $d=2$, $c_2^{(2)} = 2$, yielding
\begin{align}
	\alpha_2^{(2)}(t) = \frac{\langle r^4(t) \rangle}{2 \langle r^2(t) \rangle^2} - 1 \geq -\tfrac{1}{2}.
\end{align}

Negative $\alpha_2(t)$ implies suppressed long displacements; positive values enhance the wings of $G(r,t)$, consistent with mobile particle clusters.

An alternative 3D non-Gaussian parameter $\gamma_2^{(3)}(t) $, according to Flenner and Szamel~\cite{FlennerSzamel:2005, Salzmann:2006}, sensitive to subdiffusive dynamics, is defined as
\begin{align}
	\gamma_2^{(3)}(t) = \frac{1}{3} \langle r^2(t) \rangle_3 \left< \frac{1}{r^2(t)} \right>_3.
\end{align}
For Gaussian $G(r,t)$ in $d=3$, $\langle 1/r^2(t) \rangle = 1/(2W(t))$, hence $\gamma_2^{(3)}(t)=0$.

Here, $\gamma_2^{(3)}(t)$ weights strongly particles which have moved less than expected for the Gaussian distribution through its $1/r^2$ moment, and particles moving farther than Gaussian ones through the $r^2$ moment. Salzman and Schweizer claim that different from $\alpha_2^{(d=3)}(t)$ this alternative non-Gaussian factor allows for detecting non-Fickean diffusion (i.e. a MSD non-linear in time) at relatively long times and hence long distances.

Note that a non-standard Gaussian parameter in two dimensions can not include a factor $\langle 1/r^2\rangle_2$  
as it is infinite for a Gaussian distribution. To this end, a non-standard Gaussian factor 
invoking the 2D-MSD $\!\big< r^2(t)\big>_2\;\!$ and $\langle 1/r\rangle_2$ can be defined analogously in two dimensions as\begin{align}
	\gamma^\text{{(d=2)}}(t)= \frac{1}{\pi}\;\!\big< r^2(t)\big>_2\;\!\left(\Big<\frac{1}{r(t)}\Big>_2\right)^2-1\,,
\end{align} 
which becomes zero for a Gaussian distributed $r$.






\begin{mcitethebibliography}{112}
	\providecommand*{\natexlab}[1]{#1}
	\providecommand*{\mciteSetBstSublistMode}[1]{}
	\providecommand*{\mciteSetBstMaxWidthForm}[2]{}
	\providecommand*{\mciteBstWouldAddEndPuncttrue}
	{\def\EndOfBibitem{\unskip.}}
	\providecommand*{\mciteBstWouldAddEndPunctfalse}
	{\let\EndOfBibitem\relax}
	\providecommand*{\mciteSetBstMidEndSepPunct}[3]{}
	\providecommand*{\mciteSetBstSublistLabelBeginEnd}[3]{}
	\providecommand*{\EndOfBibitem}{}
	\mciteSetBstSublistMode{f}
	\mciteSetBstMaxWidthForm{subitem}
	{(\emph{\alph{mcitesubitemcount}})}
	\mciteSetBstSublistLabelBeginEnd{\mcitemaxwidthsubitemform\space}
	{\relax}{\relax}
	
	\bibitem[Liu and Xi(2019)]{yliu:2019}
	Y.~Liu and Y.~Xi, \emph{Curr. Opin. Colloid Interface Sci.}, 2019,
	\textbf{39}, 123--136\relax
	\mciteBstWouldAddEndPuncttrue
	\mciteSetBstMidEndSepPunct{\mcitedefaultmidpunct}
	{\mcitedefaultendpunct}{\mcitedefaultseppunct}\relax
	\EndOfBibitem
	\bibitem[Ruiz-Franco and Zaccarelli(2021)]{ruiz:2021}
	J.~Ruiz-Franco and E.~Zaccarelli, \emph{Annu. Rev. Condens. Matter Phys.},
	2021, \textbf{12}, 51--70\relax
	\mciteBstWouldAddEndPuncttrue
	\mciteSetBstMidEndSepPunct{\mcitedefaultmidpunct}
	{\mcitedefaultendpunct}{\mcitedefaultseppunct}\relax
	\EndOfBibitem
	\bibitem[Stradner \emph{et~al.}(2004)Stradner, Sedgwick, Cardinaux, Poon,
	Egelhaaf, and Schurtenberger]{strad:2004}
	A.~Stradner, H.~Sedgwick, F.~Cardinaux, W.~C. Poon, S.~U. Egelhaaf and
	P.~Schurtenberger, \emph{Nature}, 2004, \textbf{432}, 492--495\relax
	\mciteBstWouldAddEndPuncttrue
	\mciteSetBstMidEndSepPunct{\mcitedefaultmidpunct}
	{\mcitedefaultendpunct}{\mcitedefaultseppunct}\relax
	\EndOfBibitem
	\bibitem[Tan \emph{et~al.}(2024)Tan, Calandrini, Dhont, and Nägele]{tan04}
	Z.~Tan, V.~Calandrini, J.~K.~G. Dhont and G.~Nägele, \emph{Soft Matter}, 2024,
	\textbf{20}, 9528--9546\relax
	\mciteBstWouldAddEndPuncttrue
	\mciteSetBstMidEndSepPunct{\mcitedefaultmidpunct}
	{\mcitedefaultendpunct}{\mcitedefaultseppunct}\relax
	\EndOfBibitem
	\bibitem[Destainville \emph{et~al.}(2018)Destainville, Manghi, and
	Cornet]{Destainville2018}
	N.~Destainville, M.~Manghi and J.~Cornet, \emph{Biomolecules}, 2018,
	\textbf{8}, 64\relax
	\mciteBstWouldAddEndPuncttrue
	\mciteSetBstMidEndSepPunct{\mcitedefaultmidpunct}
	{\mcitedefaultendpunct}{\mcitedefaultseppunct}\relax
	\EndOfBibitem
	\bibitem[Wasnik \emph{et~al.}(2015)Wasnik, Wingreen, and
	Mukhopadhyay]{wasnik:2015}
	V.~Wasnik, N.~S. Wingreen and R.~Mukhopadhyay, \emph{PLoS One}, 2015,
	\textbf{10}, 1--13\relax
	\mciteBstWouldAddEndPuncttrue
	\mciteSetBstMidEndSepPunct{\mcitedefaultmidpunct}
	{\mcitedefaultendpunct}{\mcitedefaultseppunct}\relax
	\EndOfBibitem
	\bibitem[Sieber \emph{et~al.}(2007)Sieber, Willig, Kutzner, Gerding-Reimers,
	Harke, Donnert, Rammner, Eggeling, Hell, Grubmüller, and Lang]{sieber:2007}
	J.~J. Sieber, K.~I. Willig, C.~Kutzner, C.~Gerding-Reimers, B.~Harke,
	G.~Donnert, B.~Rammner, C.~Eggeling, S.~W. Hell, H.~Grubmüller and T.~Lang,
	\emph{Science}, 2007, \textbf{317}, 1072--1076\relax
	\mciteBstWouldAddEndPuncttrue
	\mciteSetBstMidEndSepPunct{\mcitedefaultmidpunct}
	{\mcitedefaultendpunct}{\mcitedefaultseppunct}\relax
	\EndOfBibitem
	\bibitem[Gurry \emph{et~al.}(2009)Gurry, Kahramanohgullari, and
	Endres]{gurry:2009}
	T.~Gurry, O.~Kahramanohgullari and R.~G. Endres, \emph{PLoS One}, 2009,
	\textbf{4}, e6148\relax
	\mciteBstWouldAddEndPuncttrue
	\mciteSetBstMidEndSepPunct{\mcitedefaultmidpunct}
	{\mcitedefaultendpunct}{\mcitedefaultseppunct}\relax
	\EndOfBibitem
	\bibitem[Godfrin \emph{et~al.}(2014)Godfrin, Valadez-Pérez, Castañeda-Priego,
	Wagner, and Liu]{godfrin:2014}
	P.~D. Godfrin, N.~E. Valadez-Pérez, R.~Castañeda-Priego, N.~J. Wagner and
	Y.~Liu, \emph{Soft Matter}, 2014, \textbf{10}, 5061--5071\relax
	\mciteBstWouldAddEndPuncttrue
	\mciteSetBstMidEndSepPunct{\mcitedefaultmidpunct}
	{\mcitedefaultendpunct}{\mcitedefaultseppunct}\relax
	\EndOfBibitem
	\bibitem[Zhuang and Charbonneau(2016)]{zhuang:2016}
	Y.~Zhuang and P.~Charbonneau, \emph{J. Phys. Chem. B}, 2016, \textbf{120},
	6178--6188\relax
	\mciteBstWouldAddEndPuncttrue
	\mciteSetBstMidEndSepPunct{\mcitedefaultmidpunct}
	{\mcitedefaultendpunct}{\mcitedefaultseppunct}\relax
	\EndOfBibitem
	\bibitem[Zhuang and Charbonneau(2016)]{charbonneau:2016}
	Y.~Zhuang and P.~Charbonneau, \emph{J. Phys. Chem. B}, 2016, \textbf{120},
	7775--7782\relax
	\mciteBstWouldAddEndPuncttrue
	\mciteSetBstMidEndSepPunct{\mcitedefaultmidpunct}
	{\mcitedefaultendpunct}{\mcitedefaultseppunct}\relax
	\EndOfBibitem
	\bibitem[Litniewski \emph{et~al.}(2025)Litniewski, Gozdz, and
	Ciach]{Litniewski:2025}
	M.~Litniewski, W.~T. Gozdz and A.~Ciach, \emph{Soft Matter}, 2025, \textbf{21},
	6801--6813\relax
	\mciteBstWouldAddEndPuncttrue
	\mciteSetBstMidEndSepPunct{\mcitedefaultmidpunct}
	{\mcitedefaultendpunct}{\mcitedefaultseppunct}\relax
	\EndOfBibitem
	\bibitem[Archer(2008)]{archer:2008pre}
	A.~J. Archer, \emph{Phys. Rev. E}, 2008, \textbf{78}, 031402\relax
	\mciteBstWouldAddEndPuncttrue
	\mciteSetBstMidEndSepPunct{\mcitedefaultmidpunct}
	{\mcitedefaultendpunct}{\mcitedefaultseppunct}\relax
	\EndOfBibitem
	\bibitem[Imperio and Reatto(2004)]{Imperio:2004}
	A.~Imperio and L.~Reatto, \emph{J. Phys. Condens. Matter}, 2004, \textbf{16},
	S3769\relax
	\mciteBstWouldAddEndPuncttrue
	\mciteSetBstMidEndSepPunct{\mcitedefaultmidpunct}
	{\mcitedefaultendpunct}{\mcitedefaultseppunct}\relax
	\EndOfBibitem
	\bibitem[Imperio and Reatto(2006)]{Imperio:2006}
	A.~Imperio and L.~Reatto, \emph{J. Chem. Phys.}, 2006, \textbf{124},
	164712\relax
	\mciteBstWouldAddEndPuncttrue
	\mciteSetBstMidEndSepPunct{\mcitedefaultmidpunct}
	{\mcitedefaultendpunct}{\mcitedefaultseppunct}\relax
	\EndOfBibitem
	\bibitem[Sciortino \emph{et~al.}(2004)Sciortino, Mossa, Zaccarelli, and
	Tartaglia]{sciortino:2004}
	F.~Sciortino, S.~Mossa, E.~Zaccarelli and P.~Tartaglia, \emph{Phys. Rev.
		Lett.}, 2004, \textbf{93}, 055701\relax
	\mciteBstWouldAddEndPuncttrue
	\mciteSetBstMidEndSepPunct{\mcitedefaultmidpunct}
	{\mcitedefaultendpunct}{\mcitedefaultseppunct}\relax
	\EndOfBibitem
	\bibitem[Campbell \emph{et~al.}(2005)Campbell, Anderson, van Duijneveldt, and
	Bartlett]{camp:2005prl}
	A.~I. Campbell, V.~J. Anderson, J.~S. van Duijneveldt and P.~Bartlett,
	\emph{Phys. Rev. Lett.}, 2005, \textbf{94}, 208301\relax
	\mciteBstWouldAddEndPuncttrue
	\mciteSetBstMidEndSepPunct{\mcitedefaultmidpunct}
	{\mcitedefaultendpunct}{\mcitedefaultseppunct}\relax
	\EndOfBibitem
	\bibitem[Klix \emph{et~al.}(2010)Klix, Royall, and Tanaka]{klix:2010}
	C.~L. Klix, C.~P. Royall and H.~Tanaka, \emph{Phys. Rev. Lett.}, 2010,
	\textbf{104}, 165702\relax
	\mciteBstWouldAddEndPuncttrue
	\mciteSetBstMidEndSepPunct{\mcitedefaultmidpunct}
	{\mcitedefaultendpunct}{\mcitedefaultseppunct}\relax
	\EndOfBibitem
	\bibitem[Schwanzer \emph{et~al.}(2016)Schwanzer, Coslovich, and
	Kahl]{schw:2016}
	D.~F. Schwanzer, D.~Coslovich and G.~Kahl, \emph{J. Phys. Condens. Matter},
	2016, \textbf{28}, 414015\relax
	\mciteBstWouldAddEndPuncttrue
	\mciteSetBstMidEndSepPunct{\mcitedefaultmidpunct}
	{\mcitedefaultendpunct}{\mcitedefaultseppunct}\relax
	\EndOfBibitem
	\bibitem[Barhoum and Yethiraj(2010)]{barhoum:2010}
	S.~Barhoum and A.~Yethiraj, \emph{J. Phys. Chem. B}, 2010, \textbf{114},
	17062--17067\relax
	\mciteBstWouldAddEndPuncttrue
	\mciteSetBstMidEndSepPunct{\mcitedefaultmidpunct}
	{\mcitedefaultendpunct}{\mcitedefaultseppunct}\relax
	\EndOfBibitem
	\bibitem[Yearley \emph{et~al.}(2013)Yearley, Zarraga, Shire, Scherer, Gokarn,
	Wagner, and Liu]{yearley:2013}
	E.~Yearley, I.~Zarraga, S.~Shire, T.~Scherer, Y.~Gokarn, N.~Wagner and Y.~Liu,
	\emph{Biophys. J.}, 2013, \textbf{105}, 720--731\relax
	\mciteBstWouldAddEndPuncttrue
	\mciteSetBstMidEndSepPunct{\mcitedefaultmidpunct}
	{\mcitedefaultendpunct}{\mcitedefaultseppunct}\relax
	\EndOfBibitem
	\bibitem[Godfrin \emph{et~al.}(2015)Godfrin, Hudson, Hong, Porcar, Falus,
	Wagner, and Liu]{godfrin:2015}
	P.~D. Godfrin, S.~D. Hudson, K.~Hong, L.~Porcar, P.~Falus, N.~J. Wagner and
	Y.~Liu, \emph{Phys. Rev. Lett.}, 2015, \textbf{115}, 228302\relax
	\mciteBstWouldAddEndPuncttrue
	\mciteSetBstMidEndSepPunct{\mcitedefaultmidpunct}
	{\mcitedefaultendpunct}{\mcitedefaultseppunct}\relax
	\EndOfBibitem
	\bibitem[Yearley \emph{et~al.}(2014)Yearley, Godfrin, Perevozchikova, Zhang,
	Falus, Porcar, Nagao, Curtis, Gawande, Taing, Zarraga, Wagner, and
	Liu]{year:2014}
	E.~Yearley, P.~Godfrin, T.~Perevozchikova, H.~Zhang, P.~Falus, L.~Porcar,
	M.~Nagao, J.~Curtis, P.~Gawande, R.~Taing, I.~Zarraga, N.~Wagner and Y.~Liu,
	\emph{Biophys. J.}, 2014, \textbf{106}, 1763--1770\relax
	\mciteBstWouldAddEndPuncttrue
	\mciteSetBstMidEndSepPunct{\mcitedefaultmidpunct}
	{\mcitedefaultendpunct}{\mcitedefaultseppunct}\relax
	\EndOfBibitem
	\bibitem[Godfrin \emph{et~al.}(2016)Godfrin, Zarraga, Zarzar, Porcar, Falus,
	Wagner, and Liu]{godf:2016}
	P.~D. Godfrin, I.~E. Zarraga, J.~Zarzar, L.~Porcar, P.~Falus, N.~J. Wagner and
	Y.~Liu, \emph{J. Phys. Chem. B}, 2016, \textbf{120}, 278--291\relax
	\mciteBstWouldAddEndPuncttrue
	\mciteSetBstMidEndSepPunct{\mcitedefaultmidpunct}
	{\mcitedefaultendpunct}{\mcitedefaultseppunct}\relax
	\EndOfBibitem
	\bibitem[Porcar \emph{et~al.}(2010)Porcar, Falus, Chen, Faraone, Fratini, Hong,
	Baglioni, and Liu]{porcar:2010}
	L.~Porcar, P.~Falus, W.-R. Chen, A.~Faraone, E.~Fratini, K.~Hong, P.~Baglioni
	and Y.~Liu, \emph{J. Phys. Chem. Lett.}, 2010, \textbf{1}, 126--129\relax
	\mciteBstWouldAddEndPuncttrue
	\mciteSetBstMidEndSepPunct{\mcitedefaultmidpunct}
	{\mcitedefaultendpunct}{\mcitedefaultseppunct}\relax
	\EndOfBibitem
	\bibitem[Liu \emph{et~al.}(2011)Liu, Porcar, Chen, Chen, Falus, Faraone,
	Fratini, Hong, and Baglioni]{yliu:2011}
	Y.~Liu, L.~Porcar, J.~Chen, W.-R. Chen, P.~Falus, A.~Faraone, E.~Fratini,
	K.~Hong and P.~Baglioni, \emph{J. Phys. Chem. B}, 2011, \textbf{115},
	7238--7247\relax
	\mciteBstWouldAddEndPuncttrue
	\mciteSetBstMidEndSepPunct{\mcitedefaultmidpunct}
	{\mcitedefaultendpunct}{\mcitedefaultseppunct}\relax
	\EndOfBibitem
	\bibitem[Nawrocki \emph{et~al.}(2017)Nawrocki, Wang, Yu, Sugita, and
	Feig]{nawrocki:2017}
	G.~Nawrocki, P.-h. Wang, I.~Yu, Y.~Sugita and M.~Feig, \emph{J. Phys. Chem.
		B}, 2017, \textbf{121}, 11072--11084\relax
	\mciteBstWouldAddEndPuncttrue
	\mciteSetBstMidEndSepPunct{\mcitedefaultmidpunct}
	{\mcitedefaultendpunct}{\mcitedefaultseppunct}\relax
	\EndOfBibitem
	\bibitem[Das \emph{et~al.}(2018)Das, Riest, Winkler, Gompper, Dhont, and
	N{\"a}gele]{das2018clustering}
	S.~Das, J.~Riest, R.~G. Winkler, G.~Gompper, J.~K. Dhont and G.~N{\"a}gele,
	\emph{Soft Matter}, 2018, \textbf{14}, 92--103\relax
	\mciteBstWouldAddEndPuncttrue
	\mciteSetBstMidEndSepPunct{\mcitedefaultmidpunct}
	{\mcitedefaultendpunct}{\mcitedefaultseppunct}\relax
	\EndOfBibitem
	\bibitem[Godfrin \emph{et~al.}(2018)Godfrin, Falus, Porcar, Hong, Hudson,
	Wagner, and Liu]{godfrin:2018}
	P.~D. Godfrin, P.~Falus, L.~Porcar, K.~Hong, S.~D. Hudson, N.~J. Wagner and
	Y.~Liu, \emph{Soft Matter}, 2018, \textbf{14}, 8570--8579\relax
	\mciteBstWouldAddEndPuncttrue
	\mciteSetBstMidEndSepPunct{\mcitedefaultmidpunct}
	{\mcitedefaultendpunct}{\mcitedefaultseppunct}\relax
	\EndOfBibitem
	\bibitem[Perdomo-Pérez \emph{et~al.}(2022)Perdomo-Pérez, Martínez-Rivera,
	Palmero-Cruz, Sandoval-Puentes, Gallegos, Lázaro-Lázaro, Valadez-Pérez,
	Torres-Carbajal, and Castañeda-Priego]{perdomo:2022}
	R.~Perdomo-Pérez, J.~Martínez-Rivera, N.~C. Palmero-Cruz, M.~A.
	Sandoval-Puentes, J.~A.~S. Gallegos, E.~Lázaro-Lázaro, N.~E.
	Valadez-Pérez, A.~Torres-Carbajal and R.~Castañeda-Priego, \emph{J. Phys.
		Condens. Matter}, 2022, \textbf{34}, 144005\relax
	\mciteBstWouldAddEndPuncttrue
	\mciteSetBstMidEndSepPunct{\mcitedefaultmidpunct}
	{\mcitedefaultendpunct}{\mcitedefaultseppunct}\relax
	\EndOfBibitem
	\bibitem[Bera \emph{et~al.}(2016)Bera, Qiao, Seifert, Burton-Pye, Olvera de~la
	Cruz, and Antonio]{bera:2016}
	M.~K. Bera, B.~Qiao, S.~Seifert, B.~P. Burton-Pye, M.~Olvera de~la Cruz and
	M.~R. Antonio, \emph{J. Phys. Chem. C}, 2016, \textbf{120}, 1317--1327\relax
	\mciteBstWouldAddEndPuncttrue
	\mciteSetBstMidEndSepPunct{\mcitedefaultmidpunct}
	{\mcitedefaultendpunct}{\mcitedefaultseppunct}\relax
	\EndOfBibitem
	\bibitem[Erlkamp \emph{et~al.}(2014)Erlkamp, Grobelny, Faraone, Czeslik, and
	Winter]{erlkamp:2014}
	M.~Erlkamp, S.~Grobelny, A.~Faraone, C.~Czeslik and R.~Winter, \emph{J. Phys.
		Chem. B}, 2014, \textbf{118}, 3310--3316\relax
	\mciteBstWouldAddEndPuncttrue
	\mciteSetBstMidEndSepPunct{\mcitedefaultmidpunct}
	{\mcitedefaultendpunct}{\mcitedefaultseppunct}\relax
	\EndOfBibitem
	\bibitem[Balbo \emph{et~al.}(2013)Balbo, Mereghetti, Herten, and
	Wade]{balbo:2013}
	J.~Balbo, P.~Mereghetti, D.-P. Herten and R.~Wade, \emph{Biophys. J.}, 2013,
	\textbf{104}, 1576--1584\relax
	\mciteBstWouldAddEndPuncttrue
	\mciteSetBstMidEndSepPunct{\mcitedefaultmidpunct}
	{\mcitedefaultendpunct}{\mcitedefaultseppunct}\relax
	\EndOfBibitem
	\bibitem[Varga and Swan(2016)]{varga:2016}
	Z.~Varga and J.~Swan, \emph{Soft Matter}, 2016, \textbf{12}, 7670--7681\relax
	\mciteBstWouldAddEndPuncttrue
	\mciteSetBstMidEndSepPunct{\mcitedefaultmidpunct}
	{\mcitedefaultendpunct}{\mcitedefaultseppunct}\relax
	\EndOfBibitem
	\bibitem[Kob \emph{et~al.}(1997)Kob, Donati, Plimpton, Poole, and
	Glotzer]{Kob:1999}
	W.~Kob, C.~Donati, S.~J. Plimpton, P.~H. Poole and S.~C. Glotzer, \emph{Phys.
		Rev. Lett.}, 1997, \textbf{79}, 2827--2830\relax
	\mciteBstWouldAddEndPuncttrue
	\mciteSetBstMidEndSepPunct{\mcitedefaultmidpunct}
	{\mcitedefaultendpunct}{\mcitedefaultseppunct}\relax
	\EndOfBibitem
	\bibitem[Reichhardt and Olson~Reichhardt(2003)]{Reichhardt:2003}
	C.~Reichhardt and C.~J. Olson~Reichhardt, \emph{Phys. Rev. Lett.}, 2003,
	\textbf{90}, 095504\relax
	\mciteBstWouldAddEndPuncttrue
	\mciteSetBstMidEndSepPunct{\mcitedefaultmidpunct}
	{\mcitedefaultendpunct}{\mcitedefaultseppunct}\relax
	\EndOfBibitem
	\bibitem[Schnyder \emph{et~al.}(2017)Schnyder, Skinner, Thorneywork, Aarts,
	Horbach, and Dullens]{SchnyderDullens:2017}
	S.~K. Schnyder, T.~O.~E. Skinner, A.~L. Thorneywork, D.~G. A.~L. Aarts,
	J.~Horbach and R.~P.~A. Dullens, \emph{Phys. Rev. E}, 2017, \textbf{95},
	032602\relax
	\mciteBstWouldAddEndPuncttrue
	\mciteSetBstMidEndSepPunct{\mcitedefaultmidpunct}
	{\mcitedefaultendpunct}{\mcitedefaultseppunct}\relax
	\EndOfBibitem
	\bibitem[Kaur and Das(2003)]{KaurDas:2003}
	C.~Kaur and S.~P. Das, \emph{Phys. Rev. E}, 2003, \textbf{67}, 051505\relax
	\mciteBstWouldAddEndPuncttrue
	\mciteSetBstMidEndSepPunct{\mcitedefaultmidpunct}
	{\mcitedefaultendpunct}{\mcitedefaultseppunct}\relax
	\EndOfBibitem
	\bibitem[Yeh \emph{et~al.}(2025)Yeh, Hatch, Sreenivasan, Bharti, Shen, Sherman,
	and Truskett]{Yeh:2025}
	C.-C.~G. Yeh, H.~W. Hatch, A.~N. Sreenivasan, B.~Bharti, V.~K. Shen, Z.~M.
	Sherman and T.~M. Truskett, \emph{J. Phys. Chem. B}, 2025, \textbf{129},
	6428--6438\relax
	\mciteBstWouldAddEndPuncttrue
	\mciteSetBstMidEndSepPunct{\mcitedefaultmidpunct}
	{\mcitedefaultendpunct}{\mcitedefaultseppunct}\relax
	\EndOfBibitem
	\bibitem[Hilger \emph{et~al.}(2018)Hilger, Masureel, and Kobilka]{hilger2018}
	D.~Hilger, M.~Masureel and B.~K. Kobilka, \emph{Nat. Struct. Mol. Biol.}, 2018,
	\textbf{25}, 4--12\relax
	\mciteBstWouldAddEndPuncttrue
	\mciteSetBstMidEndSepPunct{\mcitedefaultmidpunct}
	{\mcitedefaultendpunct}{\mcitedefaultseppunct}\relax
	\EndOfBibitem
	\bibitem[Saffman and Delbr{\"u}ck(1975)]{Saffman1975}
	P.~G. Saffman and M.~Delbr{\"u}ck, \emph{Proc. Natl. Acad. Sci. U.S.A.}, 1975,
	\textbf{72}, 3111--3113\relax
	\mciteBstWouldAddEndPuncttrue
	\mciteSetBstMidEndSepPunct{\mcitedefaultmidpunct}
	{\mcitedefaultendpunct}{\mcitedefaultseppunct}\relax
	\EndOfBibitem
	\bibitem[Ramadurai \emph{et~al.}(2009)Ramadurai, Holt, Krasnikov, van~den
	Bogaart, Killian, and Poolman]{ramadurai2009}
	S.~Ramadurai, A.~Holt, V.~Krasnikov, G.~van~den Bogaart, J.~A. Killian and
	B.~Poolman, \emph{J. Am. Chem. Soc.}, 2009, \textbf{131}, 12650--12656\relax
	\mciteBstWouldAddEndPuncttrue
	\mciteSetBstMidEndSepPunct{\mcitedefaultmidpunct}
	{\mcitedefaultendpunct}{\mcitedefaultseppunct}\relax
	\EndOfBibitem
	\bibitem[Gao \emph{et~al.}(2025)Gao, Shen, Komura, Hu, Shen, and
	Hu]{hu:2025pnas}
	J.~Gao, Y.~Shen, S.~Komura, W.~Hu, L.~Shen and J.~Hu, \emph{Proc. Natl. Acad.
		Sci. U.S.A.}, 2025, \textbf{122}, e2503203122\relax
	\mciteBstWouldAddEndPuncttrue
	\mciteSetBstMidEndSepPunct{\mcitedefaultmidpunct}
	{\mcitedefaultendpunct}{\mcitedefaultseppunct}\relax
	\EndOfBibitem
	\bibitem[Javanainen \emph{et~al.}(2017)Javanainen, Martinez-Seara, Metzler, and
	Vattulainen]{Vattulainen:2017}
	M.~Javanainen, H.~Martinez-Seara, R.~Metzler and I.~Vattulainen, \emph{J.
		Phys. Chem. Lett.}, 2017, \textbf{8}, 4308--4313\relax
	\mciteBstWouldAddEndPuncttrue
	\mciteSetBstMidEndSepPunct{\mcitedefaultmidpunct}
	{\mcitedefaultendpunct}{\mcitedefaultseppunct}\relax
	\EndOfBibitem
	\bibitem[Pesch\'e and N\"agele(2000)]{naegele2000wall}
	R.~Pesch\'e and G.~N\"agele, \emph{Phys. Rev. E}, 2000, \textbf{62},
	5432--5443\relax
	\mciteBstWouldAddEndPuncttrue
	\mciteSetBstMidEndSepPunct{\mcitedefaultmidpunct}
	{\mcitedefaultendpunct}{\mcitedefaultseppunct}\relax
	\EndOfBibitem
	\bibitem[Pesché \emph{et~al.}(2001)Pesché, Kollmann, and
	Nägele]{naegele2001jcp}
	R.~Pesché, M.~Kollmann and G.~Nägele, \emph{J. Chem. Phys.}, 2001,
	\textbf{114}, 8701--8707\relax
	\mciteBstWouldAddEndPuncttrue
	\mciteSetBstMidEndSepPunct{\mcitedefaultmidpunct}
	{\mcitedefaultendpunct}{\mcitedefaultseppunct}\relax
	\EndOfBibitem
	\bibitem[N{\"a}gele \emph{et~al.}(2002)N{\"a}gele, Banchio, Kollmann, and
	Pesch{\'e}]{naegele2002}
	G.~N{\"a}gele, A.~J. Banchio, M.~Kollmann and R.~Pesch{\'e}, \emph{Mol.
		Phys.}, 2002, \textbf{100}, 2921--2933\relax
	\mciteBstWouldAddEndPuncttrue
	\mciteSetBstMidEndSepPunct{\mcitedefaultmidpunct}
	{\mcitedefaultendpunct}{\mcitedefaultseppunct}\relax
	\EndOfBibitem
	\bibitem[Bleibel \emph{et~al.}(2017)Bleibel, Dom{\'\i}nguez, and
	Oettel]{bleibel2017onset}
	J.~Bleibel, A.~Dom{\'\i}nguez and M.~Oettel, \emph{Phys. Rev. E}, 2017,
	\textbf{95}, 032604\relax
	\mciteBstWouldAddEndPuncttrue
	\mciteSetBstMidEndSepPunct{\mcitedefaultmidpunct}
	{\mcitedefaultendpunct}{\mcitedefaultseppunct}\relax
	\EndOfBibitem
	\bibitem[Pel{\'a}ez \emph{et~al.}(2018)Pel{\'a}ez, Usabiaga, Panzuela, Xiao,
	Delgado-Buscalioni, and Donev]{pelaez2018hydrodynamic}
	R.~P. Pel{\'a}ez, F.~B. Usabiaga, S.~Panzuela, Q.~Xiao, R.~Delgado-Buscalioni
	and A.~Donev, \emph{J. Stat. Mech.}, 2018, \textbf{2018}, 063207\relax
	\mciteBstWouldAddEndPuncttrue
	\mciteSetBstMidEndSepPunct{\mcitedefaultmidpunct}
	{\mcitedefaultendpunct}{\mcitedefaultseppunct}\relax
	\EndOfBibitem
	\bibitem[Panzuela and Delgado-Buscalioni(2018)]{Panzuela18}
	S.~Panzuela and R.~Delgado-Buscalioni, \emph{Phys. Rev. Lett.}, 2018,
	\textbf{121}, 048101\relax
	\mciteBstWouldAddEndPuncttrue
	\mciteSetBstMidEndSepPunct{\mcitedefaultmidpunct}
	{\mcitedefaultendpunct}{\mcitedefaultseppunct}\relax
	\EndOfBibitem
	\bibitem[Mackay \emph{et~al.}(2024)Mackay, Marbach, Sprinkle, and
	Thorneywork]{Alice:2024}
	E.~K.~R. Mackay, S.~Marbach, B.~Sprinkle and A.~L. Thorneywork, \emph{Phys.
		Rev. X}, 2024, \textbf{14}, 041016\relax
	\mciteBstWouldAddEndPuncttrue
	\mciteSetBstMidEndSepPunct{\mcitedefaultmidpunct}
	{\mcitedefaultendpunct}{\mcitedefaultseppunct}\relax
	\EndOfBibitem
	\bibitem[Carter \emph{et~al.}(2025)Carter, Mackay, Sprinkle, Thorneywork, and
	Marbach]{alice:2025}
	A.~Carter, E.~K.~R. Mackay, B.~Sprinkle, A.~L. Thorneywork and S.~Marbach,
	\emph{Soft Matter}, 2025, \textbf{21}, 3991--4002\relax
	\mciteBstWouldAddEndPuncttrue
	\mciteSetBstMidEndSepPunct{\mcitedefaultmidpunct}
	{\mcitedefaultendpunct}{\mcitedefaultseppunct}\relax
	\EndOfBibitem
	\bibitem[Chamorro-Burgos and Dom\'{\i}nguez(2026)]{Chamorro-Burgos:2026}
	M.~Chamorro-Burgos and A.~Dom\'{\i}nguez, \emph{Phys. Rev. E}, 2026,
	\textbf{113}, 055402\relax
	\mciteBstWouldAddEndPuncttrue
	\mciteSetBstMidEndSepPunct{\mcitedefaultmidpunct}
	{\mcitedefaultendpunct}{\mcitedefaultseppunct}\relax
	\EndOfBibitem
	\bibitem[Panzuela \emph{et~al.}(2017)Panzuela, Pel{\'a}ez, and
	Delgado-Buscalioni]{panzuela2017collective}
	S.~Panzuela, R.~P. Pel{\'a}ez and R.~Delgado-Buscalioni, \emph{Phys. Rev. E},
	2017, \textbf{95}, 012602\relax
	\mciteBstWouldAddEndPuncttrue
	\mciteSetBstMidEndSepPunct{\mcitedefaultmidpunct}
	{\mcitedefaultendpunct}{\mcitedefaultseppunct}\relax
	\EndOfBibitem
	\bibitem[Franosch \emph{et~al.}(2011)Franosch, Grimm, Belushkin, Mor, Foffi,
	Forr{\'o}, and Jeney]{franosch2011resonances}
	T.~Franosch, M.~Grimm, M.~Belushkin, F.~M. Mor, G.~Foffi, L.~Forr{\'o} and
	S.~Jeney, \emph{Nature}, 2011, \textbf{478}, 85--88\relax
	\mciteBstWouldAddEndPuncttrue
	\mciteSetBstMidEndSepPunct{\mcitedefaultmidpunct}
	{\mcitedefaultendpunct}{\mcitedefaultseppunct}\relax
	\EndOfBibitem
	\bibitem[Belushkin \emph{et~al.}(2011)Belushkin, Winkler, and
	Foffi]{winkler2011backtracking}
	M.~Belushkin, R.~Winkler and G.~Foffi, \emph{J. Phys. Chem. B}, 2011,
	\textbf{115}, 14263--14268\relax
	\mciteBstWouldAddEndPuncttrue
	\mciteSetBstMidEndSepPunct{\mcitedefaultmidpunct}
	{\mcitedefaultendpunct}{\mcitedefaultseppunct}\relax
	\EndOfBibitem
	\bibitem[han(2013)]{hansen:2013}
	J.-P. Hansen and I.~R. McDonald, \emph{Theory of Simple Liquids} (4th ed.),
	Academic Press, Oxford, 2013\relax
	\mciteBstWouldAddEndPuncttrue
	\mciteSetBstMidEndSepPunct{\mcitedefaultmidpunct}
	{\mcitedefaultendpunct}{\mcitedefaultseppunct}\relax
	\EndOfBibitem
	\bibitem[Ladd \emph{et~al.}(1995)Ladd, Gang, Zhu, and Weitz]{ladd1995pre}
	A.~J.~C. Ladd, H.~Gang, J.~X. Zhu and D.~A. Weitz, \emph{Phys. Rev. E}, 1995,
	\textbf{52}, 6550--6572\relax
	\mciteBstWouldAddEndPuncttrue
	\mciteSetBstMidEndSepPunct{\mcitedefaultmidpunct}
	{\mcitedefaultendpunct}{\mcitedefaultseppunct}\relax
	\EndOfBibitem
	\bibitem[Ladd \emph{et~al.}(1995)Ladd, Gang, Zhu, and Weitz]{ladd1995prl}
	A.~J.~C. Ladd, H.~Gang, J.~X. Zhu and D.~A. Weitz, \emph{Phys. Rev. Lett.},
	1995, \textbf{74}, 318--321\relax
	\mciteBstWouldAddEndPuncttrue
	\mciteSetBstMidEndSepPunct{\mcitedefaultmidpunct}
	{\mcitedefaultendpunct}{\mcitedefaultseppunct}\relax
	\EndOfBibitem
	\bibitem[Bakker and Lowe(2002)]{bakker2002role}
	A.~Bakker and C.~Lowe, \emph{J. Chem. Phys.}, 2002, \textbf{116},
	5867--5876\relax
	\mciteBstWouldAddEndPuncttrue
	\mciteSetBstMidEndSepPunct{\mcitedefaultmidpunct}
	{\mcitedefaultendpunct}{\mcitedefaultseppunct}\relax
	\EndOfBibitem
	\bibitem[Dom{\'\i}nguez(2014)]{dominguez2014signature}
	A.~Dom{\'\i}nguez, \emph{Phys. Rev. E}, 2014, \textbf{90}, 062314\relax
	\mciteBstWouldAddEndPuncttrue
	\mciteSetBstMidEndSepPunct{\mcitedefaultmidpunct}
	{\mcitedefaultendpunct}{\mcitedefaultseppunct}\relax
	\EndOfBibitem
	\bibitem[Charbonneau and Reichman(2007)]{char:2007}
	P.~Charbonneau and D.~R. Reichman, \emph{Phys. Rev. E}, 2007, \textbf{75},
	050401\relax
	\mciteBstWouldAddEndPuncttrue
	\mciteSetBstMidEndSepPunct{\mcitedefaultmidpunct}
	{\mcitedefaultendpunct}{\mcitedefaultseppunct}\relax
	\EndOfBibitem
	\bibitem[Mani \emph{et~al.}(2014)Mani, Lechner, Kegel, and Bolhuis]{mani:2014}
	E.~Mani, W.~Lechner, W.~K. Kegel and P.~G. Bolhuis, \emph{Soft Matter}, 2014,
	\textbf{10}, 4479--4486\relax
	\mciteBstWouldAddEndPuncttrue
	\mciteSetBstMidEndSepPunct{\mcitedefaultmidpunct}
	{\mcitedefaultendpunct}{\mcitedefaultseppunct}\relax
	\EndOfBibitem
	\bibitem[Sciortino \emph{et~al.}(2005)Sciortino, Tartaglia, and
	Zaccarelli]{zacca:2005}
	F.~Sciortino, P.~Tartaglia and E.~Zaccarelli, \emph{J. Phys. Chem. B}, 2005,
	\textbf{109}, 21942--21953\relax
	\mciteBstWouldAddEndPuncttrue
	\mciteSetBstMidEndSepPunct{\mcitedefaultmidpunct}
	{\mcitedefaultendpunct}{\mcitedefaultseppunct}\relax
	\EndOfBibitem
	\bibitem[Zahn and Maret(2000)]{Zahn:2000}
	K.~Zahn and G.~Maret, \emph{Phys. Rev. Lett.}, 2000, \textbf{85},
	3656--3659\relax
	\mciteBstWouldAddEndPuncttrue
	\mciteSetBstMidEndSepPunct{\mcitedefaultmidpunct}
	{\mcitedefaultendpunct}{\mcitedefaultseppunct}\relax
	\EndOfBibitem
	\bibitem[Haghgooie and Doyle(2005)]{Haghgooie:2005}
	R.~Haghgooie and P.~S. Doyle, \emph{Phys. Rev. E}, 2005, \textbf{72},
	011405\relax
	\mciteBstWouldAddEndPuncttrue
	\mciteSetBstMidEndSepPunct{\mcitedefaultmidpunct}
	{\mcitedefaultendpunct}{\mcitedefaultseppunct}\relax
	\EndOfBibitem
	\bibitem[Lin \emph{et~al.}(2006)Lin, Zheng, and Trimper]{Lin:2006}
	S.~Z. Lin, B.~Zheng and S.~Trimper, \emph{Phys. Rev. E}, 2006, \textbf{73},
	066106\relax
	\mciteBstWouldAddEndPuncttrue
	\mciteSetBstMidEndSepPunct{\mcitedefaultmidpunct}
	{\mcitedefaultendpunct}{\mcitedefaultseppunct}\relax
	\EndOfBibitem
	\bibitem[Kelleher \emph{et~al.}(2017)Kelleher, Guerra, Hollingsworth, and
	Chaikin]{Kelleher:2017}
	C.~P. Kelleher, R.~E. Guerra, A.~D. Hollingsworth and P.~M. Chaikin,
	\emph{Phys. Rev. E}, 2017, \textbf{95}, 022602\relax
	\mciteBstWouldAddEndPuncttrue
	\mciteSetBstMidEndSepPunct{\mcitedefaultmidpunct}
	{\mcitedefaultendpunct}{\mcitedefaultseppunct}\relax
	\EndOfBibitem
	\bibitem[Thorneywork \emph{et~al.}(2018)Thorneywork, Abbott, Aarts, Keim, and
	Dullens]{alice:2018}
	A.~L. Thorneywork, J.~L. Abbott, D.~G. A.~L. Aarts, P.~Keim and R.~P.~A.
	Dullens, \emph{J. Phys. Condens. Matter}, 2018, \textbf{30}, 104003\relax
	\mciteBstWouldAddEndPuncttrue
	\mciteSetBstMidEndSepPunct{\mcitedefaultmidpunct}
	{\mcitedefaultendpunct}{\mcitedefaultseppunct}\relax
	\EndOfBibitem
	\bibitem[Halperin and Nelson(1978)]{halpe:1978}
	B.~I. Halperin and D.~R. Nelson, \emph{Phys. Rev. Lett.}, 1978, \textbf{41},
	121--124\relax
	\mciteBstWouldAddEndPuncttrue
	\mciteSetBstMidEndSepPunct{\mcitedefaultmidpunct}
	{\mcitedefaultendpunct}{\mcitedefaultseppunct}\relax
	\EndOfBibitem
	\bibitem[Nelson and Halperin(1979)]{Nelson1979}
	D.~R. Nelson and B.~I. Halperin, \emph{Phys. Rev. B}, 1979, \textbf{19},
	2457--2484\relax
	\mciteBstWouldAddEndPuncttrue
	\mciteSetBstMidEndSepPunct{\mcitedefaultmidpunct}
	{\mcitedefaultendpunct}{\mcitedefaultseppunct}\relax
	\EndOfBibitem
	\bibitem[Bialk\'e \emph{et~al.}(2012)Bialk\'e, Speck, and L\"owen]{bialk:2012}
	J.~Bialk\'e, T.~Speck and H.~L\"owen, \emph{Phys. Rev. Lett.}, 2012,
	\textbf{108}, 168301\relax
	\mciteBstWouldAddEndPuncttrue
	\mciteSetBstMidEndSepPunct{\mcitedefaultmidpunct}
	{\mcitedefaultendpunct}{\mcitedefaultseppunct}\relax
	\EndOfBibitem
	\bibitem[Z\"ottl and Stark(2014)]{zottl2014}
	A.~Z\"ottl and H.~Stark, \emph{Phys. Rev. Lett.}, 2014, \textbf{112},
	118101\relax
	\mciteBstWouldAddEndPuncttrue
	\mciteSetBstMidEndSepPunct{\mcitedefaultmidpunct}
	{\mcitedefaultendpunct}{\mcitedefaultseppunct}\relax
	\EndOfBibitem
	\bibitem[Theers \emph{et~al.}(2018)Theers, Westphal, Qi, Winkler, and
	Gompper]{theers2018clustering}
	M.~Theers, E.~Westphal, K.~Qi, R.~G. Winkler and G.~Gompper, \emph{Soft
		Matter}, 2018, \textbf{14}, 8590--8603\relax
	\mciteBstWouldAddEndPuncttrue
	\mciteSetBstMidEndSepPunct{\mcitedefaultmidpunct}
	{\mcitedefaultendpunct}{\mcitedefaultseppunct}\relax
	\EndOfBibitem
	\bibitem[Pasupalak \emph{et~al.}(2020)Pasupalak, Yan-Wei, Ni, and
	Pica~Ciamarra]{Pasupalak:2020}
	A.~Pasupalak, L.~Yan-Wei, R.~Ni and M.~Pica~Ciamarra, \emph{Soft Matter},
	2020, \textbf{16}, 3914--3920\relax
	\mciteBstWouldAddEndPuncttrue
	\mciteSetBstMidEndSepPunct{\mcitedefaultmidpunct}
	{\mcitedefaultendpunct}{\mcitedefaultseppunct}\relax
	\EndOfBibitem
	\bibitem[Wang \emph{et~al.}(2012)Wang, Kuo, Bae, and Granick]{Granick:2012}
	B.~Wang, J.~Kuo, S.~C. Bae and S.~Granick, \emph{Nat. Mater.}, 2012,
	\textbf{11}, 481--485\relax
	\mciteBstWouldAddEndPuncttrue
	\mciteSetBstMidEndSepPunct{\mcitedefaultmidpunct}
	{\mcitedefaultendpunct}{\mcitedefaultseppunct}\relax
	\EndOfBibitem
	\bibitem[Fuchs \emph{et~al.}(1998)Fuchs, G\"otze, and Mayr]{Fuchs:1998}
	M.~Fuchs, W.~G\"otze and M.~R. Mayr, \emph{Phys. Rev. E}, 1998, \textbf{58},
	3384--3399\relax
	\mciteBstWouldAddEndPuncttrue
	\mciteSetBstMidEndSepPunct{\mcitedefaultmidpunct}
	{\mcitedefaultendpunct}{\mcitedefaultseppunct}\relax
	\EndOfBibitem
	\bibitem[Banchio(1999)]{banchio1999diffusion}
	A.~Banchio, 1999\relax
	\mciteBstWouldAddEndPuncttrue
	\mciteSetBstMidEndSepPunct{\mcitedefaultmidpunct}
	{\mcitedefaultendpunct}{\mcitedefaultseppunct}\relax
	\EndOfBibitem
	\bibitem[Malevanets and Kapral(1999)]{kap99}
	A.~Malevanets and R.~Kapral, \emph{J. Chem. Phys.}, 1999, \textbf{110},
	8605--8613\relax
	\mciteBstWouldAddEndPuncttrue
	\mciteSetBstMidEndSepPunct{\mcitedefaultmidpunct}
	{\mcitedefaultendpunct}{\mcitedefaultseppunct}\relax
	\EndOfBibitem
	\bibitem[Malevanets and Kapral(2000)]{kap00}
	A.~Malevanets and R.~Kapral, \emph{J. Chem. Phys.}, 2000, \textbf{112},
	7260--7269\relax
	\mciteBstWouldAddEndPuncttrue
	\mciteSetBstMidEndSepPunct{\mcitedefaultmidpunct}
	{\mcitedefaultendpunct}{\mcitedefaultseppunct}\relax
	\EndOfBibitem
	\bibitem[Kapral(2008)]{kapral_review}
	R.~Kapral, \emph{Adv. Chem. Phys.}, 2008, \textbf{140}, 89--146\relax
	\mciteBstWouldAddEndPuncttrue
	\mciteSetBstMidEndSepPunct{\mcitedefaultmidpunct}
	{\mcitedefaultendpunct}{\mcitedefaultseppunct}\relax
	\EndOfBibitem
	\bibitem[Gompper \emph{et~al.}(2009)Gompper, Ihle, Kroll, and Winkler]{MPCD}
	G.~Gompper, T.~Ihle, D.~Kroll and R.~Winkler, \emph{Adv. Polym. Sci.}, 2009,
	\textbf{221}, 1--87\relax
	\mciteBstWouldAddEndPuncttrue
	\mciteSetBstMidEndSepPunct{\mcitedefaultmidpunct}
	{\mcitedefaultendpunct}{\mcitedefaultseppunct}\relax
	\EndOfBibitem
	\bibitem[Z{\"o}ttl and Stark(2018)]{zoettl:2018}
	A.~Z{\"o}ttl and H.~Stark, \emph{Eur. Phys. J. E}, 2018, \textbf{41}, 61\relax
	\mciteBstWouldAddEndPuncttrue
	\mciteSetBstMidEndSepPunct{\mcitedefaultmidpunct}
	{\mcitedefaultendpunct}{\mcitedefaultseppunct}\relax
	\EndOfBibitem
	\bibitem[Tan \emph{et~al.}(2025)Tan, Peters, and Stark]{tan05}
	Z.~Tan, J.~I.~U. Peters and H.~Stark, \emph{New J. Phys.}, 2025, \textbf{27},
	064401\relax
	\mciteBstWouldAddEndPuncttrue
	\mciteSetBstMidEndSepPunct{\mcitedefaultmidpunct}
	{\mcitedefaultendpunct}{\mcitedefaultseppunct}\relax
	\EndOfBibitem
	\bibitem[Huang \emph{et~al.}(2015)Huang, Varghese, Gompper, and
	Winkler]{huang2015}
	C.-C. Huang, A.~Varghese, G.~Gompper and R.~G. Winkler, \emph{Phys. Rev. E},
	2015, \textbf{91}, 013310\relax
	\mciteBstWouldAddEndPuncttrue
	\mciteSetBstMidEndSepPunct{\mcitedefaultmidpunct}
	{\mcitedefaultendpunct}{\mcitedefaultseppunct}\relax
	\EndOfBibitem
	\bibitem[Tan \emph{et~al.}(2017)Tan, Yang, and Ripoll]{tan01}
	Z.~Tan, M.~Yang and M.~Ripoll, \emph{Soft Matter}, 2017, \textbf{13},
	7283--7291\relax
	\mciteBstWouldAddEndPuncttrue
	\mciteSetBstMidEndSepPunct{\mcitedefaultmidpunct}
	{\mcitedefaultendpunct}{\mcitedefaultseppunct}\relax
	\EndOfBibitem
	\bibitem[Tan \emph{et~al.}(2019)Tan, Yang, and Ripoll]{tan02}
	Z.~Tan, M.~Yang and M.~Ripoll, \emph{Phys. Rev. Applied}, 2019, \textbf{11},
	054004\relax
	\mciteBstWouldAddEndPuncttrue
	\mciteSetBstMidEndSepPunct{\mcitedefaultmidpunct}
	{\mcitedefaultendpunct}{\mcitedefaultseppunct}\relax
	\EndOfBibitem
	\bibitem[Tan \emph{et~al.}(2021)Tan, Calandrini, Dhont, Nägele, and
	Winkler]{tan03}
	Z.~Tan, V.~Calandrini, J.~K.~G. Dhont, G.~Nägele and R.~G. Winkler, \emph{Soft
		Matter}, 2021, \textbf{17}, 7978--7990\relax
	\mciteBstWouldAddEndPuncttrue
	\mciteSetBstMidEndSepPunct{\mcitedefaultmidpunct}
	{\mcitedefaultendpunct}{\mcitedefaultseppunct}\relax
	\EndOfBibitem
	\bibitem[Ihle and Kroll(2001)]{ihl01}
	T.~Ihle and D.~M. Kroll, \emph{Phys. Rev. E}, 2001, \textbf{63}, 020201\relax
	\mciteBstWouldAddEndPuncttrue
	\mciteSetBstMidEndSepPunct{\mcitedefaultmidpunct}
	{\mcitedefaultendpunct}{\mcitedefaultseppunct}\relax
	\EndOfBibitem
	\bibitem[Westphal \emph{et~al.}(2014)Westphal, Singh, Huang, Gompper, and
	Winkler]{Westphal:2014}
	E.~Westphal, S.~Singh, C.-C. Huang, G.~Gompper and R.~Winkler, \emph{Comput.
		Phys. Commun.}, 2014, \textbf{185}, 495--503\relax
	\mciteBstWouldAddEndPuncttrue
	\mciteSetBstMidEndSepPunct{\mcitedefaultmidpunct}
	{\mcitedefaultendpunct}{\mcitedefaultseppunct}\relax
	\EndOfBibitem
	\bibitem[Qi \emph{et~al.}(2020)Qi, Annepu, Gompper, and Winkler]{Qi:2020}
	K.~Qi, H.~Annepu, G.~Gompper and R.~G. Winkler, \emph{Phys. Rev. Res.}, 2020,
	\textbf{2}, 033275\relax
	\mciteBstWouldAddEndPuncttrue
	\mciteSetBstMidEndSepPunct{\mcitedefaultmidpunct}
	{\mcitedefaultendpunct}{\mcitedefaultseppunct}\relax
	\EndOfBibitem
	\bibitem[Theers \emph{et~al.}(2016)Theers, Westphal, Gompper, and
	Winkler]{theers16friction}
	M.~Theers, E.~Westphal, G.~Gompper and R.~G. Winkler, \emph{Phys. Rev. E},
	2016, \textbf{93}, 032604\relax
	\mciteBstWouldAddEndPuncttrue
	\mciteSetBstMidEndSepPunct{\mcitedefaultmidpunct}
	{\mcitedefaultendpunct}{\mcitedefaultseppunct}\relax
	\EndOfBibitem
	\bibitem[Padding \emph{et~al.}(2005)Padding, Wysocki, L{\"o}wen, and
	Louis]{padding2005stick}
	J.~Padding, A.~Wysocki, H.~L{\"o}wen and A.~Louis, \emph{J. Phys. Condens.
		Matter}, 2005, \textbf{17}, S3393\relax
	\mciteBstWouldAddEndPuncttrue
	\mciteSetBstMidEndSepPunct{\mcitedefaultmidpunct}
	{\mcitedefaultendpunct}{\mcitedefaultseppunct}\relax
	\EndOfBibitem
	\bibitem[Theers \emph{et~al.}(2016)Theers, Westphal, Gompper, and
	Winkler]{theers2016modeling}
	M.~Theers, E.~Westphal, G.~Gompper and R.~G. Winkler, \emph{Soft Matter},
	2016, \textbf{12}, 7372--7385\relax
	\mciteBstWouldAddEndPuncttrue
	\mciteSetBstMidEndSepPunct{\mcitedefaultmidpunct}
	{\mcitedefaultendpunct}{\mcitedefaultseppunct}\relax
	\EndOfBibitem
	\bibitem[Lamura \emph{et~al.}(2001)Lamura, Gompper, Ihle, and Kroll]{Lamura01}
	A.~Lamura, G.~Gompper, T.~Ihle and D.~M. Kroll, \emph{EPL}, 2001, \textbf{56},
	319\relax
	\mciteBstWouldAddEndPuncttrue
	\mciteSetBstMidEndSepPunct{\mcitedefaultmidpunct}
	{\mcitedefaultendpunct}{\mcitedefaultseppunct}\relax
	\EndOfBibitem
	\bibitem[Hecht \emph{et~al.}(2005)Hecht, Harting, Ihle, and
	Herrmann]{hecht2005}
	M.~Hecht, J.~Harting, T.~Ihle and H.~J. Herrmann, \emph{Phys. Rev. E}, 2005,
	\textbf{72}, 011408\relax
	\mciteBstWouldAddEndPuncttrue
	\mciteSetBstMidEndSepPunct{\mcitedefaultmidpunct}
	{\mcitedefaultendpunct}{\mcitedefaultseppunct}\relax
	\EndOfBibitem
	\bibitem[Toledano \emph{et~al.}(2009)Toledano, Sciortino, and
	Zaccarelli]{toledano:2009}
	J.~C.~F. Toledano, F.~Sciortino and E.~Zaccarelli, \emph{Soft Matter}, 2009,
	\textbf{5}, 2390--2398\relax
	\mciteBstWouldAddEndPuncttrue
	\mciteSetBstMidEndSepPunct{\mcitedefaultmidpunct}
	{\mcitedefaultendpunct}{\mcitedefaultseppunct}\relax
	\EndOfBibitem
	\bibitem[Brito \emph{et~al.}(2020)Brito, Carignano, and Marconi]{Brito:2020}
	M.~E. Brito, M.~A. Carignano and V.~I. Marconi, \emph{Sci. Rep.}, 2020,
	\textbf{10}, 3971\relax
	\mciteBstWouldAddEndPuncttrue
	\mciteSetBstMidEndSepPunct{\mcitedefaultmidpunct}
	{\mcitedefaultendpunct}{\mcitedefaultseppunct}\relax
	\EndOfBibitem
	\bibitem[Bergenholtz and Wagner(2019)]{berg:2019}
	J.~Bergenholtz and N.~J. Wagner, \emph{EPL}, 2019, \textbf{126}, 38002\relax
	\mciteBstWouldAddEndPuncttrue
	\mciteSetBstMidEndSepPunct{\mcitedefaultmidpunct}
	{\mcitedefaultendpunct}{\mcitedefaultseppunct}\relax
	\EndOfBibitem
	\bibitem[Chan and Halle(1984)]{Chan:1984}
	D.~Y. Chan and B.~Halle, \emph{J. Colloid Interface Sci.}, 1984, \textbf{102},
	400--409\relax
	\mciteBstWouldAddEndPuncttrue
	\mciteSetBstMidEndSepPunct{\mcitedefaultmidpunct}
	{\mcitedefaultendpunct}{\mcitedefaultseppunct}\relax
	\EndOfBibitem
	\bibitem[Kramers(1940)]{Kramers:1940}
	H.~Kramers, \emph{Physica}, 1940, \textbf{7}, 284--304\relax
	\mciteBstWouldAddEndPuncttrue
	\mciteSetBstMidEndSepPunct{\mcitedefaultmidpunct}
	{\mcitedefaultendpunct}{\mcitedefaultseppunct}\relax
	\EndOfBibitem
	\bibitem[Flenner and Szamel(2005)]{FlennerSzamel:2005}
	E.~Flenner and G.~Szamel, \emph{Phys. Rev. E}, 2005, \textbf{72}, 011205\relax
	\mciteBstWouldAddEndPuncttrue
	\mciteSetBstMidEndSepPunct{\mcitedefaultmidpunct}
	{\mcitedefaultendpunct}{\mcitedefaultseppunct}\relax
	\EndOfBibitem
	\bibitem[Saltzman and Schweizer(2006)]{Salzmann:2006}
	E.~J. Saltzman and K.~S. Schweizer, \emph{Phys. Rev. E}, 2006, \textbf{74},
	061501\relax
	\mciteBstWouldAddEndPuncttrue
	\mciteSetBstMidEndSepPunct{\mcitedefaultmidpunct}
	{\mcitedefaultendpunct}{\mcitedefaultseppunct}\relax
	\EndOfBibitem
	\bibitem[Wang \emph{et~al.}(2009)Wang, Anthony, Bae, and Granick]{wang:2009}
	B.~Wang, S.~M. Anthony, S.~C. Bae and S.~Granick, \emph{Proc. Natl. Acad.
		Sci. U.S.A.}, 2009, \textbf{106}, 15160--15164\relax
	\mciteBstWouldAddEndPuncttrue
	\mciteSetBstMidEndSepPunct{\mcitedefaultmidpunct}
	{\mcitedefaultendpunct}{\mcitedefaultseppunct}\relax
	\EndOfBibitem
	\bibitem[Barkai and Burov(2020)]{Barkai:2020}
	E.~Barkai and S.~Burov, \emph{Phys. Rev. Lett.}, 2020, \textbf{124},
	060603\relax
	\mciteBstWouldAddEndPuncttrue
	\mciteSetBstMidEndSepPunct{\mcitedefaultmidpunct}
	{\mcitedefaultendpunct}{\mcitedefaultseppunct}\relax
	\EndOfBibitem
	\bibitem[Rusciano \emph{et~al.}(2022)Rusciano, Pastore, and
	Greco]{Pastore:2022}
	F.~Rusciano, R.~Pastore and F.~Greco, \emph{Phys. Rev. Lett.}, 2022,
	\textbf{128}, 168001\relax
	\mciteBstWouldAddEndPuncttrue
	\mciteSetBstMidEndSepPunct{\mcitedefaultmidpunct}
	{\mcitedefaultendpunct}{\mcitedefaultseppunct}\relax
	\EndOfBibitem
	\bibitem[Metzler(2020)]{Metzler:2020}
	R.~Metzler, \emph{Eur. Phys. J. Spec. Top.}, 2020, \textbf{229},
	711--728\relax
	\mciteBstWouldAddEndPuncttrue
	\mciteSetBstMidEndSepPunct{\mcitedefaultmidpunct}
	{\mcitedefaultendpunct}{\mcitedefaultseppunct}\relax
	\EndOfBibitem
	\bibitem[Chubynsky and Slater(2014)]{Chubynsky:2014}
	M.~V. Chubynsky and G.~W. Slater, \emph{Phys. Rev. Lett.}, 2014, \textbf{113},
	098302\relax
	\mciteBstWouldAddEndPuncttrue
	\mciteSetBstMidEndSepPunct{\mcitedefaultmidpunct}
	{\mcitedefaultendpunct}{\mcitedefaultseppunct}\relax
	\EndOfBibitem
	\bibitem[Chechkin \emph{et~al.}(2017)Chechkin, Seno, Metzler, and
	Sokolov]{Chechkin:2017}
	A.~V. Chechkin, F.~Seno, R.~Metzler and I.~M. Sokolov, \emph{Phys. Rev. X},
	2017, \textbf{7}, 021002\relax
	\mciteBstWouldAddEndPuncttrue
	\mciteSetBstMidEndSepPunct{\mcitedefaultmidpunct}
	{\mcitedefaultendpunct}{\mcitedefaultseppunct}\relax
	\EndOfBibitem
	\bibitem[Riest and Nägele(2015)]{riest2015salr}
	J.~Riest and G.~Nägele, \emph{Soft Matter}, 2015, \textbf{11},
	9273--9280\relax
	\mciteBstWouldAddEndPuncttrue
	\mciteSetBstMidEndSepPunct{\mcitedefaultmidpunct}
	{\mcitedefaultendpunct}{\mcitedefaultseppunct}\relax
	\EndOfBibitem
	\bibitem[Riest \emph{et~al.}(2018)Riest, Nägele, Liu, Wagner, and
	Godfrin]{ries:2018}
	J.~Riest, G.~Nägele, Y.~Liu, N.~J. Wagner and P.~D. Godfrin, \emph{J. Chem.
		Phys.}, 2018, \textbf{148}, 065101\relax
	\mciteBstWouldAddEndPuncttrue
	\mciteSetBstMidEndSepPunct{\mcitedefaultmidpunct}
	{\mcitedefaultendpunct}{\mcitedefaultseppunct}\relax
	\EndOfBibitem
	\bibitem[{J\"{u}lich Supercomputing Centre}(2018)]{jureca}
	{J\"{u}lich Supercomputing Centre}, \emph{JLSRF}, 2018, \textbf{4}, A132\relax
	\mciteBstWouldAddEndPuncttrue
	\mciteSetBstMidEndSepPunct{\mcitedefaultmidpunct}
	{\mcitedefaultendpunct}{\mcitedefaultseppunct}\relax
	\EndOfBibitem
\end{mcitethebibliography}
\bibliographystyle{rsc} 
\providecommand*{\mcitethebibliography}{\thebibliography}
\csname @ifundefined\endcsname{endmcitethebibliography}
{\let\endmcitethebibliography\endthebibliography}{}

\end{document}